\documentclass[a4paper,11pt]{article}
\pdfoutput=1 % if your are submitting a pdflatex (i.e. if you have
\usepackage{jcappub} % for details on the use of the package, please
\usepackage[T1]{fontenc}
\usepackage[skip=10pt plus1pt, indent=40pt]{parskip}
\usepackage[utf8]{inputenc}
\usepackage{amssymb,amsmath,amsfonts}%,amsart}
\usepackage{bm}
\usepackage{longtable}
\usepackage{verbatim}
\usepackage{color}
\usepackage{mdframed}
\usepackage{soul}
\usepackage{mathrsfs,esint}
\usepackage{xcolor}
\usepackage{latexsym}
\usepackage{booktabs}
\usepackage{graphicx}
\usepackage{slashed, cancel}
\usepackage{caption}
\usepackage{subcaption}
\usepackage{framed}
\usepackage{mdframed}
\usepackage{datetime}
\newdateformat{mydate}{\THEDAY{ }\monthname[\THEMONTH]{ }\THEYEAR}
\usepackage{tikz}
\usepackage{breqn}
\usepackage{bm}
\usepackage{simplewick} % For Wick contractions
\allowdisplaybreaks
\usepackage{calc}
\usepackage[titletoc,title]{appendix}
\newcommand{\vardbtilde}[1]{\tilde{\raisebox{0pt}[0.85\height]{$\tilde{#1}$}}}
\graphicspath{ {./images/} }
\usepackage{hyperref}
\hypersetup{colorlinks, citecolor=bluscuro, linkcolor=black, urlcolor=bluscuro}
\definecolor{rossos}{cmyk}{0,1,1,0.55}
\definecolor{bluscuro}{rgb}{0.15, 0.2, .85}
\definecolor{bluchiaro}{cmyk}{1,.3,0.,0.1}

\newcounter{lastnote}

\newcommand{\be}{\begin{equation}}
\newcommand{\ee}{\end{equation}}
\newcommand{\bea}{\begin{eqnarray}}
\newcommand{\eea}{\end{eqnarray}}
\newcommand{\beq}{\begin{equation}}
\newcommand{\eeq}{\end{equation}}
\def\beqa{\begin{eqnarray}}

\def\eeqa{\end{eqnarray}}

\def\lsim{\mathrel{\rlap{\lower4pt\hbox{\hskip0.5pt$\sim$}}
   \raise1pt\hbox{$<$}}}         %less than or approx. symbol
\def\gsim{\mathrel{\rlap{\lower4pt\hbox{\hskip0.5pt$\sim$}}
   \raise1pt\hbox{$>$}}}         %greater than or approx. symbol

\title{\boldmath An Analytical Study of the Primordial Gravitational-Wave-Induced Contribution to the Large-Scale Structure of the Universe}

\author[1,2]{Pritha Bari}
\author[1,2,3]{Daniele Bertacca}
\author[1,2,3]{Nicola Bartolo}
\author[1,2] {Angelo Ricciardone}
\author[4,5]{Serena Giardiello}
\author[1,2,3,6]{Sabino Matarrese}

\affiliation[1]{\it Dipartimento di Fisica e Astronomia ``G. Galilei",
Universit\`a degli Studi di Padova, via Marzolo 8, I-35131 Padova, Italy}
\affiliation[2]{\it INFN, Sezione di Padova,
via Marzolo 8, I-35131 Padova, Italy}
\affiliation[3]{{\it INAF - Osservatorio Astronomico di Padova, I-35122 Padova, Italy}}
\affiliation[4]{\it Dipartimento di Fisica e Scienze della Terra, Università di Ferrara, Polo Scientifico e Tecnologico - Edificio C Via Saragat, 1, I-44122, Ferrara, Italy}
\affiliation[5]{\it Istituto Nazionale di Fisica Nucleare, sezione di Ferrara, Polo Scientifico e Tecnologico - Edificio C Via Saragat, 1, I-44122, Ferrara, Italy}
\affiliation[6]{\it Gran Sasso Science Institute, I-67100 L’Aquila, Italy}
\emailAdd{pbari@pd.infn.it}
\emailAdd{daniele.bertacca@pd.infn.it}
\emailAdd{nicola.bartolo@pd.infn.it}
\emailAdd{angelo.ricciardone@pd.infn.it}
\emailAdd{serena.giardiello@unife.it}
\emailAdd{sabino.matarrese@pd.infn.it}

\abstract{ The imprint of gravitational waves (GWs) on large-scale structures (LSS) is a useful and promising way to detect or to constrain them. Tensor fossils have been largely studied in the literature as an indirect way to detect primordial GWs. In this paper we analyze a new effect induced by primordial GWs: a correction to the density contrast of the underlying matter distribution of LSS, as well as its radiation counterpart, induced by the energy density fluctuation of the gravitational radiation. We perform our derivation of the full analytical solution of the density contrast for waves entering the horizon during radiation dominance. We account for two phases in the radiation era, depending on the main contributor to the perturbed energy density of the Universe. By comparing the density contrast of cold dark matter and radiation --
  sourced by linear gravitational waves only -- we conclude that the former overcomes the latter at some time in the radiation era, a behaviour analogous to their linear counterpart. Then we conclude by discussing the case of density perturbations produced by GWs entering the Hubble radius during the matter era as well as their evolution in the late dark-energy dominated phase.}
\makeatletter
\gdef\@fpheader{}
\makeatother

\begin{document}
\maketitle
\flushbottom
\section{Introduction}

 The recent groundbreaking detection of gravitational waves (GWs) \cite{LIGOScientific:2016aoc} gave a boost to the observational search for the same, followed by an inevitable push on the theoretical side of research directions related to the study of GWs.
Apart from the resolved astrophysical sources like compact binaries (e.g. see \cite{Buonanno:2014aza} and references therein), there are two more expected contribution: the stochastic background of astrophysical GWs, arising from the coherent superposition of GWs from unresolved astrophysical sources \cite{Ferrari:1998jf,Phinney:2001di,Regimbau:2011rp}, and the background of cosmological GWs, produced via some early Universe phenomena, such as e.g. inflation \cite{Guzzetti:2016mkm}, phase transitions \cite{Kamionkowski:1993fg} etc. A cosmological stochastic GW background is a potentially observable smoking gun of inflation \cite{Guth:1980zm,Starobinsky:1980te,Lyth:1998xn}, 
which is so far the most successful theory to explain the origin of most cosmological observables \cite{Planck:2018nkj,Planck:2018vyg}. 
Due to their feeble interaction with all matter components, GWs from inflation carry pristine information about the early Universe below the Planck scale, unreachable by any other means. Searching for the  primordial GWs background has become a major focus in cosmology, for references see
 \cite{Maggiore:1999vm,Guzzetti:2016mkm,Watanabe:2006qe,Sakamoto:2022nth, CMB-S4:2020lpa,Campeti:2020xwn,Flauger:2020qyi} and the references therein. At the moment, a tight constraint on the tensor-to-scalar ratio $r$ ($<0.032$) has been put on their amplitude on CMB scales through the joint observation of Planck, BICEP2/Keck, and  WMAP \cite{Tristram:2021tvh}. Adding LIGO-Virgo-KAGRA data to those obtained from CMB scales, a tighter constraint ($r<0.028$) has been obtained recently \cite{Galloni:2022mok}.
 The next-generation ground-based CMB experiment CMB-S4 aims to optimize the constraint on $r\,(<0.001)$ at $95\%$ Confidence Level (CL) \cite{CMB-S4:2020lpa}. Before CMB-S4, LiteBIRD \cite{LiteBIRD:2022cnt} and Simons Observatory \cite{SimonsObservatory:2018koc} plan to set an upper limit $r<0.002$ and  $r<0.01$ respectively.

Apart from CMB and interferometer experiments, recently primordial GWs have also been sought in theoretical studies through their imprint on large-scale-structures (LSS). For example, long-wavelength tensor perturbations (tensor fossils) are believed to induce local quadrupolar anisotropic signatures in the  otherwise statistically isotropic two-point correlation function of the mass distribution or the galaxies through scalar-scalar-tensor interaction \cite{Masui:2010cz, Jeong:2012df,Dai:2013kra,Dimastrogiovanni:2014ina,Dimastrogiovanni:2019bfl}.
GWs can also have projection effects due to the perturbation of the space-time on the galaxy distribution \cite{Schmidt:2012ne}. The presence of GWs perturbs the photon geodesics,  
and hence the observed angular positions and redshifts of the galaxies, which in turn modifies the observed galaxy density (e.g., see \cite{Jeong:2012nu}).
Perturbed photon geodesics also modify the observed flux of a given source, inducing additional fluctuations in the galaxy density through magnification bias \cite{Jeong:2012nu}. 
These projection effects, along with the intrinsic alignment (alignment of galaxy orientation with large-scale tidal field) induced by the tidal effect of GWs \cite{Schmidt:2013gwa} lead to a correlation of galaxy ellipticities.  Finally, only GWs (and not scalar modes at linear order) contribute to the parity-odd B-mode component, and thus
acts as a probe to search for gravitational waves \cite{Dodelson:2003bv,Dodelson:2010qu,Schmidt:2012nw,Schmidt:2012ne}.

Another effect of  GWs on LSS was analysed in a recent preceding paper \cite{PhysRevLett.129.091301}: GWs, produced in the early Universe, can source matter perturbation upon re-entering the horizon, which are statistically independent from the linear matter perturbation, and can give rise to observable effects in the matter power spectrum. In particular, \cite{PhysRevLett.129.091301}  showed that GW energy density fluctuations  generate an additional correction to the matter density contrast. The possibility of constraining $r$ through an accurate observation of the scalar modes was also pointed out in \cite{Nakamura:2008db}.
This mechanism
was first proposed and analyzed in  \cite{1967PThPh..37..831T,1972PThPh..47..416T,Matarrese_1998}
and can be considered the opposite effect to that in which  gravitational waves are induced by linear scalars:
large amplitude scalar perturbations, upon entering the horizon, source GWs, and as  the scalar perturbations are the most dominant ones at the first order, one can expect an observable GWs background if the source scalar perturbations are enhanced. This approach has been studied in detail over the years, e.g. see \cite{Matarrese_1998,Matarrese:1996pp,Bartolo_2010,Nakamura:2004rm,Baumann_2007,Inomata_2020,Yuan_2020, Ananda_2007,Kohri:2018awv,Domenech:2020kqm,Espinosa_2018,Saito_2009, 2021Univ....7..398D}. 
Finally, it is worth mentioning that the scalar induced GWs 
can be used to probe the primordial black holes \cite{Bartolo:2019zvb,Bartolo:2018evs,Saito:2008jc,Garcia-Bellido:2017aan}.

In \cite{PhysRevLett.129.091301}, the treatment was limited to the matter dominated era, i.e. only to the scales which entered the horizon after matter-radiation equality, and included a correction  considering late-time dark energy dominance.
 Due to the fact that these modes are statistically independent of standard adiabatic density perturbations, they can be studied separately. However, to understand and interpret the effect properly, it is necessary to extend the study to smaller scales. 
In this paper, we consider the same  effect in the radiation domination era, taking into account the density perturbation modes entering the horizon since the end of inflation to the matter epoch.  
Finally, we show the expression of tensor-induced CDM and radiation contrast modes entering the Hubble radius during the era of matter and their evolution in the late phase dominated by dark energy. It is important to note that our previous paper \cite{PhysRevLett.129.091301} focused on the matter power spectrum of tensor-induced-scalar modes during the matter-dominated era. In contrast, the current study specifically examines the sub-horizon evolution of these modes during the radiation-dominated and dark energy-dominated epoch and does not address the matter power spectrum issue.

 It is to be noted that \cite{Wang:2019zhj} discussed the second-order perturbations in synchronous gauge for the scalar-tensor and tensor-tensor couplings for radiation domination, but their study assumed that in the whole radiation regime, radiation is the main component in both background and perturbation. The same limitation can be observed in \cite{Doring:2021gue}, which studied the contribution to CDM density contrast sourced by tensor perturbations produced via a phase transition in radiation domination. In this paper we provide a complete solution of second-order density contrast, sourced by only linear GWs, taking into account the whole radiation epoch, up to matter-radiation equality. Then we proceed to study the phenomenon in late times, when the contribution of dark energy to the background energy density has grown to be significant. Our study is complete and fully analytical, leading way to a future numerical treatment of the problem.
 
 The correction to the density contrast sourced by GWs can be an indirect probe of GWs, and in the case of non-detection, it can help to constrain the amplitude of the same. Future LSS surveys
such as Euclid \cite{Euclid:2019clj}, DESI \cite{DESI:2018ymu}, SPHEREx \cite{2016arXiv160607039D}, SKA \cite{SKA:2018ckk}, Roman Space Telescope \cite{2021arXiv211103081R} and Vera Rubin Observatory (LSST)\cite{2020arXiv200907653V} are extremely good candidates for this purpose.

We would like to address an additional point that has come up during the course of our research. Although not directly related to our main research focus, we understand the value of briefly discussing this topic to provide a comprehensive perspective. The question pertains to whether our tensor-induced-scalar has a non-zero correlation with the linear scalars. There might be a confusion regarding Maldacena's work \cite{Maldacena:2002vr}, (see also \cite{Gangui:1993tt,Acquaviva:2002ud}), which shows both a non-vanishing scalar-tensor-tensor and tensor-scalar-scalar bispectrum. However, it is important to note that Maldacena's work was conducted within an interaction picture in the  single-field inflation, leading to non-Gaussian initial conditions, and the scalar-tensor-tensor correlation corresponds to a tensor four-point function, in the language of the standard perturbation theory. In \cite{PhysRevLett.129.091301}, we had considered Gaussian initial perturbations, which is a very standard approach in perturbation theory. Consequently, we found that the correlation between our tensor-induced-scalar and linear scalar perturbations is zero, similar to the well-studied case of scalar-induced scalars at second order, which exhibit no correlation with linear tensors; similarly, Maldacena's mixed bispectra are not included in various studies on the non-linear gravitational evolution of scalar perturbations (e.g. in the framework of the Effective Field Theory of LSS). Here, however, we do not need to care about Gaussian/non-Gaussian initial conditions, as we are not concerned about the correlations.
 
\begin{figure}
    \includegraphics[width=16cm]{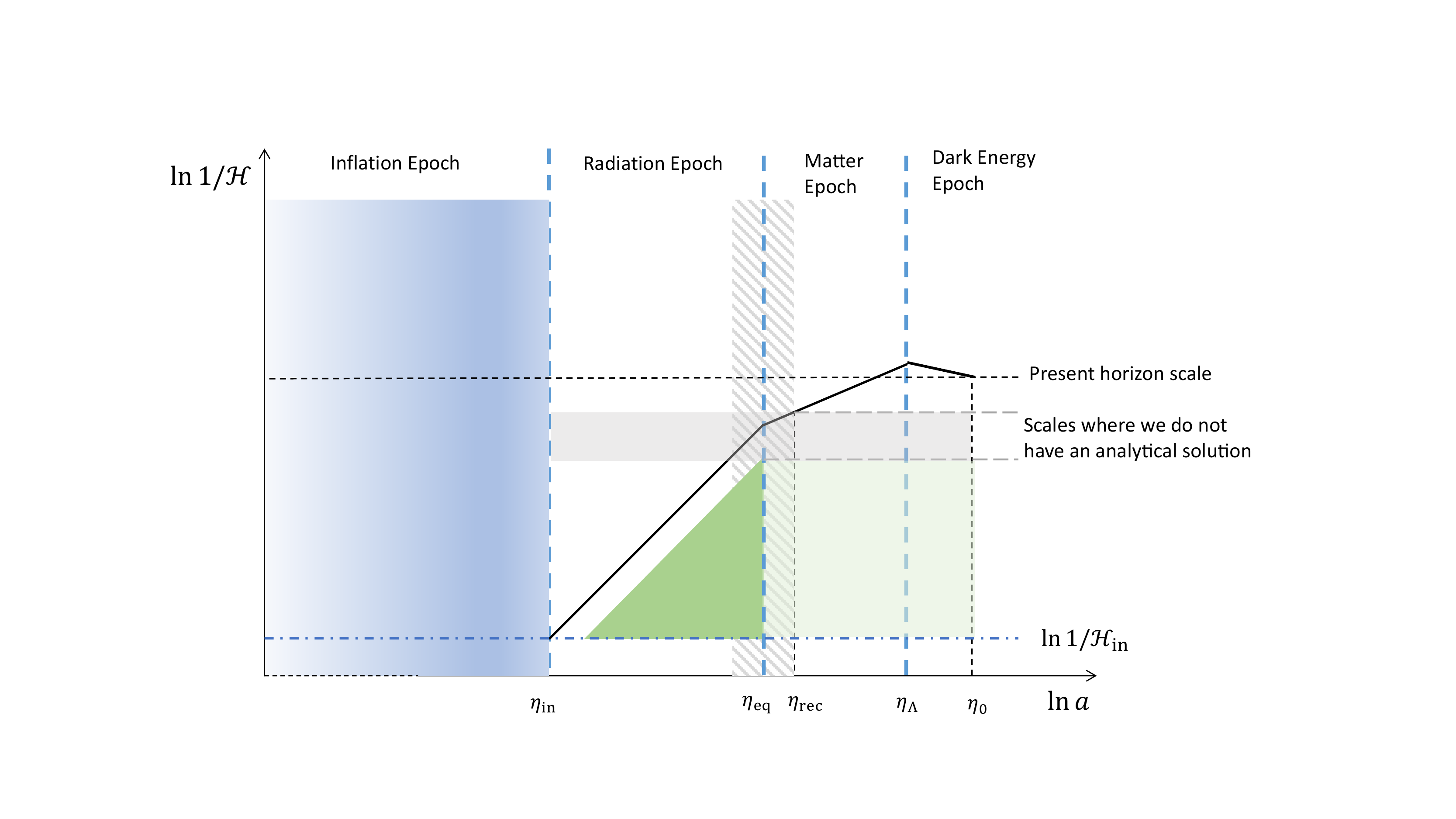}
    \caption{We show a plot of $\ln{1/\cal H}$ versus $\ln a$ in the different epochs analyzed in this work, separated by a blue vertical dashed line.
 The solid black curve indicates the evolution of the modes that cross the horizon from the end of inflation until today, the black dashed line shows the extrapolation of the present horizon scale.
The green region highlights sub-horizon scales during the radiation epoch in which Einstein’s field equations are governed by the matter perturbations generated only by linear gravitational waves (discussed in Section \ref{subrd}). 
The area above the green region indicates all modes during the deep radiation epoch (for details see Section \ref{purerd}).
The light-green area describes these sub-horizon scales, related to the  green region described above, during matter and dark energy epochs (discussed in Section \ref{matt}). 
The gray shaded area denotes those scales where we do not have an analytical solution. The  dashed-dot blue line shows the horizon scale at the end of inflation.} 
    \label{Figure}
\end{figure}

The paper is structured as follows: In section \ref{Section2}, we define the perturbations, and discuss the tensor-sourced scalar perturbations for a Universe where radiation and cold dark matter (CDM) both are present. In section \ref{purerd}, evolution in a deep radiation dominated regime is discussed. We follow the evolution towards matter-radiation equality in section \ref{subrd}, and the full solution in the end of radiation era is presented in \ref{match}.
Section \ref{matt} is dedicated to the study of perturbations produced by gravitational waves entering the Hubble radius during the matter era as well as their evolution in the late Dark Energy dominated phase.
Finally, we summarize in \ref{disc}. In Fig. \ref{Figure} we  graphically show all the epochs and scales studied and analyzed in the paper.

\section{Tensor-sourced scalar perturbations} 
\label{Section2}

\subsection{Perturbations in the metric and matter components} 
We begin by introducing the notation and conventions used for metric and matter perturbations. We consider a flat Friedmann-Lema\^itre-Robertson-Walker (FLRW) space-time, which is described by the metric: $d s^2= a^2(\eta)\left[-d\eta^2+d x^2\right]$, where $\eta$ is the conformal time, and $a(\eta)$ the scale factor. Here we assume that  $c=\hslash=1$ throughout this paper.

In the previous work \cite{PhysRevLett.129.091301}, we focused only on a Universe dominated by cold dark matter (CDM) and a cosmological constant.
The absence of a pressure gradient in the matter sector allowed us to apply directly a synchronous, time-orthogonal and comoving (with CDM) gauge, e.g. see also \cite{Matarrese_1998}. 
Here, due to the presence of the contribution of radiation, in principle, we need to be more general.
Hence, in this paper, we chose to start  with comoving and time-orthogonal gauge with CDM
by choosing $\delta g_{0i}$ to be zero and doing the calculation in the rest frame of CDM.  However, as is shown later, in our specific choice of perturbations, which depends only on the linear tensor contribution, our gauge becomes synchronous again.

In this  time-orthogonal  gauge,
a perturbed flat FLRW metric becomes \cite{Kodama:1985bj}
\begin{equation}\label{line}
    d s^2= a^2(\eta)\left[-\big(1+2\psi\big) d\eta^2+\gamma_{ij}\big(\bm{x},\eta\big)d x^i d x^j\right],
\end{equation}
 where the spatial  metric $\gamma_{ij}$ contains  second order scalar and linear tensor modes. Here we ignore the linear scalar and vector modes 
 because they are statistically independent with tensor-sourced scalars modes, and we can in principle set  them to zero by hand. At second order we have scalar, vector and tensor contributions
  whose  governing equations have source terms 
  quadratic in the first order perturbations with respect to scalars, 
 mixing linear scalars and tensors (i.e., ``tensor fossils'',  tensor-induced
vector and tensor modes \cite{Matarrese_1998,Jeong:2012df}) and, finally, scalar modes originating from linear tensors. The last source term is the contribution we are interested in and will consider in the main text of the paper (see also \cite{Zhang:2017hiu,Carrilho_2016}). (For a focused study of the second order vector and tensor contributions see Appendix \ref{Appendix-pert}.)
Since the tensor-sourced-scalar modes are statistically independent of linear scalar modes, we are allowed to deal with them separately.  In the following, the decomposition of the metric components is shown \footnote{In general, for any perturbation X and Y,  $X=X^{(2)}/2$, and $XY=X^{(1)}Y^{(1)}$, 
as we are considering perturbative terms up to second order. }
\begin{align}\label{metdec}
    \psi&= \frac{\psi^{(2)}}{2},\\
    \begin{split}
    \gamma_{ij}&= \delta_{ij}+ \gamma_{ij}^{(1)}+\frac{ \gamma_{ij}^{(2)}}{2},\\
    &=  \delta_{ij}+ \chi_{ij}^{(1)}+\frac{1}{2}\big(-2\phi^{(2)}\delta_{ij}+\chi^{(2)}_{ij}\big),\\
    &=  \delta_{ij}+ \chi_{ij}^{(1)}-\phi^{(2)}\delta_{ij}+\frac{1}{2} D_{ij}\chi^{(2)||},
    \end{split}\\
    \gamma^{ij}&= \delta^{ij}- \chi^{ij(1)}+\phi^{(2)}\delta^{ij}-\frac{1}{2} D^{ij}\chi^{(2)||}+\chi^{ik(1)}{\chi_{k}}^{j(1)},
\end{align}
where $\chi_{ij}^{(1)}$, from here on $\chi_{ij}$, is the linear tensor perturbation, and our scalar perturbations (at second order) are $\psi^{(2)}$, $\phi^{(2)}$, and $\chi^{(2)||}$. $D_{ij}$ is defined as $ \partial_i \partial_j - (1/3) \nabla^2 \delta_{ij}$.  In Appendix \ref{Appendix-pert}, briefly, we consider the vector and tensor contributions of the metric and set the equations that allow us to find their dynamics.

In this paper, the matter component of the Universe consists of a mixture of an irrotational dust, with which the observer is comoving,
and the radiation. Note that the perturbations arising in the energy-momentum tensor of the matter components are solely sourced by the contribution linked to the primordial GWs.
As both of them are perfect fluids (here, we are making the assumption that the coupling between baryons and radiation is neglected. We aim to analyse how much significance the effect of this coupling might pose in a future work.),
their energy-momentum tensor is given by: $T_{\mu \nu}=\big(\rho+p\big)u_\mu u_\nu+pg_{\mu \nu} $. Here $\rho_{\rm r}$, $\rho_{\rm m}$, and $p_{\rm r}$  are the energy density of radiation and matter respectively, and pressure of radiation (here we are assuming $p_{\rm m}=0$), and ${u_{\rm r}}^\mu$ and ${u_{\rm m}}^\mu$ are their respective four-velocities, normalised as $u_\mu u^\mu=-1$.  Its components are, for matter
\begin{align}\label{Tdec}
\begin{split}
     {u_{\rm m}}_0&=-a \left(1+\psi\right),\\
    {u_{\rm m}}^0&= {1 \over a}\left(1-\psi\right),\\
    {u_{\rm m}}^i&=0,\\
\end{split}
\end{align}
and for radiation
\begin{align}
\begin{split}
    {u_{\rm r}}_0&=-a\left(1+\psi\right),\\
    {u_{\rm r}}^0&= {1 \over a}\left(1-\psi\right),\\
    {u_{\rm r}}_i&=a\, {v_{\rm r }}_i =a\, {v_{\rm r}}_{,i},
    \\{u_{\rm r}}^i&=\frac{1}{a}\,{v_{\rm r}}^i  =\frac{1}{a}\,{v_{\rm r}}^{,i}\,.
\end{split}
\end{align}
Here  $_{, i}=\partial_i$ is used to indicate a derivative w.r.t. $x^i$, and $v_{\rm r}$ is the velocity potential of radiation. 

In the next sections we discuss the conservation equation of radiation and CDM. Then, we analyse Einstein's field equations which allow to study these contributions both during the radiation and CDM dominated epochs of the Universe. Precisely, using the characteristic scale $k_{\rm eq} \sim 1/(100 {\rm Mpc})$ defined by the comoving size of the Hubble horizon at matter-radiation equality,
our treatment of modes with $k>k_{\rm eq}$ is divided into two phases: i) start with the modes entering the horizon at very early times, at the beginning of radiation domination ii) follow their subhorizon evolution where they travel through matter-radiation equality towards matter domination.
The large scale modes ($k<k_{\rm eq}$), which enter the horizon during matter domination, are already discussed in \cite{PhysRevLett.129.091301}.  However, in order to have a complete picture of primordial GW contribution, in the Section \ref{matt} we analyse and discuss also these solutions.

\subsection{Conservation equations}

As the radiation and CDM components interact only gravitationally,
their energy–momentum tensors satisfy the conservation laws ${T^{\alpha \beta}}_{;\beta}=0$ separately. 
For $\alpha=0$ and $\alpha=i$, we get continuity and momentum conservation equation respectively. The next two subsections will be devoted to the derivation of the contribution of tensor-scalar perturbations within the CDM and radiation component.

\subsubsection{Conservation equation for matter}\label{conmat}

Assuming that the observer is comoving with the CDM component, the  energy-momentum tensor of a pressure-free matter is
\begin{align}
\begin{split}
    T^{00}_{\rm m}&= \rho_{\rm m} u_{\rm m}^0 u_{\rm m}^0=\frac{\overline{\rho}_{\rm m}\left(1+\delta_{\rm m}\right)}{a^2}\left(1-2\psi\right),\\
     T^{ij}_{\rm m}&= T^{0i}_{\rm m}=0\,,
\end{split}
\end{align}
where $\delta_{\rm m}= ({\rho}_{\rm m}-\overline{\rho}_{\rm m})/\overline{\rho}_{\rm m}$ is the density contrast of the matter.
Energy–momentum conservation gives evolution equations for the density contrast at second order
\begin{equation}\label{contim}
   {\delta_{\rm m}^{(2)}}'-\gamma^{(1)ik} {\gamma_{ki}^{(1)}}'+\frac{1}{2}\delta^{ik} {\gamma_{ki}^{(2)}}'=0,
\end{equation}
where $'$ indicates derivative w.r.t. $\eta$.
Considering only tensor contribution at the first order, the evolution second-order density contrast reads
\begin{align}\label{deltam-evol}
    {\delta_{\rm m}^{(2)}}'&= \frac{1}{2}{\Big(\chi^{ij}\chi_{ij}+6\phi^{(2)}\Big)}'\,,
    \end{align}
and, consequently, we have
    \begin{align}\label{deltam0}
     \delta_{\rm m}^{(2)}&= \frac{1}{2}\Big(\chi^{ij}\chi_{ij}-\chi^{ij}_0\chi_{0ij}\Big)+3\left(\phi^{(2)}-\phi_0^{(2)}\right)+\delta_{\rm m 0}^{(2)}\,.
\end{align}
Here the subscript `$0$' denotes the value of the variable at the initial time, i.e. the end of inflation.

Following the analysis made in Appendix \ref{phi0} we can set $\phi_0^{(2)}$ and $\delta_{\rm m 0}^{(2)}$ equal to zero, and we can simply rewrite Eq. \eqref{deltam0} in the following way
\begin{equation}\label{deltam}
    \delta_{\rm m}^{(2)}= \frac{1}{2}\Big(\chi^{ij}\chi_{ij}-\chi^{ij}_0\chi_{0ij}\Big)+3\phi^{(2)}.
\end{equation}
Note that this expression has exactly the same form as that obtained in \cite{PhysRevLett.129.091301}, where the tensor-sourced matter perturbation in the comoving (with CDM) and synchronous gauge, during the epoch of matter domination was studied (see also \cite{Matarrese_1998,Zhang:2017hiu}).

From the momentum conservation for matter 
\begin{equation}
    \partial^i \psi^{(2)} +2\partial^i\psi^{(1)}\delta_{\rm m}^{(1)}-4 \psi^{(1)}\partial^i \psi^{(1)} =0,
\end{equation}
we observe that $\psi^{(2)}$ can only be sourced by the first-order scalar modes. This means that, in our purpose, $\psi$ can be safely ignored.
Then, although started from the time-orthogonal gauge, our system of equations can directly be written in the synchronous gauge.
This is the first result of the paper.
\subsubsection{Conservation equation for radiation}

According to the discussion in the last section, $\psi$ can be ignored from hereon. The components of the energy-momentum tensor of radiation are
\begin{align}
\begin{split}
    T^{00}_{\rm r}&=  \frac{\overline{\rho}_{\rm r}(1+\delta_{\rm r})}{a^2},\\
     T^{0i}_{\rm r}&=\frac{4\overline{\rho}_{\rm r}}{3a^2}{v_{\rm r}}^{,i},\\
  T^{ij}_{\rm r}&=\frac{\overline{\rho}_{\rm r}(1+\delta_{\rm r})}{3a^2}\gamma^{ij}.
\end{split}
\end{align}
From the continuity equation we have
\begin{equation}\label{radc}
   {\delta_{\rm r}^{(2)}}'-\frac{4}{3}\chi^{(1)ij}{\chi_{ij}^{(1)}}'+\frac{4}{3}\nabla^2 v_{\rm r}^{(2)}-4{\phi^{(2)}}'=0.
\end{equation}
Whereas the momentum conservation equation gives
\begin{equation}\label{momr}
     4 {v_{\rm r}^{(2)}}'+ \delta_{\rm r}^{(2)}=0.
\end{equation}
\subsection{Einstein equations}\label{radmatt}

After the end of inflation, even though radiation dominates the energy density of the background, as it decays faster than CDM, it is toppled by the latter as the main contributor of the energy density of the Universe at the matter-radiation equality. As a result, towards the end of radiation domination, $\overline{\rho}_{\rm m}$ can not be ignored anymore. 

Then, there is another aspect in this study that should not be overlooked.
During the end of the radiation era, as in the linear case, it is possible that the perturbative contribution of the CDM component could be greater than that of the radiation. 
In this work we will also probe this possibility and accurately analyze the trend of each component
 both during the radiation epoch and during the matter-radiation equality.
 
In this section, we mainly focus on tensor-sourced CDM perturbation $\delta_{\rm m}^{(2)}$  evolution in presence of a perturbed radiation component, considering adiabatic perturbations only. 
Using the metric \eqref{metdec} and stress-energy tensor decomposition \eqref{Tdec}, and keeping in mind the discussion in \ref{conmat} that $\psi$ in our case effectively vanishes, we have the second order Einstein equations. $00$-th, $0i$-th, and $ij$-th 
Einstein equations become, respectively
\begin{align}\label{all00}
&\nabla^2\phi^{(2)}+\frac{1}{2} \chi^{ij}\nabla^2\chi_{ij}-3\mathcal{H}{\phi^{(2)}}'+\frac{1}{6}\nabla^2 \nabla^2 \chi^{||(2)}-\frac{1}{8}{\chi^{ij}}'{\chi}'_{ij}-\mathcal{H}\chi^{ij}{\chi}'_{ij}+\frac{3}{8}\chi^{kl,i}\chi_{kl,i}\nonumber\\&-\frac{1}{4}\chi^{ik,l}\chi_{li,k}=4\pi G a^2 (\overline{\rho}_{\rm m} \delta_{\rm m}^{(2)}+\overline{\rho}_{\rm r}\delta_{\rm r}^{(2)}),
        \end{align}        
\begin{equation}\label{all0i}
{\phi^{(2)}}'_{,i}-\frac{1}{2}\chi^{jk}{\chi}'_{ki,j}+\frac{1}{4}D_{ij}{\chi^{||(2),j}}'+\frac{1}{2}\chi^{jk}{\chi}'_{jk,i}+\frac{1}{4}{\chi^{jk}}'\chi_{jk,i}=-\frac{16\pi G a^2}{3} \overline{\rho}_{\rm r} v^{(2)}_{{\rm r},i},
\end{equation}
\begin{align}\label{allij}
&\frac{1}{4}D_{ij}{\chi^{||(2)}}''+\frac{\mathcal{H}}{2}D_{ij}{\chi^{||(2)}}'+\frac{1}{12}\nabla^2 D_{ij}\chi^{||(2)}-\frac{1}{18}\nabla^2 \nabla^2 \chi^{||(2)}\delta_{ij}\nonumber\\&+2\mathcal{H}{\phi^{(2)}}'\delta_{ij}+{\phi^{(2)}}''\delta_{ij}+\frac{1}{2}D_{ij}\phi^{(2)}-\frac{1}{3}\nabla^2 \phi^{(2)}\delta_{ij}
    -\frac{1}{2}\chi^{kl}(\chi_{lj,ik} +\chi_{il,jk}-\chi_{ij,lk}-\chi_{kl,ij})\nonumber\\& +\frac{1}{4}{\chi^{kl}}_{,j}\chi_{kl,i}-\frac{1}{2}\chi^{jk,l}\chi_{li,k}+\frac{1}{2}\chi^{jk,l}\chi_{ki,l}-\frac{3}{8}\chi^{kl,p}\chi_{kl,p}\delta_{ij}+\frac{1}{4}\chi^{kp,l}\chi_{lp,k}\delta_{ij}\nonumber\\&-\frac{1}{2}{\chi^k_j}'{\chi_{ki}}'+\frac{3}{8}{\chi^{kl}}'{\chi_{kl}}'\delta_{ij}=\frac{4 \pi G\overline{\rho}_{\rm r} a^2}{3} \delta_{\rm r}^{(2)}\delta_{ij}\,.
\end{align}
In the Einstein equations, we have also used the evolution equation of the linear GWs
\begin{align}\label{chi_ij-1}
    {\chi_{ij}}''+2\mathcal{H}{\chi_{ij}}'- \nabla^2 \chi_{ij}=0\,.
\end{align}
Let us stress again that here we are ignoring any effect related to the anisotropic stress tensor.
  Decomposing (\ref{allij}) into a trace equation and a trace-less one, the trace part becomes
 \begin{align}\label{allijtr}
& {\phi^{(2)}}''+ 2\mathcal{H}{\phi^{(2)}}'-\frac{1}{3}\nabla^2\phi^{(2)}-\frac{1}{18}\nabla^2 \nabla^2\chi^{||(2)} -\frac{1}{8}\chi^{kl,i}\chi_{kl,i}+\frac{1}{12}\chi^{ik,l}\chi_{li,k} \nonumber\\& +\frac{5}{24}{\chi^{kl}}'{\chi_{kl}}'+\frac{1}{6}\chi^{kl}\nabla^2\chi_{kl}=\frac{4 \pi G\overline{\rho}_{\rm r} a^2}{3} \delta_{\rm r}^{(2)}.
        \end{align} 
In the next sections, we split the treatment in two regimes, first, immediately after inflation, the deep radiation one, and second, when the Universe evolves towards matter-radiation equality. As we will discuss, these two regimes have different dynamics, depending on the dominant contributor to the background matter component as well as to the perturbation content. According to what is found in \cite{PhysRevLett.129.091301}, we are only interested in the sub-horizon evolution, as the GWs radiation sourcing our second order perturbations exists only there.

We note that \cite{Wang:2019zhj} discusses tensor-sourced scalars in radiation domination without the sub-horizon assumption, and it only focuses on the first of the two phases stated above. We will show below a full solution of density contrast, for modes evolving (always in the sub-horizon) throughout radiation domination, comprising of the contributions from both the phases.
However let us emphasise that, for completeness, in Appendix \ref{vrd}, we have redone the general analysis for the deep radiation-dominated Universe for all scales.
\section{Einstein equations in the deep radiation-dominated Universe} \label{purerd}

Although there are two components of stress-energy tensor (radiation and CDM) in radiation era, in the very early stages of radiation domination, the ratio of energy density of radiation to that of CDM component is too high, and the Einstein equations have only radiation (and not CDM) perturbations (sourced by tensors) on the matter side. The continuity equation (\ref{deltam}) can be used to retrieve CDM perturbation from the potential, obtained as a solution of Einstein equations. 

In this phase of evolution, discarding the CDM perturbation, \eqref{all00}, \eqref{all0i}, \eqref{allij} become
 \begin{align}\label{00th}
           &\nabla^2\phi^{(2)}+\frac{1}{2} \chi^{ij}\nabla^2\chi_{ij}-3\mathcal{H}{\phi^{(2)}}'+\frac{1}{6}\nabla^2 \nabla^2 \chi^{||(2)}-\frac{1}{8}{\chi^{ij}}'{\chi_{ij}}'-\mathcal{H}\chi^{ij}{\chi_{ij}}'+\frac{3}{8}\chi^{kl,i}\chi_{kl,i}\nonumber\\&-\frac{1}{4}\chi^{ik,l}\chi_{li,k}=\frac{3\mathcal{H}^2}{2}\delta_{\rm r}^{(2)}\,,
        \end{align}
        \begin{equation}\label{0ith}
           {\phi^{(2)}_{,i}}'-\frac{1}{2}\chi^{jk}{\chi_{ki,j}}'+\frac{1}{4}D_{ij}{\chi^{||(2),j}}'+\frac{1}{2}\chi^{jk}{\chi_{jk,i}}'+\frac{1}{4}{\chi^{jk}}'\chi_{jk,i}=-2\mathcal{H}^2 v^{(2)}_{{\rm r},i}\,,
        \end{equation}
        
        \begin{align}\label{ijth}
        & \frac{1}{4}D_{ij}{\chi^{||(2)}}''+\frac{\mathcal{H}}{2}D_{ij}{\chi^{||(2)}}'+\frac{1}{12}\nabla^2 D_{ij}\chi^{||(2)}-\frac{1}{18}\nabla^2 \nabla^2 \chi^{||(2)}\delta_{ij}\nonumber\\&+2\mathcal{H}{\phi^{(2)}}'\delta_{ij}+{\phi^{(2)}}''\delta_{ij}+\frac{1}{2}D_{ij}\phi^{(2)}-\frac{1}{3}\nabla^2 \phi^{(2)}\delta_{ij}
    -\frac{1}{2}\chi^{kl}(\chi_{lj,ik} +\chi_{il,jk}-\chi_{ij,lk}-\chi_{kl,ij})\nonumber\\& +\frac{1}{4}{\chi^{kl}}_{,j}\chi_{kl,i}-\frac{1}{2}\chi^{jk,l}\chi_{li,k}+\frac{1}{2}\chi^{jk,l}\chi_{ki,l}-\frac{3}{8}\chi^{kl,p}\chi_{kl,p}\delta_{ij}+\frac{1}{4}\chi^{kp,l}\chi_{lp,k}\delta_{ij}\nonumber\\&-\frac{1}{2}{\chi^{k}}'_j {\chi_{ki}}'+\frac{3}{8}\chi^{kl'}{\chi_{kl}}'\delta_{ij}=\frac{\mathcal{H}^2}{2} \delta_{\rm r}^{(2)}\delta_{ij}.
        \end{align}
      
       Let us point out that $ \delta_{\rm m}^{(2)} \ll  \delta_{\rm r}^{(2)}$ 
        is not assumed in this period; rather we take on that $\delta^{(2)}\rho_{\rm m}$  is negligible w.r.t. $\delta^{(2)}\rho_{\rm r}$ . This assumption is only valid for the regime discussed in this section. 
        The trace part becomes
 \begin{align}\label{tr}
     & {\phi^{(2)}}''+ 2\mathcal{H}{\phi^{(2)}}'-\frac{1}{3}\nabla^2\phi^{(2)}-\frac{1}{18}\nabla^2 \nabla^2\chi^{||(2)} -\frac{1}{8}\chi^{kl,i}\chi_{kl,i}+\frac{1}{12}\chi^{ik,l}\chi_{li,k}+\frac{5}{24}{\chi^{kl}}' {\chi_{kl}}'\nonumber\\&+\frac{1}{6}\chi^{kl}\nabla^2\chi_{kl}=\frac{\mathcal{H}^2}{2} \delta_{\rm r}^{(2)}.
 \end{align}
 Trace-less part of (\ref{ijth}) gives
 \begin{align}\label{trl}
   &  D_{ij}\phi^{(2)}+\frac{1}{2} D_{ij}{\chi^{||(2)}}''+\mathcal{H}D_{ij}{\chi^{||(2)}}'+ \frac{1}{6}\nabla^2 D_{ij}\chi^{||(2)}\nonumber\\
   & -\chi^{kl}(\chi_{lj,ik} +\chi_{il,jk}-\chi_{ij,lk}-\chi_{kl,ij}) +\frac{1}{2}{\chi^{kl}}_{,j}\chi_{kl,i}-\chi^{jk,l}\chi_{li,k}+\chi^{jk,l}\chi_{ki,l}\nonumber\\
    & -{\chi^{k}_j}' {\chi_{ki}}' -\frac{1}{3}\chi^{kl}\nabla^2\chi_{kl}\delta_{ij}+\frac{1}{3}{\chi^{kl}}'{\chi_{kl}}' \delta_{ij}-\frac{1}{2}\chi^{kl,p}\chi_{kl,p}\delta_{ij}+\frac{1}{3}\chi^{kp,l}\chi_{lp,k}\delta_{ij}=0.
 \end{align}
 Replacing $\nabla^2 \nabla^2\chi^{||(2)}$ from (\ref{tr}) in (\ref{00th}),
 and using (\ref{radc}) and (\ref{momr}), we get a third order differential equation of radiation velocity potential
\begin{equation}\label{vthird}
    {v^{(2)}_{\rm r}}'''+\mathcal{H}  {v^{(2)}_{\rm r}}''-4\mathcal{H}^2 {v^{(2)}_{\rm r}}'-\frac{1}{3}\nabla^2  {v^{(2)}_{\rm r}}'-\frac{\mathcal{H}}{3}\nabla^2 v^{(2)}_{\rm r}+\frac{1}{6}{\chi^{kl}}'{\chi_{kl}}'=0.
\end{equation}
In Fourier space\footnote{Throughout this work, we use two notations for Fourier space representation of a generic variable $X(\bm{x})$: $X_{{\bm k}}$, or $X(\bm{k},\eta)$. Both are equivalent expressions.} the same equation turns out 
\begin{equation}\label{veq}
    {v^{(2)}_{{\rm r}{\bm k}}}'''+\mathcal{H}  {v^{(2)}_{{\rm r} {\bm k}}}''+\Big(\frac{k^2}{3}-4\mathcal{H}^2\Big)
    {v^{(2)}_{{\rm r}\bm{k}}}'+\frac{\mathcal{H}k^2}{3} v^{(2)}_{{\rm r}\bm{k}}= S_{\bm{k}}.
\end{equation}
where $S_{\bm{k}}$ is the Fourier transformation of the source term $-(1/6){\chi^{kl}}'{\chi_{kl}}'$, i.e.
\begin{equation}\label{S_k-eta}
    S_{\bm{k}}(\eta)=-\frac{1}{6}\sum\limits_{\sigma,\sigma'}\int \frac{d^3\bm{q}}{(2\pi)^3} A_{\sigma'} (\bm{q})A_{\sigma} \left(\bm{k}-\bm{q}\right)\epsilon^{\sigma'}_{ij}(\bm{\hat{q}})\epsilon^{\sigma ij}\left(\widehat{\bm{k}-\bm{q}}\right) \mathcal{T}'\left(q,\eta\right)\mathcal{T}'\left(|\bm{k}-\bm{q}|,\eta\right)
\end{equation} 
and the real space tensors have been defined in the following way
\begin{equation}\label{defchiij}
    \chi_{ij}(\bm{x},\eta)= \frac{1}{(2\pi)^3} \int d^3\bm{k} e^{i\bm{k}.\bm{x}} \chi_\sigma (\bm{k}, \eta)\epsilon^\sigma_{ij}(\bm{\hat{k}}).
\end{equation}
Here we are separating the amplitude $ \chi_\sigma (\bm{k}, \eta)$, which includes the time evolution, from the polarisation tensor $\epsilon^\sigma_{ij}(\bm{\hat{k}}) $. Then this amplitude is further split into the transfer function $\mathcal{T}\left(\bm{k},\eta \right)$  and a stochastic zero-mean variable $A_\sigma (\bm{k})$
\begin{equation} \label{def-calT}
    \chi_\sigma (\bm{k}, \eta)= A_\sigma (\bm{k})\mathcal{T}\left(k,\eta\right),
\end{equation}
where $A_\sigma (\bm{k})$ is characterised by the following auto-correlation function,  
\begin{equation}\label{aa}
     \langle A_{\sigma_1}\big(\bm{k}_1\big)\, A_{\sigma_2}\big(\bm{k}_2\big)\rangle= \frac{2^4 \pi^5}{k_1^3} \delta^3\big(\bm{k}_1+\bm{k}_2\big)\, \delta_{\sigma_1 \sigma_2}\, \Delta^2_{\sigma}\big(k_1\big)
\end{equation}
and  $\Delta^2_\sigma(k)$ is the dimension-less power-spectrum for each GW polarisation. The time evolution of the GWs in given by the transfer function \cite{Watanabe:2006qe}
\begin{equation}\label{transfer_function}
     \mathcal{T}\left(k,\eta\right)=
     \begin{cases}
     j_0\left(k\eta\right) & \quad {\rm for~ } \eta <\eta_{\rm eq},~ k>k_{\rm eq}\,,\\
     \frac{\eta_{\rm eq}}{\eta} \left[A_{\rm GW}(k) j_1\left(k\eta\right)+B_{\rm GW}(k) y_1\left(k\eta\right) \right] & \quad {\rm for~ } \eta_\Lambda \gg \eta >\eta_{\rm eq}, ~k>k_{\rm eq}\,,\\
    \frac{3j_1\left(k\eta\right)}{k\eta} & \quad \text{for~any $\eta \ll \eta_\Lambda$},~ k<k_{\rm eq}\,.
     \end{cases}
\end{equation}
Here $\eta_{\rm eq}$ and $k_{\rm eq}$  correspond respectively to the conformal time and wavenumber of the modes entering the horizon at matter-radiation equality, and $A_{\rm GW}(k)$ and $B_{\rm GW}(k)$ are suitable coefficients, obtained by equating the first and the second line of Eq. \eqref{transfer_function} and their first derivatives  at the matter-radiation equality (see the complete derivation in \cite{Watanabe:2006qe}).
Let us point out that $k_{\rm eq}$ has already been defined above. 
Note that, we define   $\eta_\Lambda$ as the conformal time in which density parameter of the matter is equal to that of the cosmological constant/Dark Energy (see also Section \ref{matt}), and in the second and third line of Eq. \eqref{transfer_function} we put ``$\eta \ll \eta_\Lambda$" because these solutions are correct up to the matter epoch\footnote{This point is not important for this section, but it will be relevant for the discussion in Section \ref{matt}.}.

Introducing a new definition $$u^{(2)}_{{\rm r} \bm{k}}={v^{(2)}_{{\rm r} \bm{k}}}'+\mathcal{H}v^{(2)}_{{\rm r} \bm{k}}={(a v^{(2)}_{{\rm r} \bm{k}})'\over a}\,,$$
we have\footnote{In principle, we can introduce another variable $\theta=\sqrt{3}/2a$, and solve the system according to the procedure demonstrated in \cite{mukhanov2005physical}. In that case, we have $\left[\theta^2(u^{(2)}_{{\rm r}\bm{k}}/\theta)'\right]'+c_s^2\theta k^2 u^{(2)}_{{\rm r}\bm{k}} =- \theta S_{\bm{k}}$.}
\begin{equation}\label{ueq}
    {u^{(2)}_{{\rm r} \bm{k}}}''+\Big(\frac{k^2}{3}-2\mathcal{H}^2\Big)u^{(2)}_{{\rm r} \bm{k}}= S_{\bm{k}},
\end{equation}
which reduces \eqref{veq} to a second-order equation.  As we are interested in the regime $k\eta \gg 1$, this further simplifies to\footnote{Without this approximation, the results for tensor-sourced scalar quantities like  potential and density contrast derived in \cite{Wang:2019zhj} can be accurately recovered with the help of the variable $ u^{(2)}_{{\rm r}\bm{k}}$. See Appendix \ref{vrd} for details.}
\begin{equation}\label{uk}
    {u^{(2)}_{{\rm r} \bm{k}}}''+\frac{k^2}{3}u^{(2)}_{{\rm r} \bm{k}}= S_{\bm{k}}.
\end{equation}
The solution to this equation is 
\begin{align}
   u^{(2)}_{{\rm r}\bm{k}}(\eta)&= A(\bm{k}) \cos{\frac{k\eta}{\sqrt{3}}} +B(\bm{k}) \sin{\frac{k\eta}{\sqrt{3}}} +\frac{\sqrt{3}}{k} \sin{\frac{k\eta}{\sqrt{3}}} \int_{\eta_{\rm in}}^\eta d\tilde{\eta}\,\cos{\frac{k\tilde{\eta}}{\sqrt{3}}}\, S_{\bm{k}}(\tilde{\eta})\nonumber \\&-\frac{\sqrt{3}}{k} \cos{\frac{k\eta}{\sqrt{3}}}  \int_{\eta_{\rm in}}^\eta d\tilde{\eta}\, \sin{\frac{k\tilde{\eta}}{\sqrt{3}}} \,S_{\bm{k}}(\tilde{\eta})\;,
\end{align}
%\begin{multline}
 %  u^{(2)}_{{\rm r}\bm{k}}= A(\bm{k}) \cos{\frac{k\eta}{\sqrt{3}}} +B(\bm{k}) \sin{\frac{k\eta}{\sqrt{3}}} +\frac{\sqrt{3}}{k} \sin{\frac{k\eta}{\sqrt{3}}} \int_{\eta_{\rm in}}^\eta d\tilde{\eta}\,\cos{\frac{k\tilde{\eta}}{\sqrt{3}}}\, S_{\bm{k}}(\tilde{\eta}) \\-\frac{\sqrt{3}}{k} \cos{\frac{k\eta}{\sqrt{3}}}  \int_{\eta_{\rm in}}^\eta d\tilde{\eta}\, \sin{\frac{k\tilde{\eta}}{\sqrt{3}}} \,S_{\bm{k}}(\tilde{\eta})\;,
%\end{multline}
where $ A(\bm{k})$ and $ B(\bm{k})$ depend on the initial conditions at $\eta=\eta_{\rm in}$. Here $\eta_{\rm in}$ is
indicating the end of inflation (i.e. the ``last scattering" surface for gravitons).

Assuming initial adiabatic conditions and considering the particular gauge that we have fixed 
(i.e. the synchronous comoving gauge), due to the fact that all modes considered here are yet to enter the horizon at initial time $\eta_{\rm in}$,  we do not have an initial source/contribution to the second order perturbation of GW density. Consequently, we can neglect the homogeneous solutions and obtain the following results
\begin{align}
    v^{(2)}_{{\rm r}\bm{k}}&=\frac{\sqrt{3} }{k a(\eta) } \int_{\eta_{\rm in}}^\eta d\tilde{\eta}\,\left[ a(\tilde{\eta})  \sin{\frac{k\tilde{\eta}}{\sqrt{3}}} \int^{\tilde{\eta}}_{\eta_{\rm in}}d\vardbtilde{\eta} \, \cos{\frac{k\vardbtilde{\eta}}{\sqrt{3}}} \,S_{\bm{k}}(\vardbtilde{\eta})   -a(\tilde{\eta})  \cos{\frac{k\tilde{\eta}}{\sqrt{3}}} \int^{\tilde{\eta}}_{\eta_{\rm in}} d\vardbtilde{\eta} \, \sin{\frac{k\vardbtilde{\eta}}{\sqrt{3}}}\, S_{\bm{k}}(\vardbtilde{\eta}) \right]\\
    %&=- \frac{\sqrt{3} }{6k^4 \eta}\sum\limits_{\sigma,\sigma'}\int \frac{d^3\bm{k}_2}{(2\pi)^3} A_{\sigma'} (\bm{k}_2)A_{\sigma} \left(|\bm{k}-\bm{k}_2|\right)\epsilon^{\sigma'}_{ij}(\bm{\hat{k}}_2)\epsilon^{\sigma ij}\left(\widehat{|\bm{k}-\bm{k}_2|}\right) \\
  % &  \int_{\eta_{\rm in}}^\eta \left(k\tilde{\eta}\right) \left(kd\tilde{\eta}\right)\left[\int^{\tilde{\eta}}_{\eta_{\rm in}}       kd\vardbtilde{\eta}\left( \sin{\frac{k\tilde{\eta}}{\sqrt{3}}}  \cos{\frac{k\vardbtilde{\eta}}{\sqrt{3}}}  - \cos{\frac{k\tilde{\eta}}{\sqrt{3}}} \sin{\frac{k\vardbtilde{\eta}}{\sqrt{3}}} \right)\mathcal{T}'\left(k_2\vardbtilde{\eta}\right)\mathcal{T}'\left(|\bm{k}-\bm{k}_2|\vardbtilde{\eta}\right) \right].
  &= \frac{\sqrt{3} }{k^3 \tau}
  \int_{\tau_{\rm in}}^\tau \tilde{\tau} d\tilde{\tau}\left[\int^{\tilde{\tau}}_{\tau_{\rm in}}       d\vardbtilde{\tau}\left( \sin{\frac{\tilde{\tau}}{\sqrt{3}}}  \cos{\frac{\vardbtilde{\tau}}{\sqrt{3}}}  - \cos{\frac{\tilde{\tau}}{\sqrt{3}}} \sin{\frac{\vardbtilde{\tau}}{\sqrt{3}}} \right)S_{\bm{k}}(\vardbtilde{\tau}/k) \right],
\end{align}
where we have defined $\tau = k\eta$. 
Here, as we explained above, we have ignored the integration constant as $ v^{(2)}_{{\rm r} \bm{k}}(\eta_{\rm in})=0$, for the reason explained above. 
Finally, using the relation \eqref{momr},
    %&=-\frac{2}{\sqrt{3}k^4 \eta^2} \sum\limits_{\sigma,\sigma'}\int \frac{d^3\bm{k}_2}{(2\pi)^3} A_{\sigma'} (\bm{k}_2)A_{\sigma} \left(|\bm{k}-\bm{k}_2|\right)\epsilon^{\sigma'}_{ij}(\bm{\hat{k}}_2)\epsilon^{\sigma ij}\left(\hat{|\bm{k}-\bm{k}_2|}\right)\\
   % &\int^{\eta} k\tilde{\eta} kd\tilde{\eta}\left[\int^{\tilde{\eta}}kd\vardbtilde{\eta}\left( \sin{\frac{k\tilde{\eta}}{\sqrt{3}}}  \cos{\frac{k\vardbtilde{\eta}}{\sqrt{3}}}  - \cos{\frac{k\tilde{\eta}}{\sqrt{3}}} \sin{\frac{k\vardbtilde{\eta}}{\sqrt{3}}} \right)\mathcal{T}'\left(k_2\vardbtilde{\eta}\right)\mathcal{T}'\left(|\bm{k}-\bm{k}_2|\vardbtilde{\eta}\right) \right] \\
   % &+\frac{2}{\sqrt{3}k^2} \sum\limits_{\sigma,\sigma'}\int \frac{d^3\bm{k}_2}{(2\pi)^3} A_{\sigma'} (\bm{k}_2)A_{\sigma} \left(|\bm{k}-\bm{k}_2|\right)\epsilon^{\sigma'}_{ij}(\bm{\hat{k}}_2)\epsilon^{\sigma ij}\left(\widehat{|\bm{k}-\bm{k}_2|}\right)\\
   % &\int^{\tilde{\eta}}kd\vardbtilde{\eta}\left( \sin{\frac{k\tilde{\eta}}{\sqrt{3}}}  \cos{\frac{k\vardbtilde{\eta}}{\sqrt{3}}}  - \cos{\frac{k\tilde{\eta}}{\sqrt{3}}} \sin{\frac{k\vardbtilde{\eta}}{\sqrt{3}}} \right)\mathcal{T}'\left(k_2\vardbtilde{\eta}\right)\mathcal{T}'\left(|\bm{k}-\bm{k}_2|\vardbtilde{\eta}\right)\\
 the radiation density perturbation turns out
 \begin{align}\label{radpertfull}
 \delta_{\rm r}^{(2)}(\bm{k},\tau) %&= \frac{4\sqrt{3}}{k^2 \tau^2} \int^{\tau}_{\tau_{\rm in}} \tilde{\tau} d\tilde{\tau}\left[\int^{\tilde{\tau}}_{\tau_{\rm in}} d\vardbtilde{\tau}\left( \sin{\frac{\tilde{\tau}}{\sqrt{3}}}  \cos{\frac{\vardbtilde{\tau}}{\sqrt{3}}}  - \cos{\frac{\tilde{\tau}}{\sqrt{3}}} \sin{\frac{\vardbtilde{\tau}}{\sqrt{3}}} \right)S_{\bm{k}}(\vardbtilde{\tau}/k)\right] \nonumber\\
  % &-\frac{4\sqrt{3} }{k^2}
   %  \int^{\tau}_{\tau_{\rm in}}d\tilde{\tau}\left( \sin{\frac{\tau}{\sqrt{3}}}  \cos{\frac{\tilde{\tau}}{\sqrt{3}}}  - \cos{\frac{\tau}{\sqrt{3}}} \sin{\frac{\tilde{\tau}}{\sqrt{3}}} \right)S_{\bm{k}}(\tilde{\tau}/k)\nonumber\\
   %  &=-\frac{4\sqrt{3} }{k^2} \left[  \sin{\frac{\tau}{\sqrt{3}}} \int^{\tau}_{\tau_{\rm in}} \cos{\frac{\tilde{\tau}}{\sqrt{3}}} S_{\bm{k}}(\tilde{\tau}) d\tilde{\tau}-  \cos{\frac{\tau}{\sqrt{3}}} \int^{\tau}_{\tau_{\rm in}} \sin{\frac{\tilde{\tau}}{\sqrt{3}}} S_{\bm{k}}(\tilde{\tau}/k) d\tilde{\tau}\right]\nonumber\\
   %  &- \frac{12 }{k^2\tau^2 }  \left[\left(\tau \cos{\frac{\tau }{\sqrt{3}}}-\sqrt{3}\sin{\frac{\tau}{\sqrt{3}}} \right) \int^{\tau}_{\tau_{\rm in}} \cos{\frac{\tilde{\tau}}{\sqrt{3}}} S_{\bm{k}}(\tilde{\tau}/k)k d \tilde{\eta}\right.\nonumber\\
%&\left.+\left(\tau \sin{\frac{\tau}{\sqrt{3}}}+\sqrt{3}\cos{\frac{\tau}{\sqrt{3}}} \right) \int^{\tau}_{\tau_{\rm in}} \sin{\frac{\tilde{\tau}}{\sqrt{3}}} S_{\bm{k}}(\tilde{\tau}/k) d \tilde{\tau} -\int^{\tau}_{\tau_{\rm in}} \tilde{\tau}S_{\bm{k}}(\tilde{\tau}/k) d \tilde{\tau}\right]\nonumber\\
&=\frac{12 }{k^2\tau^2 } \int^{\tau}_{\tau_{\rm in}}  d \tilde{\tau}\,\tilde{\tau}S_{\bm{k}}(\tilde{\tau}/k) \nonumber\\
&-\frac{4\sqrt{3} }{k^2}\left[  \left(1-\frac{3}{\tau^2}\right)\sin{\frac{\tau}{\sqrt{3}}}+\frac{\sqrt{3}}{\tau}\cos{\frac{\tau}{\sqrt{3}}}\right]\int^{\tau}_{\tau_{\rm in}} d \tilde{\tau} \,\cos{\frac{\tilde{\tau}}{\sqrt{3}}}\, S_{\bm{k}}(\tilde{\tau}/k)\nonumber\\
&+\frac{4\sqrt{3} }{k^2} \left[  \left(1-\frac{3}{\tau^2}\right)\cos{\frac{\tau}{\sqrt{3}}}-\frac{\sqrt{3}}{\tau}\sin{\frac{\tau}{\sqrt{3}}}\right]\int^{\tau}_{\tau_{\rm in}} d \tilde{\tau} \,\sin{\frac{\tilde{\tau}}{\sqrt{3}}}\, S_{\bm{k}}(\tilde{\tau}/k)\,.
\end{align}
On sub-Hubble scales, this expression can  be further simplified. 
In fact, for $\tau (= k\eta) \gg 1$, we find
%\begin{multline}\label{radpertshort}
%\delta_{\rm r}^{(2)}(\bm{k},\tau)=
%      -\frac{4\sqrt{3} }{k^2} \left[\sin{\frac{\tau}{\sqrt{3}}}+\frac{\sqrt{3}}{\tau}\cos{\frac{\tau}{\sqrt{3}}}\right]\int^{\tau}_{\tau_{\rm in}} \cos{\frac{\tilde{\tau}}{\sqrt{3}}} S_{\bm{k}}(\tilde{\tau})d \tilde{\tau} \\
%      +\frac{4\sqrt{3} }{k^2} \left[\cos{\frac{\tau}{\sqrt{3}}}-\frac{\sqrt{3}}{\tau}\sin{\frac{\tau}{\sqrt{3}}}\right]\int^{\tau}_{\tau_{\rm in}}\sin{\frac{\tilde{\tau}}{\sqrt{3}}} S_{\bm{k}}(\tilde{\tau}) d \tilde{\tau} .
%\end{multline}
\begin{align}\label{radpertshort}
     \delta_{\rm r}^{(2)}(\bm{k},\tau)&=\frac{12 }{k^2\tau^2 } \int^{\tau}_{\tau_{\rm in}}d \tilde{\tau}\, \tilde{\tau}S_{\bm{k}}(\tilde{\tau})  
      -\frac{4\sqrt{3} }{k^2} \left[\sin{\frac{\tau}{\sqrt{3}}}+\frac{\sqrt{3}}{\tau}\cos{\frac{\tau}{\sqrt{3}}}\right]\int^{\tau}_{\tau_{\rm in}}d \tilde{\tau}\, \cos{\frac{\tilde{\tau}}{\sqrt{3}}}\, S_{\bm{k}}(\tilde{\tau}) \nonumber\\
     & +\frac{4\sqrt{3} }{k^2} \left[\cos{\frac{\tau}{\sqrt{3}}}-\frac{\sqrt{3}}{\tau}\sin{\frac{\tau}{\sqrt{3}}}\right]\int^{\tau}_{\tau_{\rm in}}d \tilde{\tau}\,\sin{\frac{\tilde{\tau}}{\sqrt{3}}}\, S_{\bm{k}}(\tilde{\tau}).
      \end{align}
Now, from (\ref{radc}), we get 
%\begin{align}\label{pot}
%    \phi^{(2)}&=-{v^{(2)}_{\rm r}}'-\frac{1}{6}\left(\mathcal{X}_{\bm{k}}-\mathcal{X}_{\bm{k}0}\right)-\frac{k^2}{3}\int_{\eta_{\rm in}}^\eta d\tilde{\eta}\, v_{\rm r} ,
%\end{align}
\begin{align}\label{pot}
    \phi^{(2)}(\bm{k},\eta)&=-{v^{(2)}_{\rm r}}'(\bm{k},\eta)-\frac{1}{6}\left[\mathcal{X}_{\bm{k}}(\eta)-\mathcal{X}_{\bm{k}}(\eta_{\rm in})\right]-\frac{k^2}{3}\int_{\eta_{\rm in}}^\eta d\tilde{\eta}\, v_{\rm r} ,
\end{align}
where $\mathcal{X}_{\bm{k}}$ is the Fourier transform of $\chi^{kl}\chi_{kl}$, i.e.
\begin{equation}
   \mathcal{X}_{\bm{k}}(\eta)=\sum\limits_{\sigma,\sigma'}\int \frac{d^3\bm{q}}{(2\pi)^3} A_{\sigma'} (\bm{q})A_{\sigma} \left(|\bm{k}-\bm{q}|\right)\epsilon^{\sigma'}_{ij}(\bm{\hat{q}})\epsilon^{\sigma ij}\left(\widehat{|\bm{k}-\bm{q}|}\right) \mathcal{T}\left(q,\eta\right)\mathcal{T}\left(|\bm{k}-\bm{q}|,\eta\right)\,.\nonumber\\
   \end{equation}
%   \DB{\Large I am concerned about the definition of $\mathcal{X}_{\bm{k}0}$.\\ It's defined in the beginning, right? for $\eta_{\rm in}$?? in my opinion we must 1) put a value, 2) write it as $\mathcal{X}_{\bm{k}}(\eta_{\rm in})$ and not $\mathcal{X}_{\bm{k}0}$... This also for all equations that we will write below!!!!}\\
(Here $v^{(2)}_{\rm r}({\bm k},\eta)=v^{(2)}_{{\rm r} \bm{k}}(\eta)$.)
Finally, using \eqref{deltam} and \eqref{pot}, we have the expression for $ \delta_{\rm m}^{(2)}$. Precisely, using 
\begin{equation}\label{vmatt}
     \delta_{\rm m}^{(2)}(\bm{k},\eta)=-3{v^{(2)}_{\rm r}}'-k^2 \int^{\eta}_{\eta_{\rm in}} d\tilde{\eta}\, v_{\rm r} \,,
    \end{equation}
we obtain
     \begin{align}
     \label{matpertfull} 
     \delta_{\rm m}^{(2)}(\bm{k},\tau)&= \frac{1}{k^2} \left(\frac{9 }{\tau^2 }-3\ln{\tau}\right)\int^{\tau}_{\tau_{\rm in}} d \tilde{\tau}\, \tilde{\tau}S_{\bm{k}}(\tilde{\tau}/k) 
    -\frac{1}{k^2} \left[ 3\sqrt{3}  \left(1-\frac{3}{\tau^2}+\ln{\tau}\right)\sin{\frac{\tau}{\sqrt{3}}}\right.\nonumber\\
    &\left.+\left(\frac{9 }{\tau^2 }-3\ln{\tau}\right)\tau\cos{\frac{\tau}{\sqrt{3}}}\right]\int^{\tau}_{\tau_{\rm in}}d \tilde{\tau}\, \cos{\frac{\tilde{\tau}}{\sqrt{3}}} S_{\bm{k}}(\tilde{\tau}/k) 
       -\frac{1}{k^2} \left[- 3\sqrt{3} \left(1-\frac{3}{\tau^2}+\ln{\tau}\right)\right.\nonumber\\
       &\left. \times \cos{\frac{\tau}{\sqrt{3}}}+\left(\frac{9 }{\tau^2 }-3\ln{\tau}\right)\tau \sin{\frac{\tau}{\sqrt{3}}}\right] \int^{\tau}_{\tau_{\rm in}}d \tilde{\tau}\, \sin{\frac{\tilde{\tau}}{\sqrt{3}}} S_{\bm{k}}(\tilde{\tau}/k)\nonumber\\
      &+\frac{\sqrt{3}}{k^2}\int^{\tau}_{\tau_{\rm in}}d\tilde{\tau}\, \tilde{\tau} \ln{\tilde{\tau}}  \left[\sin{\frac{\tilde{\tau}}{\sqrt{3}}} \int^{\tilde{\tau}}_{\tau_{\rm in}}d\vardbtilde{\tau}\, \cos{\frac{\vardbtilde{\tau}}{\sqrt{3}}} S_{\bm{k}}(\vardbtilde{\tau}/k)- \cos{\frac{\tilde{\tau}}{\sqrt{3}}} \int^{\tilde{\tau}}_{\tau_{\rm in}} d\vardbtilde{\tau}\, \sin{\frac{\vardbtilde{\tau}}{\sqrt{3}}} S_{\bm{k}}(\vardbtilde{\tau}/k)\right] .
\end{align}
Imposing again that $\tau=k\eta \gg 1$, it becomes
\begin{align} \label{matpertshort} 
\delta_{\rm m}^{(2)}(\bm{k},\tau)&= -\frac{3}{k^2}\ln{\tau}\int^{\tau}_{\tau_{\rm in}}d \tilde{\tau}\, \tilde{\tau}\, S_{\bm{k}}(\tilde{\tau}/k)  \nonumber \\
    & - \frac{1}{k^2}\left[ 3\sqrt{3} \left(1+\ln{\tau}\right)\sin{\frac{\tau}{\sqrt{3}}}
   -3\tau\ln{\tau}\cos{\frac{\tau}{\sqrt{3}}}\right]\int^{\tau}_{\tau_{\rm in}}d \tilde{\tau}\, \cos{\frac{\tilde{\tau}}{\sqrt{3}}} S_{\bm{k}}(\tilde{\tau}/k) \nonumber \\
  & +\frac{1}{k^2} \left[ 3\sqrt{3} \left(1+\ln{\tau}\right)\cos{\frac{\tau}{\sqrt{3}}}+3\tau\ln{\tau} \sin{\frac{\tau}{\sqrt{3}}}\right] \int^{\tau}_{\tau_{\rm in}} d \tilde{\tau}\,\sin{\frac{\tilde{\tau}}{\sqrt{3}}} S_{\bm{k}}(\tilde{\tau}/k)\nonumber\\
     & +\frac{\sqrt{3}}{k^2}\int^{\tau}_{\tau_{\rm in}}d\tilde{\tau}\, \tilde{\tau} \ln{\tilde{\tau}}  \left[\sin{\frac{\tilde{\tau}}{\sqrt{3}}} \int^{\tilde{\tau}}_{\tau_{\rm in}}d\vardbtilde{\tau}\, \cos{\frac{\vardbtilde{\tau}}{\sqrt{3}}} S_{\bm{k}}(\vardbtilde{\tau}/k)- \cos{\frac{\tilde{\tau}}{\sqrt{3}}} \int^{\tilde{\tau}}_{\tau_{\rm in}} d\vardbtilde{\tau}\, \sin{\frac{\vardbtilde{\tau}}{\sqrt{3}}} S_{\bm{k}}(\vardbtilde{\tau}/k)\right] \,.
      \end{align}
In addition, the potential can be obtained using \eqref{deltam}. Then it reduces to
\begin{comment}
\begin{align}
     \phi^{(2)}(\bm{k},\tau)&= -\frac{1}{k^2}\ln{\tau}\int^{\tau}_{\tau_{\rm in}}d \tilde{\tau}\, \tilde{\tau}\, S_{\bm{k}}(\tilde{\tau}/k) -\frac{1}{6}\left(\mathcal{X}_{\bm{k}}-\mathcal{X}_{\bm{k}0}\right)\nonumber\\
    & - \frac{1}{k^2}\left[\sqrt{3}  \left(1+\ln{\tau}\right)\sin{\frac{\tau}{\sqrt{3}}}
   -\tau \ln{\tau}\cos{\frac{\tau}{\sqrt{3}}}\right]\int^{\tau}_{\tau_{\rm in}}d \tilde{\tau}\, \cos{\frac{\tilde{\tau}}{\sqrt{3}}} S_{\bm{k}}(\tilde{\tau}/k) \nonumber\\
  & + \frac{1}{k^2}\left[\sqrt{3} \left(1+\ln{\tau}\right)\cos{\frac{\tau}{\sqrt{3}}}+\tau \ln{\tau} \sin{\frac{\tau}{\sqrt{3}}}\right] \int^{\tau}_{\tau_{\rm in}}d \tilde{\tau}\, \sin{\frac{\tilde{\tau}}{\sqrt{3}}} S_{\bm{k}}(\tilde{\tau}/k)\nonumber\\
     & +\frac{1}{\sqrt{3}k^2}\int^{\tau}_{\tau_{\rm in}} \tilde{\tau} \ln{\tilde{\tau}}  \left[\sin{\frac{\tilde{\tau}}{\sqrt{3}}} \int^{\tilde{\tau}}_{\tau_{\rm in}}d\vardbtilde{\tau}\, \cos{\frac{\vardbtilde{\tau}}{\sqrt{3}}} S_{\bm{k}}(\vardbtilde{\tau}/k)- \cos{\frac{\tilde{\tau}}{\sqrt{3}}} \int^{\tilde{\tau}}_{\tau_{\rm in}} d\vardbtilde{\tau}\, \sin{\frac{\vardbtilde{\tau}}{\sqrt{3}}} S_{\bm{k}}(\vardbtilde{\tau}/k)\right] d\tilde{\tau}\,. 
\end{align}
\end{comment}
\begin{align}\label{phideeprd}
     \phi^{(2)}(\bm{k},\tau)&= -\frac{1}{k^2}\ln{\tau}\int^{\tau}_{\tau_{\rm in}}d \tilde{\tau}\, \tilde{\tau}\, S_{\bm{k}}(\tilde{\tau}/k) -\frac{1}{6}\left(\mathcal{X}_{\bm{k}}(\eta)-\mathcal{X}_{\bm{k}}(\eta_{\rm in})\right)\nonumber\\
    & - \frac{1}{k^2}\left[\sqrt{3}  \left(1+\ln{\tau}\right)\sin{\frac{\tau}{\sqrt{3}}}
   -\tau \ln{\tau}\cos{\frac{\tau}{\sqrt{3}}}\right]\int^{\tau}_{\tau_{\rm in}}d \tilde{\tau}\, \cos{\frac{\tilde{\tau}}{\sqrt{3}}} S_{\bm{k}}(\tilde{\tau}/k) \nonumber\\
  & + \frac{1}{k^2}\left[\sqrt{3} \left(1+\ln{\tau}\right)\cos{\frac{\tau}{\sqrt{3}}}+\tau \ln{\tau} \sin{\frac{\tau}{\sqrt{3}}}\right] \int^{\tau}_{\tau_{\rm in}}d \tilde{\tau}\, \sin{\frac{\tilde{\tau}}{\sqrt{3}}} S_{\bm{k}}(\tilde{\tau}/k)\nonumber\\
     & +\frac{1}{\sqrt{3}k^2}\int^{\tau}_{\tau_{\rm in}} \tilde{\tau} \ln{\tilde{\tau}}  \left[\sin{\frac{\tilde{\tau}}{\sqrt{3}}} \int^{\tilde{\tau}}_{\tau_{\rm in}}d\vardbtilde{\tau}\, \cos{\frac{\vardbtilde{\tau}}{\sqrt{3}}} S_{\bm{k}}(\vardbtilde{\tau}/k)- \cos{\frac{\tilde{\tau}}{\sqrt{3}}} \int^{\tilde{\tau}}_{\tau_{\rm in}} d\vardbtilde{\tau}\, \sin{\frac{\vardbtilde{\tau}}{\sqrt{3}}} S_{\bm{k}}(\vardbtilde{\tau}/k)\right] d\tilde{\tau}\,. 
\end{align}
In this section we have discussed the epoch where the matter perturbation $\delta^{(2)}\rho_{\rm m}$ is negligible compared to its radiation counterpart, $\delta^{(2)}\rho_{\rm r}$. Looking at \eqref{radpertshort} and \eqref{matpertshort} one can realise that, as time progresses, %both the background and % 
the contribution of the CDM perturbation begins to become dominant with respect to that of the radiation. 
Let us explain this point in more detail. 
Comparing the absolute value of each additive term  of in Eqs. (\ref{radpertshort}) and (\ref{matpertshort}), we note that, for example, terms proportional to \[\int^{\tau}_{\tau_{\rm in}}d \tilde{\tau}\, \cos{\frac{\tilde{\tau}}{\sqrt{3}}}S_{\bm{k}}(\tilde{\tau}/k), \quad \int^{\tau}_{\tau_{\rm in}}d \tilde{\tau}\, \sin{\frac{\tilde{\tau}}{\sqrt{3}}}S_{\bm{k}}(\tilde{\tau}/k) \quad {\rm or} \quad \int^{\tau}_{\tau_{\rm in}}d \tilde{\tau}\, \tilde{\tau}S_{\bm{k}}(\tilde{\tau}/k)  \] 
have an extra multiplicative factor which is proportional to $\tau$, $\ln{\tau}$ and/or $\tau \ln{\tau}$ in $\delta_{\rm m}^{(2)}$, which are missing in  $\delta_{\rm r}^{(2)}$.
Then, $\delta_{\rm r}^{(2)}$ has extra terms which decays like $$\sim \frac{1}{\tau} \cos{\frac{\tilde{\tau}}{\sqrt{3}}} \quad {\rm or} \quad \sim \frac{1}{\tau} \sin{\frac{\tilde{\tau}}{\sqrt{3}}},$$ 
which are absent in $\delta_{\rm m}^{(2)}$. 
 Finally, Eq. \eqref{matpertshort} includes other additional contributions (they are integrals that contain complicated sine or cosine functions that multiply the terms proportional to $\ln{\tau}$. Note that we are in the regime $\tau \gg 1$). These extra terms could also  cause a faster growth of $\delta_{\rm m}^{(2)}$ w.r.t $\delta_{\rm r}^{(2)}$. In conclusion, these facts suggest the existence of a suitable time $\eta$ and scale (through $k$) in which $\delta^{(2)}\rho_{\rm m}$ becomes of the same order as $\delta^{(2)}\rho_{\rm r}$. 

Now, let us define the following new quantity 
\begin{equation}\label{F} 
     F(\bm{k},\eta)\equiv \frac{\delta^{(2)}\rho_{\rm m} (\bm{k},\eta)}{\delta^{(2)}\rho_{\rm r}(\bm{k},\eta)}=\frac{\bar \rho_{\rm m}(\eta)}{\bar \rho_{\rm r}(\eta)} \frac{\delta^{(2)}_{\rm m}(\bm{k},\eta)}{ \delta^{(2)}_{\rm r}(\bm{k},\eta)}\,.
\end{equation}
 
When $\delta^{(2)}\rho_{\rm m}$ becomes the same order as $\delta^{(2)}\rho_{\rm r}$, i.e. $F\simeq O(1)$, the perturbative contribution linked to the matter begins to overcome that of radiation even if $\bar \rho_{\rm m}$ is smaller than $\bar \rho_{\rm r}$.
Therefore we are entering a new phase of dynamical evolution in which the time-time component of Einstein's field equations is governed by the matter perturbations $\delta^{(2)}_{\rm m}$. 

A correct setting of the initial condition of this new phase will also be discussed in detail in the next section.
However, let us stress that the analysis in the subsequent sections will require a matching between the solutions of the two phases at given $\eta=\eta_\alpha$, where $\eta_\alpha$ refers to the conformal time in which
$F(\bm{k},\eta_\alpha)\equiv \alpha \simeq O(1)$, for fixed value of $\bm{k}$ and for $\tau\gg 1$. Here $\alpha$ is a suitable value which sets the initial condition for the solutions at $\eta\ge \eta_\alpha$. Also from the discussion  made above, for $\eta>\eta_{\alpha}$, $F$ has a value larger than $\alpha$, according to the respective evolution of matter and radiation density perturbation.

Before concluding this section, let us add a final comment. As we pointed out above, in order to have this  new phase, during radiation epoch, we need that $k\eta  \gg 1$. Therefore if $k\eta  \gsim 1$ %or $\eta \lesssim 1/k$
there is a concrete possibility that this phase cannot start during radiation epoch. %mainly when $\eta \lesssim 1/k_{\rm eq}$ (Here we have defined $k_{\rm eq}  \simeq 1 / \eta_{\rm eq}$). 
However, even if this new phase does not exist, for scales around $k_{\rm eq}$ and at $1/k \lesssim \eta \lesssim 1/k_{\rm eq}$ (here we have defined $k_{\rm eq}  \simeq 1 / \eta_{\rm eq}$), it is possible that matter contribution could also be non-negligible and, in this case, we should consider both matter and radiation contributions in Einstein field equations. Therefore our analytical prescription cannot work and a numerical analysis is needed.

%\PB{The emergence of this `new phase' within radiation domination is a new result in this paper. Along with this, Eqs. \eqref{radpertshort}, \eqref{matpertshort}, and \eqref{phideeprd} can be considered the second result in the paper.}
The emergence of this `new phase' at $ k \eta \gg 1 $, within the domination of radiation, is another completely new result which has never been considered in the previous literature. In this case. Eqs. \eqref{radpertshort}, \eqref{matpertshort}, and \eqref{phideeprd} can be considered the second result in the paper.

\section{Sub-horizon evolution towards matter-radiation equality}\label{subrd}
In the previous section, we saw that during the deep radiation epoch, although the matter perturbation is determined by the potential sourced by primordial GWs,  it does not affect the potential itself.
Now, when we approach the second phase of
evolution of perturbations, for modes entering the Hubble radius during the epoch of radiation, $\bar \rho_{\rm m} \delta^{(2)}_{\rm m}$ grows sufficiently to surpass $\bar \rho_{\rm r} \delta^{(2)}_{\rm r}$ as the main contributor to Einstein's field equations, although $ \bar \rho_{\rm m} \ll \bar \rho_{\rm r} $ is still maintained. Here below we see precisely how we can achieve the second order differential equation governing $\delta^{(2)}_{\rm m}$ evolution, in other words a new Meszaros equation due to GWs contribution.

As the epoch approaches towards matter-radiation equality, $\bar \rho_{\rm m}$ can not be completely ignored anymore, and it is convenient to use the usual variable $y=a/a_{\rm eq}$
as an evolution variable instead of $\eta$ and/or $a$. Here $a_{\rm eq}=a(\eta_{\rm eq})$ is  the value of the scale factor when $\bar \rho_{\rm m} (a_{\rm eq}) = \bar \rho_{\rm r}(a_{\rm eq})$. 
Trivially, this implies that
$$y={a \over a_{\rm eq}}={\bar \rho_{\rm m} \over \bar \rho_{\rm r}}\;.$$
Using $y$, the background dynamics can be described by the Friedmann equations in the following way
 \begin{align}\label{hy}
 \mathcal{H}^2=\mathcal{H}_{\rm eq}^2 \frac{y+1}{2 y^2}=k_{\rm eq}^2\frac{y+1}{2 y^2} \quad \quad {\rm and} \quad \quad 
 \mathcal{H'}= - k_{\rm eq}^2\frac{2+y}{4 y^2}\;,
\end{align}
where
\[k_{\rm eq}\equiv \mathcal{H}_{\rm eq}=a_{\rm eq} H_{\rm eq}\,,\quad \quad \mathcal{H}_{\rm eq}^2={8\pi G\over 3}\bar \rho_{\rm eq} a_{\rm eq}^2\, \quad \quad {\rm and} \quad \quad \bar \rho_{\rm eq}=2\bar\rho_{\rm m} (a_{\rm eq})\,.\]

In terms of $y$, Einstein equations \eqref{all00}, \eqref{all0i} and \eqref{allij} become, respectively
  \begin{align} \label{sub00r}
           & -3\mathcal{H}^2y\frac{d\phi^{(2)}}{dy} + \nabla^2\phi^{(2)} +\frac{1}{6}\nabla^2 \nabla^2 \chi^{||(2)} -\frac{1}{8}\left(\mathcal{H}y\right)^2\frac{d\chi^{ij}}{dy}\frac{d\chi^{ij}}{dy}-\mathcal{H}^2 y\chi^{ij}\frac{d\chi^{ij}}{dy}+\frac{3}{8}\chi^{kl,i}\chi_{kl,i}  \nonumber \\
           & -\frac{1}{4}\chi^{ik,l}\chi_{li,k}+\frac{1}{2} \chi^{ij}\nabla^2\chi_{ij}=\frac{3\mathcal{H}^2}{2(1+y)}\left(y\delta_{\rm m}^{(2)}+\delta_{\rm r}^{(2)}\right),
          \end{align}
\begin{equation}\label{sub0ir}
 \frac{d {\phi^{(2)}}_{,i}}{dy}-\frac{1}{2}\chi^{jk}\frac{d\chi_{ki,j}}{dy}+\frac{1}{4}D_{ij}\frac{d\chi^{||(2),j}}{dy}+\frac{1}{2}\chi^{jk}\frac{d\chi_{jk,i}}{dy}+\frac{1}{4}\frac{d\chi^{jk}}{dy}\chi_{jk,i}=-\frac{2\mathcal{H}}{y(1+y)}  v^{(2)}_{{\rm r},i},
\end{equation}
\begin{align}\label{subijr}
& \mathcal{H}y \left[\mathcal{H}y \frac{d^2}{dy^2}+  \left(\mathcal{H}+y \frac{d\mathcal{H}}{dy}\right)\frac{d}{dy}\right] \left(\frac{1}{4}D_{ij}\chi^{||(2)}+\phi^{(2)}\delta_{ij}\right) 
 +\frac{\mathcal{H}^2y}{2}D_{ij}\frac{d\chi^{||(2)}}{dy}+\frac{1}{12}\nabla^2 D_{ij}\chi^{||(2)}\nonumber\\&-\frac{1}{18}\nabla^2 \nabla^2 \chi^{||(2)}\delta_{ij}
 +2\mathcal{H}^2y\frac{d\phi^{(2)}}{dy}\delta_{ij}+\frac{1}{2}D_{ij}\phi^{(2)}-\frac{1}{3}\nabla^2 \phi^{(2)}\delta_{ij}
    -\frac{1}{2}\chi^{kl}\left(\chi_{lj,ik} +\chi_{il,jk}\right.\nonumber\\&\left.-\chi_{ij,lk}-\chi_{kl,ij}\right) +\frac{1}{4}{\chi^{kl}}_{,j}\chi_{kl,i}-\frac{1}{2}\chi^{jk,l}\chi_{li,k}+\frac{1}{2}\chi^{jk,l}\chi_{ki,l}-\frac{3}{8}\chi^{kl,p}\chi_{kl,p}\delta_{ij}+\frac{1}{4}\chi^{kp,l}\chi_{lp,k}\delta_{ij}\nonumber\\
    &-\frac{\mathcal{H}^2y^2}{2}\frac{d\chi^{k}_j}{dy} \frac{d\chi_{ki}}{dy} +\frac{3\mathcal{H}^2y^2}{8}\frac{d\chi^{kl}}{dy} \frac{d\chi_{kl}}{dy}\delta_{ij}=\frac{\mathcal{H}^2}{2(1+y)}\delta_{\rm r}^{(2)}\delta_{ij}.
\end{align}  
Its trace part is
\begin{align}\label{subijtr}
     &\mathcal{H}y \left[\mathcal{H}y \frac{d^2}{dy^2}+  \left(\mathcal{H}+y \frac{d\mathcal{H}}{dy}\right)\frac{d}{dy}\right] \phi^{(2)}+ 2\mathcal{H}^2y\frac{d\phi^{(2)}}{dy}-\frac{1}{3}\nabla^2\phi^{(2)}-\frac{1}{18}\nabla^2 \nabla^2\chi^{||(2)} \nonumber\\&-\frac{1}{8}\chi^{kl,i}\chi_{kl,i}+\frac{1}{12}\chi^{ik,l}\chi_{li,k}+\frac{5\mathcal{H}^2y^2}{24}\frac{d\chi^{kl}}{dy} \frac{d\chi_{kl}}{dy}+\frac{1}{6}\chi^{kl}\nabla^2\chi_{kl}=\frac{\mathcal{H}^2}{2(1+y)}\delta_{\rm r}^{(2)}.
 \end{align}
 Taking $\nabla^2 \nabla^2 \chi^{||(2)}$  from (\ref{sub00r}), and putting it in (\ref{subijtr}), we get
 \begin{align} \label{Master_eq}
  & \left[\mathcal{H}^2y^2 \frac{d^2}{dy^2}+\left(2\mathcal{H}^2y +\mathcal{H}y^2 \frac{d\mathcal{H}}{dy}\right)\frac{d}{dy}\right] \phi^{(2)} +\frac{\mathcal{H}^2y^2}{6}\frac{d\chi^{kl}}{dy} \frac{d\chi_{kl}}{dy}- \frac{1}{3} \mathcal{H}^2 y\chi^{ij}\frac{d\chi^{ij}}{dy} +\frac{1}{3}\chi^{ij}\nabla^2\chi_{ij} \nonumber \\
  &     
   = \frac{\mathcal{H}^2}{2(1+y)}\left(y\delta_{\rm m}^{(2)}+2\delta_{\rm r}^{(2)}\right).
 \end{align} 
 A comment is in order here. From Eq. \eqref{Master_eq} we can see  that for $y\delta_{\rm m}^{(2)} \gg 2\delta_{\rm r}^{(2)}$, i.e. $\bar \rho_{\rm m}\delta_{\rm m}^{(2)} =\delta \rho_{\rm m}^{(2)}\gg 2\bar \rho_{\rm r}\delta_{\rm r}^{(2)}= 2 \delta \rho_{\rm r}^{(2)}$, we can safely neglect  $\delta_{\rm r}^{(2)}$. (Here note the factor $2$ in front of $\delta \rho_{\rm r}^{(2)}$.) % from  \eqref{Master_eq}. 
 Now, according to the discussion at the end of the previous section, after $\delta \rho_{\rm m}^{(2)}$ surpasses $\delta \rho_{\rm r}^{(2)}$, i.e. when $F\simeq O(1)$, the matter perturbation keeps increasing, 
 and  we can no longer neglect this contribution.
 At the same time, we can actually discard $\delta_{\rm r}^{(2)}$ from evolution equations only for $F\simeq 2$, see Eq. \eqref{Master_eq}.
 (Here, for the sake of simplicity of the calculation we will do below
 and without any loss of generality, we are assuming that $F$ is positive.)
 This implies that, at a fixed scale (e.g. at a given  value of $k$),  the beginning of this new phase is indicated by the following range of values
 $1\lsim F (\eta_\alpha, k)=\alpha \lsim 2$.
 In other words, we have to set the initial condition at $\eta=\eta_\alpha$ imposing $\alpha$ between $1$ and $2$. However, with our present approach, which is only analytical, we cannot be more precise. In order to know exactly the value of $\alpha$, a numerical treatment would be necessary, but that is beyond the scope of this work. (Note that $\alpha$ should also depend on the cosmological parameters and the matter component of the Universe.)
 In Fig. \ref{Figure} the green and light-green regions, for $\eta \le \eta_{\rm rec}$, highlight the modes and particular period that we analyze in this section (in other words, at each sub-horizon mode with $k\ge k_{\rm eq}$ and $\eta\ge \eta_\alpha$ up to recombination). The reason why we consider $\eta$ only up to the time of recombination will be discussed in Section \ref{matt}.
 
 Now, using the definition of $F$, see Eq. \eqref{F}, 
 we are setting the initial conditions %at the time 
 $ y_\alpha = a (\eta_\alpha) / a_{\rm eq} $ which is defined, in implicit manner, from the following relation
 \begin{equation}\label{yM}
    y_\alpha \delta^{(2)}_{\rm m} (y_\alpha)=\alpha \delta^{(2)}_{\rm r} (y_\alpha)\,.
\end{equation}
Now, introducing again $y$ as variable and  combining Eqs. \eqref{deltam} and \eqref{chi_ij-1}, it easy to see that the  second-order equation which determines the evolution of the matter perturbation $\delta_{\rm m}^{(2)}$, in the second phase of radiation domination (i.e. for $y\ge y_\alpha$), is 
\begin{equation}\label{mesz}
      \frac{d^2\delta_{\rm m}^{(2)}}{dy^2} +\frac{2+3y}{2y(y+1)}\frac{d\delta_{\rm m}^{(2)}}{dy}-\frac{3}{2y(y+1)}\delta_{\rm m}^{(2)}=\frac{1}{2} \frac{d \chi^{ij}}{dy}\frac{d \chi_{ij}}{dy}\,.
\end{equation}
This is  a retelling of the Meszaros equation 
\cite{Meszaros:1974tb} 
(e.g., for the derivation, see also \cite{dodelson2021modern,Weinberg:2002kg}). In particular, the left side  is exactly the same Meszaros equation, the governing equation of the evolution of subhorizon matter perturbation, albeit having source term quadratic in tensors on the right hand side.  
This is not surprising, considering the fact that  $\delta_{\rm m}^{(2)}$ replaces $\delta_{\rm r}^{(2)}$ as the source of Einstein equations in this phase, a behaviour similar to their linear counterpart. 
Obviously, for the two solutions to the homogeneous equation, we find the same of Meszaros
\begin{itemize}
    \item $D_1(y)=y+\frac{2}{3}$,
    \item $D_2(y)=D_1(y)\ln{\frac{\sqrt{1+y}+1}{\sqrt{1+y}-1}}-2\sqrt{1+y}$\;.
\end{itemize}
It should be noted that these solutions are correct both during the epoch of radiation $y \ll 1$ (when $y\ge y_\alpha$) and of matter (for $y \gg 1$).
Indeed, in matter era, they go as $y$ and $y^{-3/2}$ respectively, whereas in radiation era (i.e., for $y_\alpha \lesssim y \ll 1$), they behave as a constant and $\ln{y}$. In conclusion, taking also into account the particular solution,  the general solution of the matter perturbation  $\delta_{m}^{(2)}$, on sub-horizon scales, takes the form
\begin{equation}\label{subdel}
    \delta_{\rm m}^{(2)}(\bm{x},y)= P_1(\bm{x}) D_1(y)+P_2(\bm{x}) D_2(y)+\frac{1}{2}\int_{y_\alpha}^y d \Tilde{y} \, G(y, \tilde{y})  \frac{d \chi^{ij}}{d\Tilde{y}}\frac{d \chi_{ij}}{d\Tilde{y}},
\end{equation}
where $ G(y, \Tilde{y}) $ is the Green's function
\begin{align}\label{Green-deltam}
    G(y, \Tilde{y})=&-\frac{1}{4}\Tilde{y}\sqrt{1+\Tilde{y}} \left[6\Big( \sqrt{1+\Tilde{y}}(2+3y)-\sqrt{1+y}(2+3\Tilde{y})\Big)\right.\nonumber\\
    &\left.-(2 + 3 \Tilde{y}) (2 + 3 y)\ln{\frac{(\sqrt{1+\Tilde{y}}+1)(\sqrt{1+y}-1)}{(\sqrt{1+\Tilde{y}}-1)(\sqrt{1+y}+1)}}\right],
\end{align}
and $P_1(\bm{x})$, $P_2(\bm{x})$ are two time-independent functions.  Let us rename the Fourier transformation of  Eq. \eqref{subdel} as $\delta_{\rm m (Tmesz)}^{(2)}(\bm{k},y)$, which can be written as 
\begin{equation}\label{subdelk}
   \delta_{\rm m (Tmesz)}^{(2)}(\bm{k},y)= P_1(\bm{k}) D_1(y)+P_2(\bm{k}) D_2(y)+\frac{1}{2}\int_{y_\alpha}^y d \Tilde{y} \, G(y, \tilde{y}) F_1(\bm{k},\tilde y),
\end{equation}
where $F_1(\bm{k},y)$ is the  Fourier transform of $d \chi^{ij}/dy\, d \chi_{ij}/dy$, i.e.
\begin{align}\label{Fone}
F_1(\bm{k},y)&=\sum\limits_{\sigma,\sigma'}\int \frac{d^3\bm{q}}{(2\pi)^3} A_{\sigma'} (\bm{q})A_{\sigma} \left(\bm{k}-\bm{q}\right)\nonumber\\&\times\epsilon^{\sigma'}_{ij}(\bm{\hat{q}})\epsilon^{\sigma ij}\left(\widehat{\bm{k}-\bm{q}}\right) \frac{d\mathcal{T}\left(q,y\right)}{d y}\frac{d\mathcal{T}\left(|\bm{k}-\bm{q}|,y\right)}{dy}. 
\end{align}
Note immediately that, using Eq. \eqref{hy}, we can easily relate the definition of $F_1$ with $S_{\bm{k}}$ in the following way
\begin{align}\label{FoneSk}
F_1(\bm{k},y) = {-12\over k^2_{\rm eq} (y+1)} S_{\bm{k}}\;.
\end{align}

Here we just make one comment.
In order to obtain the full solution, which describes the evolution of $\delta_{\rm m}^{(2)}$ at all epochs, from the deep radiation to CDM era, at a given $k > k_{\rm eq}$, we need to know the value of $P_1$, $P_2$ (or their corresponding Fourier transformations).
As we are aware, the above solution, Eq. \eqref{subdelk},
is valid only for $y \ge y_\alpha$.
At $y=y_\alpha$, we have
\[ \delta_{\rm m (Tmesz)}^{(2)}(\bm{k},y_\alpha)=   P_1(\bm{k}) D_1(y_\alpha)+P_2(\bm{k})D_2(y_\alpha)\,. \]
Consequently, we should analyse the matching conditions at $y = y_\alpha$. Then we need the solution of matter perturbation and its derivative obtained both during deep radiation epoch and in this second phase of radiation domination.
The next section will be devoted to the study of the initial condition of Eq. \eqref{subdel}.

Before concluding this section, in order to have a complete picture of the dynamics at these scales, it is also useful to get an expression for the radiation density contrast (for the complete derivation, see Appendix \ref{appdeltar})
\begin{align}\label{subdeltar}
   \delta_{\rm r (Tmesz)}^{(2)}(\bm{k},y) = & A_{\rm r}(\bm{k}) \cos{\left(2\sqrt{\frac{2}{3}}\frac{k}{k_{\rm eq}}\sqrt{1+y}\right)}+B_{\rm r}(\bm{k}) \sin{\left(2\sqrt{\frac{2}{3}}\frac{k}{k_{\rm eq}}\sqrt{1+y}\right)}\nonumber\\& 
    + \frac{k_{\rm eq}}{4k}\sqrt{\frac{3}{2}}\int_{y_\alpha}^y \frac{d \Tilde{y}}{\sqrt{1+y}}  \, \mathcal{Q}_{\bm{k}}(\sqrt{1+\tilde{y}})\left[\sin{\left(2\sqrt{\frac{2}{3}}\frac{k}{k_{\rm eq}}\sqrt{1+y}\right)}\cos{\left(2\sqrt{\frac{2}{3}}\frac{k}{k_{\rm eq}}\sqrt{1+\tilde{y}}\right)}\right.\nonumber\\&\left.-\cos{\left(2\sqrt{\frac{2}{3}}\frac{k}{k_{\rm eq}}\sqrt{1+y}\right)}\sin{\left(2\sqrt{\frac{2}{3}}\frac{k}{k_{\rm eq}}\sqrt{1+\tilde{y}}\right)} \right],
\end{align}
where 
\begin{align}
    \mathcal{Q}_{\bm{k}}(\sqrt{1+y})&= \frac{4}{3} \Bigg\{2P_1(\bm{k})+{1 \over 3 y^2}\Bigg[4(2-3y)\sqrt{1+y}+6y^2\ln\left(\frac{2+y+2\sqrt{1+y}}{y}\right)\Bigg] P_2(\bm{k})\Bigg\}\nonumber\\&
    -\frac{1}{2}\Bigg\{
    {1 \over 3 y^2}\Bigg[4(2-3y)\sqrt{1+y}+6y^2\ln\left(\frac{2+y+2\sqrt{1+y}}{y}\right)\Bigg]
   % \frac{4(2-3y)\sqrt{1+y}+6y^2\ln{\frac{2+y+2\sqrt{1+y}}{y}}}{3 y^2}
   \nonumber \\& \times  \int_{y_\alpha}^y d \tilde y \, \tilde{y} (2+3\tilde{y})\, \sqrt{1+\tilde{y}} ~ F_1(\bm{k}, \tilde y)
    + 2\int_{y_\alpha}^y d\tilde y \, \tilde{y} \sqrt{1+\tilde{y}}\left[6\sqrt{1+\tilde{y}}\right.
    \nonumber\\&\left. +(2+3\tilde{y})\ln\left(\frac{2+\Tilde{y}-2\sqrt{1+\Tilde{y}}}{\Tilde{y}}\right)\right] F_1(\bm{k}, \tilde{y})
    -\frac{16(1+y)}{3}F_1(\bm{k},\tilde y)\Bigg\}\,
\end{align}
where the naming of the variable in Eq. \eqref{subdeltar} is done in analogy its CDM counterpart. 
The coefficients $A_{\rm r}(\bm{k}), B_{\rm r}(\bm{k})$ can be obtained in a similar way as  the coefficients $P_1(\bm{k})$ and $P_2(\bm{k})$ of $ \delta_{\rm m}^{(2)}$ (see the complete expression in Appendix \ref{appdeltar}). 
Finally,
using \eqref{deltam} and \eqref{subdel} we can obtain the potential $\phi^{(2)}$. In fact we find
\begin{align}\label{phinewp}
    \phi^{(2)}(\bm{x},y)&= \frac{1}{3}\left[P_1(\bm{x}) D_1(y)+P_2(\bm{x}) D_2(y)+\frac{1}{2}\int_{y_\alpha}^y d \Tilde{y} \, G(y, \tilde{y}) \frac{d \chi^{ij}}{d\tilde y}\frac{d \chi_{ij}}{d\tilde y}\right]\nonumber\\&
    -\frac{1}{6}\left(\chi^{ij}\chi_{ij}-\chi^{ij}_0\chi_{0ij}\right) \,,
\end{align}
in configuration space, where $G(y, \tilde{y})$ has already been defined in Eq. \eqref{Green-deltam}. Still note that here 
$\chi^{ij}_0= \chi^{ij}(\bm{x},\eta_{\rm in})$. In Fourier space, $\phi^{(2)}$ becomes
\begin{align} 
    \phi^{(2)}(\bm{k},y)&= \frac{1}{3}\left[P_1(\bm{k}) D_1(y)+P_2(\bm{k}) D_2(y)+\frac{1}{2}\int_{y_\alpha}^y d \tilde{y} \, G(y, \tilde{y})F_1(\bm{k},\tilde{y})\right]\nonumber\\&
    -\frac{1}{6}\left(   \mathcal{X}_{\bm{k}}(\eta)-\mathcal{X}_{\bm{k}}(\eta_{\rm in})\right) \,.
\end{align}
Let us emphasize that the third result of this paper are Eqs. \eqref{subdel} along with \eqref{Green-deltam}, \eqref{subdeltar}, and \eqref{phinewp}.

\section{Determining the coefficients of the density contrast}\label{match}

The coefficients of the homogeneous parts of the solution $P_1(\bm{x})$ and $P_2(\bm{x})$ can be obtained by matching the solutions from section \ref{purerd} and section \ref{subrd}. 
As we discussed in the previous section, the perturbation and its derivatives have to be continuous throughout evolution. This implies that they must be matched at a particular time $\eta_\alpha$ which can be easily linked to variable $y_\alpha$, i.e. when the new phase is starting. 
This matching conditions will be of the form
\begin{align}\label{deltamatch}
  \delta_{\rm m (DRe)}^{(2)}(\bm{k},\tau_\alpha)&=  \delta_{\rm m (Tmesz)}^{(2)}(\bm{k},y_\alpha),\\
  \label{deltadotmatch}
     {\left({d \over d y}{ \delta}_{\rm m (DRe)}^{(2)}(\bm{k},\tau)\right)}\Bigg|_{\tau_\alpha}&=    {\left({d \over d y}{\delta}_{\rm m (Tmesz)}^{(2)}(\bm{k},y)\right)}\Bigg|_{y_\alpha},
\end{align}
where $\tau_\alpha=k\eta_\alpha$.
Here we have called with $ \delta_{\rm m (DRe)}^{(2)}(\bm{k},\tau)$ the matter perturbation solution during the deep radiation era, while 
$ \delta_{\rm m (Tmesz)}^{(2)}(\bm{k},y)$ is the solution obtained by the tensor-induced Meszaros equation. (The last definition was already mentioned in the previous section.)
Before  this matching, $ \delta_{\rm m (DRe)}^{(2)}(\bm{k},\tau)$ must be re-expressed in terms of the dynamic variable $y$.

Following \cite{mukhanov2005physical}, writing the scale factor as $a=a_{\rm eq} (\xi^2+2\xi)$, where $\xi=\eta/\eta_{*}$, with $\eta_{*}$ being $\eta_{\rm eq}/(\sqrt{2}-1)$, in deep radiation era we have $\xi \ll 1$ and we can write $a(\xi \ll 1)=2a_{\rm eq} \xi$.
Thus $\eta$ can easily be related to $y$ by the relation $\eta=\eta_* y/2$ and we find $\tau=k\eta_*y/2$. Then  \eqref{matpertshort} can be expressed as 
\begin{align}\label{deltamy}
 \delta_{\rm m (DRe)}^{(2)}(\bm{k},y)&= -\frac{3\eta_*^2}{4}\ln\left({\frac{k\eta_*y}{2}}\right)\int^{y}_{y_{\rm in}} d \tilde{y} \,\tilde{y}\, S_{\bm{k}}\left({\eta_*\tilde{y} \over 2}\right)  
    -\Bigg\{ \frac{3\sqrt{3}\eta_* }{2k} \left[1+\ln\left({\frac{k\eta_*y}{2}}\right)\right]\sin\left({\frac{k\eta_*y}{2\sqrt{3}}}\right)\nonumber\\&
     -\frac{3\eta_*^2 y}{4}\ln\left({\frac{k\eta_*y}{2}}\right)\cos\left({\frac{k\eta_*y}{2\sqrt{3}}}\right)\Bigg\}\int^{y}_{y_{\rm in}}d \tilde{y} \,\cos\left({\frac{k\eta_*\tilde{y}}{2\sqrt{3}}}\right) S_{\bm{k}}\left({\eta_*\tilde{y} \over 2}\right)\nonumber \\&
       + \Bigg\{ \frac{3\sqrt{3} \eta_*}{2k} \left[1+\ln\left({\frac{k\eta_*y}{2}}\right)\right]\cos\left({\frac{k\eta_*y}{2\sqrt{3}}}\right)
       +\frac{3\eta_*^2 y}{4}\ln\left({\frac{k\eta_*y}{2}}\right)\sin\left({\frac{k\eta_*y}{2\sqrt{3}}}\right)\Bigg\}%\int^{y}_{y_{\rm in}}\sin\left({\frac{k\eta_*\tilde{y}}{2\sqrt{3}}}\right) S_{\bm{k}}(\tilde{y})d \tilde{y}
       \nonumber\\&
      \times \int^{y}_{y_{\rm in}}d \tilde{y} \,\sin\left({\frac{k\eta_*\tilde{y}}{2\sqrt{3}}}\right) S_{\bm{k}}\left({\eta_*\tilde{y} \over 2}\right) +\frac{\sqrt{3}\eta_*^3k}{8} \int^{y}_{y_{\rm in}} d \tilde{y} \, \Bigg\{\tilde{y} \ln\left({\frac{k\eta_*y}{2}}\right)\nonumber\\& \times \Bigg[\sin\left({\frac{k\eta_*\tilde{y}}{2\sqrt{3}}}\right) \int^{\tilde{y}}_{y_{\rm in}}d\vardbtilde{y} \, \cos\left({\frac{k\eta_*\vardbtilde{y}}{2\sqrt{3}}} \right) S_{\bm{k}}\left({\eta_*\vardbtilde{y} \over 2}\right) 
      \nonumber\\&- \cos\left({\frac{k\eta_*\tilde{y}}{2\sqrt{3}}}\right) \int^{\tilde{y}}_{y_{\rm in}}d\vardbtilde{y}\, \sin\left({\frac{k\eta_*\vardbtilde{y}}{2\sqrt{3}}}\right) S_{\bm{k}}\left({\eta_*\vardbtilde{y} \over 2}\right) \Bigg]\Bigg\} \;.
\end{align}
Now we can finally match \eqref{deltamy} with \eqref{subdel}. The condition \eqref{deltamatch} can now be written as 
\begin{align}\label{deltamatch1}
 &-\frac{3\eta_*^2}{4}\ln\left({\frac{k\eta_*y}{2}}\right)\int^{y}_{y_{\rm in}} d \tilde{y} \, \tilde{y}\, S_{\bm{k}}\left({\eta_*\tilde{y} \over 2}\right) 
    -\Bigg\{ \frac{3\sqrt{3}\eta_* }{2k} \left[1+\ln\left({\frac{k\eta_*y}{2}}\right)\right]\sin\left({\frac{k\eta_*y}{2\sqrt{3}}}\right)\nonumber\\&
     -\frac{3\eta_*^2 y}{4}\ln\left({\frac{k\eta_*y}{2}}\right)\cos\left({\frac{k\eta_*y}{2\sqrt{3}}}\right)\Bigg\}\int^{y}_{y_{\rm in}}d \tilde{y}\, \cos\left({\frac{k\eta_*\tilde{y}}{2\sqrt{3}}}\right) S_{\bm{k}}\left({\eta_*\tilde{y} \over 2}\right) \nonumber \\&
       + \Bigg\{ \frac{3\sqrt{3} \eta_*}{2k} \left[1+\ln\left({\frac{k\eta_*y}{2}}\right)\right]\cos\left({\frac{k\eta_*y}{2\sqrt{3}}}\right)
       %\right.\\
       %\left.
       +\frac{3\eta_*^2 y}{4}\ln\left({\frac{k\eta_*y}{2}}\right)\sin\left({\frac{k\eta_*y}{2\sqrt{3}}}\right)\Bigg\}%\int^{y}_{y_{\rm in}}\sin\left({\frac{k\eta_*\tilde{y}}{2\sqrt{3}}}\right) S_{\bm{k}}(\tilde{y})d \tilde{y}
      \nonumber \\&
      \times \int^{y}_{y_{\rm in}}d \tilde{y}\, \sin\left({\frac{k\eta_*\tilde{y}}{2\sqrt{3}}}\right) S_{\bm{k}}\left({\eta_*\tilde{y} \over 2}\right)  +\frac{\sqrt{3}\eta_*^3k}{8} \int^{y}_{y_{\rm in}} d\tilde{y}\, \Bigg\{\tilde{y} \ln\left({\frac{k\eta_*y}{2}}\right)\nonumber\\& \times \Bigg[\sin\left({\frac{k\eta_*\tilde{y}}{2\sqrt{3}}}\right) \int^{\tilde{y}}_{y_{\rm in}}d\vardbtilde{y}\, \cos\left({\frac{k\eta_*\vardbtilde{y}}{2\sqrt{3}}} \right) S_{\bm{k}}\left({\eta_*\vardbtilde{y} \over 2}\right)  
     - \cos\left({\frac{k\eta_*\tilde{y}}{2\sqrt{3}}}\right) \nonumber\\& \times \int^{\tilde{y}}_{y_{\rm in}}d\vardbtilde{y}\, \sin\left({\frac{k\eta_*\vardbtilde{y}}{2\sqrt{3}}}\right) S_{\bm{k}}\left({\eta_*\vardbtilde{y} \over 2}\right) \Bigg]\Bigg\} \nonumber\\&
      =P_1(\bm{k}) D_1(y_\alpha)+P_2(\bm{k}) D_2(y_\alpha),
\end{align}
and \eqref{deltadotmatch} gives
\begin{align}\label{deltamatch2}
   & -\frac{3\eta_*}{2ky_\alpha}\int^{y_\alpha}_{y_{\rm in}}d \tilde{y}\, \tilde{y}\,S_{\bm{k}}\left({\eta_*\tilde{y} \over 2}\right)\nonumber \\&
    -\left[ \frac{3\sqrt{3}}{k^2y_\alpha}\sin{\frac{k\eta_*y_\alpha}{2\sqrt{3}}}+\frac{3\eta_*^2 }{4}\cos{\frac{k\eta_*y_\alpha}{2\sqrt{3}}}-\frac{3\eta_* }{2k}\cos{\frac{k\eta_*y_\alpha}{2\sqrt{3}}}\right]\int^{y_\alpha}_{y_{\rm in}}d \tilde{y} \, \cos{\frac{k\eta_*\tilde{y}}{2\sqrt{3}}} \,S_{\bm{k}}\left({\eta_*\tilde{y} \over 2}\right) \nonumber\\&
       - \left[- \frac{3\sqrt{3}}{k^2y_\alpha}\cos{\frac{k\eta_*y_\alpha}{2\sqrt{3}}}+\frac{3\eta_*^2 }{4}\sin{\frac{k\eta_*y_\alpha}{2\sqrt{3}}}-\frac{3\eta_* }{2k}\sin{\frac{k\eta_*y_\alpha}{2\sqrt{3}}}\right]\int^{y}_{y_{\rm in}}d \tilde{y}\,\sin{\frac{k\eta_*\tilde{y}}{2\sqrt{3}}} \,S_{\bm{k}}\left({\eta_*\tilde{y} \over 2}\right) \nonumber \\&
       = P_1(\bm{k}) + P_2(\bm{k}) \left(- \frac{2(1+3y_\alpha)}{3y_\alpha\sqrt{1+y_\alpha}}+\ln{\frac{\sqrt{1+y_\alpha}+1}{\sqrt{1+y_\alpha}-1}}\right).
\end{align}
Multiplying \eqref{deltamatch2} by $D_1(y_\alpha)$ and subtracting it from \eqref{deltamatch1}, we have 
\begin{align}\label{P2}
     P_2(\bm{k})&= \frac{9y_\alpha\sqrt{1+y_\alpha}}{2(2+3y_\alpha)} \left[-\frac{3\eta_*^2}{4}\ln{\frac{k\eta_*y_\alpha}{2}}\int^{y_\alpha}_{y_{\rm in}} d \tilde{y}\, \tilde{y}\,S_{\bm{k}}\left({\eta_*\tilde{y} \over 2}\right)
    -\left[ \frac{3\sqrt{3}\eta_* }{2k} \right.\right.\nonumber\\&
    \left.\left. \times \left(1+\ln{\frac{k\eta_*y_\alpha}{2}}\right)\sin{\frac{k\eta_*y_\alpha}{2\sqrt{3}}}-\frac{3\eta_*^2 y_\alpha}{4}\ln{\frac{k\eta_*y_\alpha}{2}}\cos{\frac{k\eta_*y_\alpha}{2\sqrt{3}}}\right]\int^{y_\alpha}_{y_{\rm in}} d \tilde{y}\,\cos{\frac{k\eta_*\tilde{y}}{2\sqrt{3}}} \,S_{\bm{k}}\left({\eta_*\tilde{y} \over 2}\right)
      \right.\nonumber\\&\left. - \left[- \frac{3\sqrt{3}\eta_*}{2k} \left(1+\ln{\frac{k\eta_*y_\alpha}{2}}\right)\cos{\frac{k\eta_*y_\alpha}{2\sqrt{3}}}-\frac{3\eta_*^2 y_\alpha}{4}\ln{\frac{k\eta_*y_\alpha}{2}}\sin{\frac{k\eta_*y_\alpha}{2\sqrt{3}}}\right] \right.\nonumber\\&\left. \times \int^{y_\alpha}_{y_{\rm in}} d \tilde{y}\, \sin{\frac{k\eta_*\tilde{y}}{2\sqrt{3}}} \,S_{\bm{k}}\left({\eta_*\tilde{y} \over 2}\right)
    +\frac{\sqrt{3}\eta_*^3k}{8} \int^{y_\alpha}_{y_{\rm in}} d \tilde{y}\, \tilde{y} \ln{\frac{k\eta_*y}{2}}\right.\nonumber \\&\left. \times  \left[\sin{\frac{k\eta_*\tilde{y}}{2\sqrt{3}}} \int^{\tilde{y}}_{y_{\rm in}}d\vardbtilde{y}\, \cos{\frac{k\eta_*\vardbtilde{y}}{2\sqrt{3}}} \,S_{\bm{k}}\left({\eta_*\vardbtilde{y} \over 2}\right)- \cos{\frac{k\eta_*\tilde{y}}{2\sqrt{3}}} \int^{\tilde{y}}_{y_{\rm in}}d\vardbtilde{y}\, \sin{\frac{k\eta_*\vardbtilde{y}}{2\sqrt{3}}} \,S_{\bm{k}}\left({\eta_*\vardbtilde{y} \over 2}\right)\right] \right.\nonumber\\&
      \left.
     % -\frac{1}{2}\int^{y_\alpha}_{y_{\rm in}} d \Tilde{y}  G(y, \tilde{y})  \frac{d \chi^{ij}}{d\Tilde{y}}\frac{d \chi_{ij}}{d\Tilde{y}}
     +\left(y_\alpha+\frac{2}{3}\right)\left( \frac{3\eta_*}{2ky}\int^{y_\alpha}_{y_{\rm in}}d \tilde{y}\, \tilde{y}\,S_{\bm{k}}\left({\eta_*\tilde{y} \over 2}\right)  \right.\right.\nonumber\\& \left.\left.
    +\left[ \frac{3\sqrt{3}}{k^2y_\alpha}\sin{\frac{k\eta_*y_\alpha}{2\sqrt{3}}}+\frac{3\eta_*^2 }{4}\cos{\frac{k\eta_*y_\alpha}{2\sqrt{3}}}-\frac{3\eta_* }{2k}\cos{\frac{k\eta_*y_\alpha}{2\sqrt{3}}}\right]\int^{y_\alpha}_{y_{\rm in}}d \tilde{y}\,\cos{\frac{k\eta_*\tilde{y}}{2\sqrt{3}}} S_{\bm{k}}\left({\eta_*\tilde{y} \over 2}\right)\right. \right.\nonumber\\& \left.
\left.       + \left[- \frac{3\sqrt{3}}{k^2y_\alpha}\cos{\frac{k\eta_*y_\alpha}{2\sqrt{3}}}+\frac{3\eta_*^2 }{4}\sin{\frac{k\eta_*y_\alpha}{2\sqrt{3}}}-\frac{3\eta_* }{2k}\sin{\frac{k\eta_*y_\alpha}{2\sqrt{3}}}\right]\int^{y}_{y_{\rm in}}d \tilde{y}\,\sin{\frac{k\eta_*\tilde{y}}{2\sqrt{3}}} S_{\bm{k}}\left({\eta_*\tilde{y} \over 2}\right)\right) \right]\;.
\end{align}
Finally, substituting the above expression in \eqref{deltamatch2}, we get the expression for $P_1(\bm{k})$
\begin{align}\label{P1}
    P_1(\bm{k})&=\left\{1+ 3y \left(1- \frac{\sqrt{1+y}}{2}\ln{\frac{\sqrt{1+y_\alpha}+1}{\sqrt{1+y_\alpha}-1}}\right)\right\} \frac{3}{2+3y}\nonumber\\& \left[-\frac{3\eta_*^2}{4}\ln{\frac{k\eta_*y_\alpha}{2}}\int^{y_\alpha}_{y_{\rm in}}  d \tilde{y}\,\tilde{y}S_{\bm{k}}\left({\eta_*\tilde{y} \over 2}\right)
    -\left[ \frac{3\sqrt{3}\eta_* }{2k} \left(1+\ln{\frac{k\eta_*y_\alpha}{2}}\right)\sin{\frac{k\eta_*y_\alpha}{2\sqrt{3}}}\right.\right.\nonumber\\&
    \left.\left.-\frac{3\eta_*^2 y_\alpha}{4}\ln{\frac{k\eta_*y_\alpha}{2}}\cos{\frac{k\eta_*y_\alpha}{2\sqrt{3}}}\right]\int^{y_\alpha}_{y_{\rm in}}d \tilde{y} \, \cos{\frac{k\eta_*\tilde{y}}{2\sqrt{3}}} S_{\bm{k}}\left({\eta_*\tilde{y} \over 2}\right)
      \right.\nonumber\\&\left. - \left[- \frac{3\sqrt{3}\eta_*}{2k} \left(1+\ln{\frac{k\eta_*y_\alpha}{2}}\right)\cos{\frac{k\eta_*y_\alpha}{2\sqrt{3}}}%\right.\right.\nonumber\\& \left.
      % \left.
      -\frac{3\eta_*^2 y_\alpha}{4}\ln{\frac{k\eta_*y_\alpha}{2}}\sin{\frac{k\eta_*y_\alpha}{2\sqrt{3}}}\right]\right.\nonumber\\&\left.\times \int^{y_\alpha}_{y_{\rm in}}d \tilde{y}\,\sin{\frac{k\eta_*\tilde{y}}{2\sqrt{3}}} S_{\bm{k}}\left({\eta_*\tilde{y} \over 2}\right)%\right.\nonumber \\&\left.
      +\frac{\sqrt{3}\eta_*^3k}{8} \int^{y_\alpha}_{y_{\rm in}}d\tilde{y}\, \tilde{y} \ln{\frac{k\eta_*y}{2}}\right.\nonumber \\&\left.\times \left[\sin{\frac{k\eta_*\tilde{y}}{2\sqrt{3}}} \int^{\tilde{y}}_{y_{\rm in}}d\vardbtilde{y}\, \cos{\frac{k\eta_*\vardbtilde{y}}{2\sqrt{3}}} S_{\bm{k}}\left({\eta_*\vardbtilde{y} \over 2}\right) - \cos{\frac{k\eta_*\tilde{y}}{2\sqrt{3}}} \int^{\tilde{y}}_{y_{\rm in}}d\vardbtilde{y}\, \sin{\frac{k\eta_*\vardbtilde{y}}{2\sqrt{3}}} S_{\bm{k}}\left({\eta_*\vardbtilde{y} \over 2}\right) \right] \right]\nonumber\\&
     % -\frac{1}{2}\int^{y_\alpha}_{y_{\rm in}} d \Tilde{y}  G(y, \tilde{y})  \frac{d \chi^{ij}}{d\Tilde{y}}\frac{d \chi_{ij}}{d\Tilde{y}}
     +3y \left(1- \frac{\sqrt{1+y}}{2}\ln{\frac{\sqrt{1+y_\alpha}+1}{\sqrt{1+y_\alpha}-1}}\right) \left[ \frac{3\eta_*}{2ky}\int^{y_\alpha}_{y_{\rm in}} d \tilde{y}\, \tilde{y}S_{\bm{k}}\left({\eta_*\tilde{y} \over 2}\right)  \right.\nonumber\\&\left.
    +\left[ \frac{3\sqrt{3}}{k^2y_\alpha}\sin{\frac{k\eta_*y_\alpha}{2\sqrt{3}}}+\frac{3\eta_*^2 }{4}\cos{\frac{k\eta_*y_\alpha}{2\sqrt{3}}}-\frac{3\eta_* }{2k}\cos{\frac{k\eta_*y_\alpha}{2\sqrt{3}}}\right]\int^{y_\alpha}_{y_{\rm in}}d \tilde{y}\,\cos{\frac{k\eta_*\tilde{y}}{2\sqrt{3}}} S_{\bm{k}}\left({\eta_*\tilde{y} \over 2}\right) \right.\nonumber\\&
\left.+\left[- \frac{3\sqrt{3}}{k^2y_\alpha}\cos{\frac{k\eta_*y_\alpha}{2\sqrt{3}}}+\frac{3\eta_*^2 }{4}\sin{\frac{k\eta_*y_\alpha}{2\sqrt{3}}}-\frac{3\eta_* }{2k}\sin{\frac{k\eta_*y_\alpha}{2\sqrt{3}}}\right]\int^{y}_{y_{\rm in}}d \tilde{y}\, \sin{\frac{k\eta_*\tilde{y}}{2\sqrt{3}}} S_{\bm{k}}\left({\eta_*\tilde{y} \over 2}\right) \right].
\end{align}
The tensor-sourced $\delta_{\rm m}^{(2)}$ is given by \eqref{subdel}, where the coefficients are shown in \eqref{P2} and \eqref{P1}.
In order to determine $y_\alpha$, we should express $\delta_{\rm r}^{(2)}$, computed in \eqref{radpertshort} as well  in terms of the variable $y$.
Indeed, it becomes
\begin{align}\label{deltary}
     \delta_{\rm r (DRe)}^{(2)}(\bm{k},y)&=\frac{12}{k^2y^2 } \int^{y}_{y_{\rm in}}d \tilde{y}\, \tilde{y}\,S_{\bm{k}}\left({\eta_*\tilde{y} \over 2}\right) 
      -\frac{2\sqrt{3}\eta_*}{k}\left[\sin{\frac{k\eta_*y}{2\sqrt{3}}}+\frac{2\sqrt{3}}{k\eta_*y}\cos{\frac{k\eta_*y}{2\sqrt{3}}}\right] \nonumber\\& \times \int^{y}_{y_{\rm in}}d \tilde{y}\, \cos{\frac{k\eta_*\tilde{y}}{2\sqrt{3}}}\, S_{\bm{k}}\left({\eta_*\tilde{y} \over 2}\right)
      +\frac{2\sqrt{3}\eta_*}{k}\left[\cos{\frac{k\eta_*y}{2\sqrt{3}}}-\frac{2\sqrt{3}}{k\eta_*y}\sin{\frac{k\eta_*y}{2\sqrt{3}}}\right] \nonumber\\&\times \int^{y}_{y_{\rm in}}d \tilde{y}\,\sin{\frac{k\eta_*\tilde{y}}{2\sqrt{3}}}\, S_{\bm{k}}\left({\eta_*\tilde{y} \over 2}\right).
\end{align}
(Here we have called with $ \delta_{\rm r (DRe)}^{(2)}(\bm{k},y)$ the radiation perturbation solution during the deep radiation era.)
Therefore, using the above relation Eq. \eqref{deltary}, taking into account of Eq. \eqref{deltamy} and the condition derived in Eq. \eqref{yM}, we can find an expression from which we can implicitly obtain the value of $y_\alpha$: 
\begin{align}
   & \frac{\sqrt{3}\eta_*y_\alpha}{4}\ln{\frac{k\eta_*y_\alpha}{2}}\int^{y_\alpha}_{y_{\rm in}}d \tilde{y}\, \tilde{y}\,S_{\bm{k}}\left({\eta_*\tilde{y} \over 2}\right)
    +\left[ \frac{3y_\alpha }{2k} \left(1+\ln{\frac{k\eta_*y_\alpha}{2}}\right)\sin{\frac{k\eta_*y_\alpha}{2\sqrt{3}}}\right.\nonumber\\&
    \left.
    -\frac{\sqrt{3}\eta_* y_\alpha^2}{4}\ln{\frac{k\eta_*y_\alpha}{2}}\cos{\frac{k\eta_*y_\alpha}{2\sqrt{3}}}\right]\int^{y_\alpha}_{y_{\rm in}}d \tilde{y}\,\cos{\frac{k\eta_*\tilde{y}}{2\sqrt{3}}}\, S_{\bm{k}}\left({\eta_*\tilde{y} \over 2}\right)
       \nonumber\\&+ \left[- \frac{3y_\alpha}{2k} \left(1+\ln{\frac{k\eta_*y_\alpha}{2}}\right)\cos{\frac{k\eta_*y_\alpha}{2\sqrt{3}}}-\frac{\sqrt{3}\eta_* y_\alpha^2}{4}\ln{\frac{k\eta_*y_\alpha}{2}}\sin{\frac{k\eta_*y_\alpha}{2\sqrt{3}}}\right]\nonumber\\& \times \int^{y_\alpha}_{y_{\rm in}}d \tilde{y}\,\sin{\frac{k\eta_*\tilde{y}}{2\sqrt{3}}}\, S_{\bm{k}}\left({\eta_*\tilde{y} \over 2}\right)
      -\frac{\eta_*^2ky_\alpha}{8} \int^{y_\alpha}_{y_{\rm in}}d \tilde{y}\, \tilde{y} \ln{\frac{k\eta_*y}{2}}\nonumber\\& \times  \left[\sin{\frac{k\eta_*\tilde{y}}{2\sqrt{3}}} \int^{\tilde{y}}_{y_{\rm in}}  d\vardbtilde{y}\,\cos{\frac{k\eta_*\vardbtilde{y}}{2\sqrt{3}}}\, S_{\bm{k}}\left({\eta_*\vardbtilde{y} \over 2}\right)- \cos{\frac{k\eta_*\tilde{y}}{2\sqrt{3}}} \int^{\tilde{y}}_{y_{\rm in}}d\vardbtilde{y}\, \sin{\frac{k\eta_*\vardbtilde{y}}{2\sqrt{3}}}\, S_{\bm{k}}\left({\eta_*\vardbtilde{y} \over 2}\right)\right]  \nonumber\\&=-\frac{4\sqrt{3}\alpha}{\eta_*k^2y_\alpha^2 } \int^{y_\alpha}_{y_{\rm in}}d \tilde{y}\, \tilde{y}\,S_{\bm{k}}\left({\eta_*\tilde{y} \over 2}\right)+\frac{2\alpha}{k}\left[\sin{\frac{k\eta_*y_\alpha}{2\sqrt{3}}}+\frac{2\sqrt{3}}{k\eta_*y_\alpha}\cos{\frac{k\eta_*y_\alpha}{2\sqrt{3}}}\right] \nonumber\\& \times \int^{y_\alpha}_{y_{\rm in}} d \tilde{y} \, \cos{\frac{k\eta_*\tilde{y}}{2\sqrt{3}}}\, S_{\bm{k}}\left({\eta_*\tilde{y} \over 2}\right)
      -\frac{2\alpha}{k}\left[\cos{\frac{k\eta_*y_\alpha}{2\sqrt{3}}}-\frac{2\sqrt{3}}{k\eta_*y_\alpha}\sin{\frac{k\eta_*y_\alpha}{2\sqrt{3}}}\right] \nonumber\\&\times \int^{y_\alpha}_{y_{\rm in}} d \tilde{y}\, \sin{\frac{k\eta_*\tilde{y}}{2\sqrt{3}}}\, S_{\bm{k}}\left({\eta_*\tilde{y} \over 2}\right).
\end{align}
As previously discussed, numerical analysis is required in order to obtain a precise and  proper value of $\alpha$. Following our analytical approach  this value should be between $1$ and $2$. 
We postpone this numerical work for the future.
Finally, the next section is devoted to the analytical analysis of the density perturbations produced by gravitational waves entering the Hubble radius during the era of matter and their evolution in the late phase dominated by dark energy.

 \section{Tensor-sourced CDM and radiation perturbation during matter and Dark Energy domination}\label{matt}

Before discussing in detail the subject of this section, a comment is in order.
Below, first of all we will recap the analysis made in \cite{PhysRevLett.129.091301} where they computed the matter density contribution when the Universe has become matter-dominated and for $k<k_{\rm eq}$. In particular, we will extend this approach for all modes considering both dark matter and dark energy period. Then we give an analytical solution of radiation perturbation for the same time period. Therefore our analysis is only for $\eta \gg \eta_{\rm eq}$ and all $k$ until today, i.e. at $\eta=\eta_0$.
Precisely, in order to be sure that we are in the CDM epoch, let us consider only the Universe for $\eta \ge \eta_{\rm rec}$, where $\eta_{\rm rec}$ is the conformal time at recombination. 
In fact, for $\eta \ge \eta_{\rm rec}$, the radiation perturbations become irrelevant with respect to those of matter and, consequently, the Universe is dominated by matter at both  background  and perturbation levels.
%for modes $k<k_{\rm eq}$. 
Finally, note that with the analytical prescription used in this work, we are not able to describe modes that enter the horizon around matter-radiation equality and for $1/\eta_{\rm rec} \le k \lsim 1/\eta_{\rm eq}$. For these scales we need a numerical approach. 
However, at the end of this section, we will provide a possible analytic prescription that could help solve this matching.

For the scalar modes entering in matter domination \cite{PhysRevLett.129.091301}, as we have also discussed in the main text, we can adopt comoving, synchronous, and time-orthogonal gauge.
Then, for $\eta \ge \eta_{\rm rec} \gg \eta_{\rm eq}$, we can discard the radiation perturbation within the Einstein field Equations. In this case, choosing comoving observers,  the deformation tensor is purely spatial and coincides with the extrinsic curvature of 
constant-time spatial hypersurfaces, i.e. $\theta^i_j= -K^i_j=\gamma^{ik}\gamma'_{kj}/2$. (Here the prime denotes the differentiation w.r.t. conformal time.) In this case, let us  consider the continuity and Raychaudhury equation \cite{Ehlers:1961xww, 2009GReGr..41..581E, Matarrese_1998} 
\begin{align}
   {\theta^{(2)}}'+{\cal H} \theta^{(2)}+2\theta^{(1)i}_j \theta^{(1)j}_i+{4 \pi G} a^2 \bar \rho_{\rm m} \delta_{\rm m}^{(2)}&=0,\\
   {\delta_{\rm m}^{(2)}}'+2  \delta_{\rm m}^{(1)}\theta^{(1)}+ \theta^{(2)}&=0,
\end{align}
where $\theta$ is the trace of $\theta^i_j$ and represents the inhomogeneous part of the volume expansion.
Taking into account that the additive term $2\delta_{\rm m}^{(1)}\theta^{(1)}$ can be discarded because it is independent of the tensor contribution, we can combine these equations to obtain the evolution equation of $\delta_{\rm m}^{(2)}$
\begin{equation}\label{evol}
   {\delta_{\rm m}^{(2)}}''+{\cal H}{\delta_{\rm m}^{(2)}}'-4\pi G a^2 \bar \rho_{\rm m}\delta_{\rm m}^{(2)}=\frac{1}{2} {\chi^{ij}}'{\chi_{ij}}'.
\end{equation}
Now, let us consider the usual relation 
$$ a^2 \bar \rho_{\rm m} = {3 \over 8 \pi G} {\cal H}^2 \Omega_{\rm m} = {3 \over 8 \pi G} {{\cal H}_0^2 \Omega_{\rm m 0} \over a}\,, $$
where  the cosmological parameter is defined as $\Omega_{\rm m} = 8 \pi G \bar \rho_{\rm m } a^2/ 3 {\cal H}^2$ and $\Omega_{\rm m 0 } = 8 \pi G \bar \rho_{\rm m 0}/ 3  {\cal H}_0^2=\Omega_{\rm m} (\eta=\eta_0)$ (here we have normalised the scale factor today as $a_0=a(\eta_0)=1$), and
 the usual background equations (that we set for $\eta \ge \eta_{\rm rec}$)
$${\cal H}^2= {{{\cal H}_0}^2 \Omega_{\rm m 0} \over a} \left(1 + R_0 a^3\right) \quad \quad {\rm and}  \quad \quad  {\cal H}'={\cal H}^2 \left(1 - {3 \over 2} \Omega_{\rm m} \right)\,,$$
where $R_0 = \Omega_{\Lambda 0 } / \Omega_{\rm m 0}=(1-\Omega_{\rm m 0})/\Omega_{\rm m 0} $. 
Then Eq. \eqref{evol}, in terms of the variable $\iota=R_0^{1/3} a$, becomes
\begin{equation}\label{evol-iota}
   \iota^2 \left(1 +\iota^3\right){d^2 \delta_{\rm m}^{(2)} \over d \iota^2}+{3 \over 2} \iota \left(1+2 \iota^3\right){d \delta_{\rm m}^{(2)} \over d \iota} -{3 \over 2}\delta_{\rm m}^{(2)}=S_{\rm m},
\end{equation}
where we have defined the source term in the following way
%\[S_{\rm m}(\xi(\eta), \eta, k)=\frac{1}{2{{\cal H}_0}^2 \Omega_{\rm m 0}} {\chi^{ij}}'\chi'_{ij}. \PB{\frac{a}{2{{\cal H}_0}^2 \Omega_{\rm m 0}} {\chi^{ij}}'\chi'_{ij}}\]
\[S_{\rm m}(\iota)=\frac{\iota^2(1+\iota^3)}{2}\frac{d\chi^{ij}}{d\iota}\frac{d \chi_{ij}}{d\iota},\]
and the following variable transformation rules have been used
\begin{align}
    \frac{d}{d\eta}%&=  \frac{d}{da} \frac{da}{d\eta}%\nonumber\\
  %  &=a'  \frac{d}{da}\nonumber\\
  %  &=a\mathcal{H} \frac{d}{da}\nonumber\\
    %&
    =\mathcal{H}_0 R_0^{1/6}\sqrt{ \Omega_{\rm m 0} \iota (1+\iota^3)} \frac{d}{d\iota},
\end{align}
and 
\begin{equation}\label{etatoiota}
     \frac{d^2}{d\eta^2}= \mathcal{H}_0^2 R_0^{1/3} \Omega_{\rm m 0}\left( \frac{ 1+4\iota^3}{2} \frac{d}{d\iota}+\iota(1+\iota^3) \frac{d^2}{d\iota^2}\right)\,.
\end{equation}
%Note that we are already in Fourier space.
Here note that the meaning of  $\iota$ can be easily related to the normalized scale factor at $\eta_\Lambda$  which is the conformal time when $\Omega_{\rm m}= \Omega_{\Lambda}$, i.e. $\iota(\eta)=a(\eta)/a(\eta_\Lambda)$ \cite{mukhanov2005physical}.
%The homogeneous solution 
The general solution in Fourier space is
\begin{align}\label{finaldelta-iota}
\delta_{\rm m}^{(2)}({\bm k},\iota) = D_-(\iota) C_-({\bm k}) + D_+ (\iota) C_+ ({\bm k}) + \int_{\iota_{\rm rec}}^\iota   d\tilde \iota \, G_{\rm m} (\iota, \tilde \iota) \, S_{\rm m}(\tilde \iota,\tilde \eta,{\bm k} )
\end{align}
where $\tilde \eta=\eta (\tilde \iota)$, $\iota_{\rm rec}=\iota( \eta_{\rm rec})$ and the source term is the Fourier transform of $S_{\rm m}(\iota)$, i.e.
\begin{align}\label{S_m-iota}
S_{\rm m}(\bm{k}, \tilde \iota,\tilde \eta)&=\frac{\tilde \iota^2(1+\tilde \iota^3)}{2} F_2(\bm{k},\tilde \iota),
%\sum\limits_{\sigma,\sigma'}\int \frac{d^3\bm{q}}{(2\pi)^3} A_{\sigma'} (\bm{q})A_{\sigma} \left(\bm{k}-\bm{q}\right)\nonumber\\&\times\epsilon^{\sigma'}_{ij}(\bm{\hat{q}})\epsilon^{\sigma ij}\left(\widehat{\bm{k}-\bm{q}}\right) \frac{d\mathcal{T}\left(q\tilde \iota \right)}{d \tilde \iota}\frac{d\mathcal{T}\left(|\bm{k}-\bm{q}|\tilde \iota\right)}{d\tilde \iota}. 
\end{align}
where the functional form of $F_2$ is defined as the  Fourier transform of $d \chi^{ij}/d\iota\, d \chi_{ij}/d\iota$, i.e.
\begin{align}\label{F2}
F_2(\bm{k},\iota)&=\sum\limits_{\sigma,\sigma'}\int \frac{d^3\bm{q}}{(2\pi)^3} A_{\sigma'} (\bm{q})A_{\sigma} \left(\bm{k}-\bm{q}\right)\nonumber\\&\times\epsilon^{\sigma'}_{ij}(\bm{\hat{q}})\epsilon^{\sigma ij}\left(\widehat{\bm{k}-\bm{q}}\right) \frac{d\mathcal{T}\left(q,\iota\right)}{d \iota}\frac{d\mathcal{T}\left(|\bm{k}-\bm{q}|,\iota\right)}{d\iota}. 
\end{align}
Obviously, we connect $F_2$ with $S_{\bm k}$ (see Eq. \eqref{S_k-eta})
\begin{align}\label{F2toSk}
F_2(\bm{k},\iota)=-{ 6 \over \mathcal{H}_0^2 R_0^{1/3} \Omega_{\rm m 0} \iota} {S_{\bm k} \over (1+\iota^3)}
\end{align}
and, therefore, we have
\begin{align}\label{SmtoSk}
S_{\rm m}({\bm k}, \tilde \iota,\tilde \eta)= -{ 3 \over \mathcal{H}_0^2 R_0^{1/3} \Omega_{\rm m 0} } \iota S_{\bm k}\;.
\end{align}

Here, $D_-$ and $D_+$ are  \cite{mukhanov2005physical}
\begin{align}
D_- (\iota) &= \sqrt{1+\iota^{-3}},\\
D_+ (\iota) &= D_-(\iota) \int_{0}^\iota d\tilde \iota \left({\tilde \iota \over 1+\tilde \iota^3}\right)^{3/2},
\end{align}
and, consequently, the Green's function is defined as
\begin{align}\label{Green-iota}
G_{\rm m} (\iota, \tilde \iota) = (1+\tilde \iota^3)\sqrt{1+\iota^{-3}}\left(\int_{0}^\iota d\tilde \iota \left({\tilde \iota \over 1+\tilde \iota^3}\right)^{3/2}-\int_{0}^{\tilde \iota} d\vardbtilde \iota \left({\vardbtilde \iota \over 1+\vardbtilde \iota^3}\right)^{3/2}\right).
\end{align}
[Here, for $D_+$, we have followed the conventions used in  \cite{mukhanov2005physical}. Consequently the integrals in $G_{\rm m}$ start at $\iota=0$.]

An important comment is in order. As we observe from Eq. 
\eqref{F2}, the source term $F_2(\iota, \bm{k})$ [and, consequently, $S_{\bm{k}}(\iota, \eta )$], which contains two $\cal T$, cannot be computed by using the analytical expressions defined in Eq. \eqref{transfer_function}, see also \cite{Watanabe:2006qe}. This is because these expressions are  obtained ignoring the contribution of the cosmological constant/Dark Energy.
In this case we need a new definition of  $\cal T$, which is not analytical, that takes into account the acceleration expansion of the Universe (note that this  very late time epoch reduces the amplitude of $h_{ij}$ for $1/\eta_0 \ge k \ge 1/ \eta_\Lambda$).

Now, as also pointed out in \cite{PhysRevLett.129.091301}, for $k \le 1/\eta_{\rm rec} < k_{\rm eq}$, we are focusing on density contrast for waves entering the horizon during  matter and dark energy domination. In this case we will not consider the homogeneous solutions. Instead for $k>k_{\rm eq}$ we need to keep all the terms in Eq. \eqref{finaldelta-iota}. For these modes we need a matching conditions at $\eta=\eta_{\rm rec}$ between the general solution Eq. \eqref{finaldelta-iota} and the one that we obtain in Section \ref{subrd}, i.e. Eq. \eqref{subdel} (e.g., see the light-green area for $\eta \ge \eta_{\rm rec}$ in Fig. \ref{Figure}). Moreover, for this match, if $\eta \to \eta^+_{\rm rec}$, $\iota \ll 1$.  In this limit, we can discard the inhomogeneous solution  and Eq. \eqref{finaldelta-iota} reads \cite{mukhanov2005physical}
\begin{align}\label{finaldelta-xi_rec}
\delta_{\rm m}^{(2)}(\iota \to \iota^+_{\rm rec},{\bf k}) = \iota^{-3/2} \, C_-({\bf k}) + {2 \over 5} \iota \, C_+({\bf k}) \;.
\end{align}
Now, using the definition of $\iota=R_0^{1/3} a$ and $y=a/a_{\rm eq}$, Eq. \eqref{finaldelta-xi_rec} can be written in the following way
\begin{align}\label{finaldelta-y_rec}
\delta_{\rm m}^{(2)}(y \to y^+_{\rm rec},{\bf k}) =  \left( a_{\rm eq}^3 R_0\right)^{-1/2} y^{-3/2} \, C_-({\bf k}) + {2 \over 5}a_{\rm eq} R_0^{1/3} y  \, C_+({\bf k}) \;.
\end{align}
Now we can impose the following matching conditions
 \begin{align}\label{deltam-match-rec}
   \delta_{\rm m (Tmesz)}^{(2)}(\bm{k},y \to y_{\rm rec}^-) \big|_{y_{\rm rec}}&=  \delta_{\rm m}^{(2)}(\bm{k},y \to y_{\rm rec}^+)\big|_{y_{\rm rec}},\\
\label{deltam-dotmatch-rec}
     {\left({d \over d y}{ \delta}_{\rm m (Tmesz)}^{(2)}(\bm{k},y\to y_{\rm rec}^-)\right)}\Bigg|_{y_{\rm rec}}&=    {\left({d \over d y}{\delta}_{\rm m }^{(2)}(\bm{k},y \to y_{\rm rec}^+)\right)}\Bigg|_{y_{\rm rec}},
\end{align}
which allow to get $C_-$ and $C_+$. Here $\delta_{\rm m (Tmesz)}^{(2)}(\bm{k},y \to y_{\rm rec}^-)$ is exactly  Eq. \eqref{subdelk} at $y\to y_{\rm rec}^-$. The matching conditions Eqs. \eqref{deltam-match-rec} and \eqref{deltam-dotmatch-rec} give us
\begin{align}\label{deltaMDmatch}
    &P_1(\bm{k}) D_1(y_{\rm rec})+P_2(\bm{k}) D_2(y_{\rm rec}) +\frac{1}{2}\int_{y_\alpha}^{y_{\rm rec}} d \Tilde{y} \, G(y, \tilde{y}) % \frac{d \chi^{ij}}{d\Tilde{y}}\frac{d \chi_{ij}}{d\Tilde{y}}
    F_1(\tilde y, \bm{k})\nonumber\\&= \left( a_{\rm eq}^3 R_0\right)^{-1/2} {y_{\rm rec}}^{-3/2} \, C_-({\bf k}) + {2 \over 5}a_{\rm eq} R_0^{1/3} y_{\rm rec}  \, C_+({\bf k}),
\end{align}
and 
\begin{align}\label{deltadotMDmatch}
    & P_1(\bm{k}) + P_2(\bm{k}) \left(- \frac{2(1+3y_{\rm rec})}{3y_{\rm rec}\sqrt{1+y_{\rm rec}}}+\ln{\frac{\sqrt{1+y_{\rm rec}}+1}{\sqrt{1+y_{\rm rec}}-1}}\right)\nonumber\\&=
   -\frac{3}{2} \left( a_{\rm eq}^3 R_0\right)^{-1/2} {y_{\rm rec}}^{-5/2} \, C_- ({\bf k}) + {2 \over 5}a_{\rm eq} R_0^{1/3}  \, C_+({\bf k})
\end{align}
respectively.
 Multiplying Eq. \eqref{deltadotMDmatch} with $y_{\rm rec}$ 
and subtracting it from  Eq. \eqref{deltaMDmatch}, we obtain 
\begin{align}
   C_- ({\bf k})&= \frac{4}{15}  \left( a_{\rm eq}^3 y_{\rm eq}^3 R_0\right)^{1/2}\left[P_1(\bm{k})+P_2(\bm{k})\left(- \frac{2}{\sqrt{1+y_{\rm rec}}}+\ln{\frac{\sqrt{1+y_{\rm rec}}+1}{\sqrt{1+y_{\rm rec}}-1}}\right)\right.\nonumber\\& \left. +\frac{3}{4}\int_{y_\alpha}^{y_{\rm rec}} d \Tilde{y} \, G(y, \tilde{y}) F_1(\tilde y, \bm{k}) %\frac{d \chi^{ij}}{d\Tilde{y}}\frac{d \chi_{ij}}{d\Tilde{y}}
   \right].
\end{align}
Similarly, multiplying Eq. \eqref{deltaMDmatch} with $3/(2y_{\rm rec})$ and adding it to  Eq. \eqref{deltadotMDmatch}, we have
\begin{align}
   C_+({\bf k})&= \frac{1}{a_{\rm eq}R_0^{1/3}}  \left[\left(\frac{5}{2}+\frac{1}{y_{\rm rec}}\right)\left(P_1(\bm{k})+P_2(\bm{k})\ln{\frac{\sqrt{1+y_{\rm rec}}+1}{\sqrt{1+y_{\rm rec}}-1}}\right)\right.\nonumber\\& \left.
   -\frac{P_2(\bm{k})}{3}\frac{11+15y_{\rm rec}}{y_{\rm rec}\sqrt{1+y_{\rm rec}}}+\frac{3}{4y_{\rm rec}}\int_{y_\alpha}^{y_{\rm rec}} d \Tilde{y} \, G(y, \tilde{y})F_1(\tilde y, \bm{k})  %\frac{d \chi^{ij}}{d\Tilde{y}}\frac{d \chi_{ij}}{d\Tilde{y}}
   \right],
\end{align}
where $P_1(\bm{k}),P_2(\bm{k})$ are given by Eq. \eqref{P1} and Eq. \eqref{P2} respectively. The form of $D_1(y),D_2(y)$ is shown in Section \ref{subrd}.

Finally, let us focus on the radiation contribution due to the primordial gravitational waves entering the horizon during 
the late time. For $\eta \gtrsim  \eta_{\rm rec}$, radiation is not the dominant contribution at both the background and perturbation levels and, as we have pointed out earlier, it can be discarded in the perturbed Einstein field equations. However its effect could be non-negligible in the CMB and we are able to find an expression of radiation density perturbation using  \eqref{deltam}, \eqref{radc}, and \eqref{momr}, which are still valid. 
Now, inserting \eqref{deltam} into \eqref{radc}, we have
\begin{equation}
     {\delta_{\rm r}^{(2)}}'+\frac{4}{3}\nabla^2 v^{(2)}_{\rm r}= \frac{4}{3} {\delta_{\rm m }^{(2)}}'.
\end{equation}
Differentiating the above equation w.r.t. time, and using \eqref{momr}, we obtain
\begin{equation}
    {\delta_{{\rm r} \bm{k}}^{(2)}}''+\frac{k^2}{3}\delta^{(2)}_{{\rm r}\bm{k}}= \frac{4}{3} {\delta_{{\rm m} \bm{k}}^{(2)}}'',
\end{equation}
which gives us an expression for $\delta_{\rm r}^{(2)}$
\begin{align}\label{deltarmd}
   \delta^{(2)}_{{\rm r}\bm{k}}(\eta)&= A_{\rm r MD}(\bm{k}) \cos{\frac{k\eta}{\sqrt{3}}} +B_{\rm r MD}(\bm{k}) \sin{\frac{k\eta}{\sqrt{3}}}\nonumber\\& +
   \frac{\sqrt{3}}{k}  \int_{\eta_{\rm rec}}^\eta d\tilde{\eta}\,\left(\sin{\frac{k\eta}{\sqrt{3}}} \cos{\frac{k\tilde{\eta}}{\sqrt{3}}}-\cos{\frac{k\eta}{\sqrt{3}}}\sin{\frac{k\tilde{\eta}}{\sqrt{3}}}  \right)\, S_{\bm{k}\rm MD}(\tilde{\eta})\;,
\end{align}
where the source term $S_{\bm{k}\rm MD}$ can be obtained by differentiating Eq. \eqref{finaldelta-iota} twice. Note that here $\eta=\eta(\iota)$, the exact functional form can be obtained by solving the Friedmann equations.
Just as discussed in the case of $ {\delta_{\rm m }^{(2)}}$, the solution can be divided for two separate range of scales. Firstly, for  $k \le 1/\eta_{\rm rec} < k_{\rm eq}$, we can ignore the homogeneous solution, and $\delta^{(2)}_{{\rm r}\bm{k}}$ is
\begin{equation}
   \delta^{(2)}_{{\rm r}\bm{k}}(\iota)=
   \frac{\sqrt{3}}{k}  \int_{\iota_{\rm rec}}^\iota d\tilde{\iota}\,\left(\sin{\frac{k\eta(\iota)}{\sqrt{3}}} \cos{\frac{k\tilde{\eta}(\tilde \iota)}{\sqrt{3}}}-\cos{\frac{k\eta(\iota)}{\sqrt{3}}}\sin{\frac{k\tilde{\eta}(\tilde \iota)}{\sqrt{3}}}  \right)\, S_{\bm{k}\rm MD}(\tilde{\iota})\;.
\end{equation}
 $S_{\bm{k}\rm MD}$ can be written in terms of $\iota$, using rule \eqref{etatoiota}, as
\begin{align}
    S_{\bm{k}\rm MD}&=\frac{4}{3} {\delta_{\rm m \bm{k}}^{(2)}}''\nonumber\\&= \frac{4}{3} \mathcal{H}_0^2 R_0^{1/3} \Omega_{\rm m 0}\left( \frac{ 1+4\iota^3}{2} \frac{d}{d\iota}+\iota(1+\iota^3) \frac{d^2}{d\iota^2}\right) \delta_{\rm m \bm{k}}^{(2)}\nonumber\\&=\frac{4}{3}\mathcal{H}_0^2 R_0^{1/3} \Omega_{\rm m 0}\left[3\iota^{-5/2}\sqrt{1+\iota^3}\int_{\iota_{\rm rec}}^\iota(1+\tilde \iota^3) S_{\rm m}(\tilde \iota,{\bf k} )\right.\nonumber\\&\times \left.\left[\int_0^\iota d\tilde \iota \left({\tilde \iota \over 1+\tilde \iota^3}\right)^{3/2}-\int_0^{\tilde \iota} d\vardbtilde \iota \left({\vardbtilde \iota \over 1+\vardbtilde \iota^3}\right)^{3/2}-\frac{\iota^{5/2}}{3\sqrt{1+\iota^3}}\right]+\iota(1+\iota^3)S_{\rm m}(\iota,{\bf k} )\right]\,,%\\&
     %+4\mathcal{H}_0^2 R_0^{1/3} \Omega_{\rm m 0} \left[\xi^{-5/2}\sqrt{1+\xi^3}\left(C_1(\bm{k}) +C_2(\bm{k}) \int_0^\xi d\tilde \xi \left({\tilde \xi \over 1+\tilde \xi^3}\right)^{3/2}\right)-\frac{C_2(\bm{k})}{3(1+\xi^3)}(1+3\xi^3)\right].
\end{align}
where Eq. \eqref{finaldelta-iota} has been used, without the homogeneous part. In the second range of scales ($k>k_{\rm eq}$) instead, the whole Eq. \eqref{finaldelta-iota} should be used to get $ S_{\bm{k}\rm MD}$
\begin{align}
    S_{\bm{k}\rm MD}(\iota)&= 4\mathcal{H}_0^2 R_0^{1/3} \Omega_{\rm m 0}\left[\iota^{-5/2}\sqrt{1+\iota^3}\left(\int_{\iota_{\rm rec}}^\iota (1+\tilde \iota^3) S_{\rm m}(\tilde \iota,{\bf k} )\right.\right.\nonumber\\& \left.\left.\times\left[\int_0^\iota d\tilde \iota \left({\tilde \iota \over 1+\tilde \iota^3}\right)^{3/2}-\int_0^{\tilde \iota} d\vardbtilde \iota \left({\vardbtilde \iota \over 1+\vardbtilde \iota^3}\right)^{3/2}-\frac{\iota^{5/2}}{3\sqrt{1+\iota^3}}\right]\right.\right.\nonumber\\& \left.\left.+\left[C_-(\bm{k}) +C_+(\bm{k}) \int_0^\iota d\tilde \iota \left({\tilde \iota \over 1+\tilde \iota^3}\right)^{3/2}\right]\right)+\frac{\iota(1+\iota^3)S_{\rm m}(\iota,{\bf k} )}{3}-\frac{C_+(\bm{k})}{3(1+\iota^3)}(1+3\iota^3)\right].
\end{align}
For the initial conditions $ A_{\rm r MD}(\bm{k}),  B_{\rm r MD}(\bm{k})$, we have to connect Eq. \eqref{deltarmd} with eq \eqref{subdeltar} at  $ \eta=\eta_{\rm rec}$. For $\eta \rightarrow \eta_{\rm rec}^-$, remembering the argument before Eq. \eqref{deltamy}, we can write $y=\xi^2=(\eta/\eta_\star)^2$, as $\xi \gg 1$. 
So Eq. \eqref{subdeltar} reads
\begin{align}
    \delta_{\rm r (Tmesz)}^{(2)} (\bm{k},\eta)&= A_{\rm r}(\bm{k}) \cos{\left(2\sqrt{\frac{2}{3}}\frac{k}{k_{\rm eq}}\sqrt{1+\left(\frac{\eta}{\eta_\star}\right)^2}\right)}+B_{\rm r}(\bm{k}) \sin{\left(2\sqrt{\frac{2}{3}}\frac{k}{k_{\rm eq}}\sqrt{1+\left(\frac{\eta}{\eta_\star}\right)^2}\right)}\nonumber\\& 
    + \frac{k_{\rm eq}}{k\eta_*^2}\sqrt{\frac{3}{2}}\int_{\eta_\alpha}^\eta d \Tilde{\eta} \, \Tilde{\eta}  \mathcal{Q}_{\bm{k}}(\tilde{\eta})\left[\sin{\left(2\sqrt{\frac{2}{3}}\frac{k}{k_{\rm eq}}\sqrt{1+(\frac{\eta}{\eta_\star})^2}\right)}\cos{\left(2\sqrt{\frac{2}{3}}\frac{k}{k_{\rm eq}}\sqrt{1+\left(\frac{\tilde \eta}{\eta_\star}\right)^2}\right)}\right.\nonumber\\&\left.-\cos{\left(2\sqrt{\frac{2}{3}}\frac{k}{k_{\rm eq}}\sqrt{1+\left(\frac{\eta}{\eta_\star}\right)^2}\right)}\sin{\left(2\sqrt{\frac{2}{3}}\frac{k}{k_{\rm eq}}\sqrt{1+\left(\frac{\tilde \eta}{\eta_\star}\right)^2}\right)} \right].
\end{align}
Applying following matching conditions
 \begin{align}\label{deltar-match-rec}
  \delta_{\rm r (Tmesz)}^{(2)}(\bm{k},\eta \to \eta_{\rm rec}^-) \big|_{\eta_{\rm rec}}&=  \delta_{\rm r }^{(2)}(\bm{k},\eta \to \eta_{\rm rec}^+)\big|_{\eta_{\rm rec}},\\
\label{deltar-dotmatch-rec}
     {\left({d \over d \eta}{ \delta}_{\rm r (Tmesz)}^{(2)}(\bm{k},\eta\to \eta_{\rm rec}^-)\right)}\Bigg|_{\eta_{\rm rec}}&=    {\left({d \over d \eta}{\delta}_{\rm r }^{(2)}(\bm{k},\eta \to \eta_{\rm rec}^+)\right)}\Bigg|_{\eta_{\rm rec}},
\end{align}
where the rhs is the solution Eq. \eqref{deltarmd},
we get, respectively,
\begin{align}\label{deltarMDmatch1}
    &A_{\rm r}(\bm{k}) \cos{\left(2\sqrt{\frac{2}{3}}\frac{k}{k_{\rm eq}}\sqrt{1+\left(\frac{\eta_{\rm rec}}{\eta_\star}\right)^2}\right)}+B_{\rm r}(\bm{k}) \sin{\left(2\sqrt{\frac{2}{3}}\frac{k}{k_{\rm eq}}\sqrt{1+\left(\frac{\eta_{\rm rec}}{\eta_\star}\right)^2}\right)}\nonumber\\& 
    + \frac{k_{\rm eq}}{k\eta_*^2}\sqrt{\frac{3}{2}}\int_{\eta_\alpha}^{\eta_{\rm rec}} d \Tilde{\eta} \, \Tilde{\eta}  \mathcal{Q}_{\bm{k}}(\tilde{\eta})\left[\sin{\left(2\sqrt{\frac{2}{3}}\frac{k}{k_{\rm eq}}\sqrt{1+\left(\frac{\eta_{\rm rec}}{\eta_\star}\right)^2}\right)}\cos{\left(2\sqrt{\frac{2}{3}}\frac{k}{k_{\rm eq}}\sqrt{1+\left(\frac{\tilde \eta}{\eta_\star}\right)^2}\right)}\right.\nonumber\\&\left.-\cos{\left(2\sqrt{\frac{2}{3}}\frac{k}{k_{\rm eq}}\sqrt{1+\left(\frac{\eta_{\rm rec}}{\eta_\star}\right)^2}\right)}\sin{\left(2\sqrt{\frac{2}{3}}\frac{k}{k_{\rm eq}}\sqrt{1+\left(\frac{\tilde \eta}{\eta_\star}\right)^2}\right)} \right]\nonumber\\&=A_{\rm r MD}(\bm{k}) \cos{\frac{k\eta_{\rm rec}}{\sqrt{3}}} +B_{\rm r MD}(\bm{k}) \sin{\frac{k\eta_{\rm rec}}{\sqrt{3}}},
\end{align}
and
\begin{align}\label{deltarMDmatch2}
    & \frac{2\sqrt{2}}{k_{\rm eq}}\frac{\eta_{\rm rec}}{\eta_*^2\sqrt{1+\left(\frac{\eta_{\rm rec}}{\eta_\star}\right)^2}} \left[-A_{\rm r}(\bm{k}) \sin{\left(2\sqrt{\frac{2}{3}}\frac{k}{k_{\rm eq}}\sqrt{1+\left(\frac{\eta_{\rm rec}}{\eta_\star}\right)^2}\right)}\right.\nonumber\\&\left.+B_{\rm r}(\bm{k}) \cos{\left(2\sqrt{\frac{2}{3}}\frac{k}{k_{\rm eq}}\sqrt{1+\left(\frac{\eta_{\rm rec}}{\eta_\star}\right)^2}\right)}\right]+\frac{2\sqrt{3}\eta_{\rm rec}}{k\eta_*^4\sqrt{1+\left(\frac{\eta_{\rm rec}}{\eta_\star}\right)^2}} \nonumber\\&
    \times \int_{\eta_\alpha}^{\eta_{\rm rec}} d \Tilde{\eta} \, \Tilde{\eta}  \mathcal{Q}_{\bm{k}}(\tilde{\eta})\left[\cos{\left(2\sqrt{\frac{2}{3}}\frac{k}{k_{\rm eq}}\sqrt{1+\left(\frac{\eta_{\rm rec}}{\eta_\star}\right)^2}\right)}\cos{\left(2\sqrt{\frac{2}{3}}\frac{k}{k_{\rm eq}}\sqrt{1+\left(\frac{\tilde \eta}{\eta_\star}\right)^2}\right)}\right.\nonumber\\&\left.+\sin{\left(2\sqrt{\frac{2}{3}}\frac{k}{k_{\rm eq}}\sqrt{1+\left(\frac{\eta_{\rm rec}}{\eta_\star}\right)^2}\right)}\sin{\left(2\sqrt{\frac{2}{3}}\frac{k}{k_{\rm eq}}\sqrt{1+\left(\frac{\tilde \eta}{\eta_\star}\right)^2}\right)} \right]\nonumber\\&=
   - A_{\rm r MD}(\bm{k}) \sin{\frac{k\eta_{\rm rec}}{\sqrt{3}}} +B_{\rm r MD}(\bm{k}) \cos{\frac{k\eta_{\rm rec}}{\sqrt{3}}}.
\end{align}

Multiplying Eq. \eqref{deltarMDmatch1} with $\sin(k\eta_{\rm rec}/\sqrt{3})$, Eq. \eqref{deltarMDmatch2} with $\cos(k\eta_{\rm rec}/\sqrt{3})$, and adding them, we get
\begin{align}
    B_{\rm r MD}(\bm{k}) &= A_{\rm r}(\bm{k})\left[\cos{\left(2\sqrt{\frac{2}{3}}\frac{k}{k_{\rm eq}}\sqrt{1+\left(\frac{\eta_{\rm rec}}{\eta_\star}\right)^2}\right)}\sin{\frac{k\eta_{\rm rec}}{\sqrt{3}}}\right.\nonumber\\&\left.-\frac{2\sqrt{2}}{k_{\rm eq}}\frac{\eta_{\rm rec}}{\eta_*^2\sqrt{1+\left(\frac{\eta_{\rm rec}}{\eta_\star}\right)^2}}\sin{\left(2\sqrt{\frac{2}{3}}\frac{k}{k_{\rm eq}}\sqrt{1+\left(\frac{\eta_{\rm rec}}{\eta_\star}\right)^2}\right)}\cos{\frac{k\eta_{\rm rec}}{\sqrt{3}}}\right]\nonumber\\&
    +B_{\rm r}(\bm{k})\left[\sin{\left(2\sqrt{\frac{2}{3}}\frac{k}{k_{\rm eq}}\sqrt{1+\left(\frac{\eta_{\rm rec}}{\eta_\star}\right)^2}\right)}\sin{\frac{k\eta_{\rm rec}}{\sqrt{3}}}\right.\nonumber\\&\left.+\frac{2\sqrt{2}}{k_{\rm eq}}\frac{\eta_{\rm rec}}{\eta_*^2\sqrt{1+\left(\frac{\eta_{\rm rec}}{\eta_\star}\right)^2}}\cos{\left(2\sqrt{\frac{2}{3}}\frac{k}{k_{\rm eq}}\sqrt{1+\left(\frac{\eta_{\rm rec}}{\eta_\star}\right)^2}\right)}\cos{\frac{k\eta_{\rm rec}}{\sqrt{3}}}\right]\nonumber\\&
    +\int_{\eta_\alpha}^{\eta_{\rm rec}} d \Tilde{\eta} \, \Tilde{\eta}  \mathcal{Q}_{\bm{k}}(\tilde{\eta})\Bigg\{\frac{k_{\rm eq}}{k\eta_*^2}\sqrt{\frac{3}{2}}\sin{\frac{k\eta_{\rm rec}}{\sqrt{3}}}\nonumber\\&\times \left[\sin{\left(2\sqrt{\frac{2}{3}}\frac{k}{k_{\rm eq}}\sqrt{1+\left(\frac{\eta_{\rm rec}}{\eta_\star}\right)^2}\right)}\cos{\left(2\sqrt{\frac{2}{3}}\frac{k}{k_{\rm eq}}\sqrt{1+\left(\frac{\tilde \eta}{\eta_\star}\right)^2}\right)}\right.\nonumber\\&\left.-\cos{\left(2\sqrt{\frac{2}{3}}\frac{k}{k_{\rm eq}}\sqrt{1+\left(\frac{\eta_{\rm rec}}{\eta_\star}\right)^2}\right)}\sin{\left(2\sqrt{\frac{2}{3}}\frac{k}{k_{\rm eq}}\sqrt{1+\left(\frac{\tilde \eta}{\eta_\star}\right)^2}\right)} \right]\nonumber\\&+\frac{2\sqrt{3}\eta_{\rm rec}}{k\eta_*^4\sqrt{1+\left(\frac{\eta_{\rm rec}}{\eta_\star}\right)^2}} \cos{\frac{k\eta_{\rm rec}}{\sqrt{3}}}\left[\cos{\left(2\sqrt{\frac{2}{3}}\frac{k}{k_{\rm eq}}\sqrt{1+\left(\frac{\eta_{\rm rec}}{\eta_\star}\right)^2}\right)}\right.\nonumber\\&\left. \times \cos{\left(2\sqrt{\frac{2}{3}}\frac{k}{k_{\rm eq}}\sqrt{1+\left(\frac{\tilde \eta}{\eta_\star}\right)^2}\right)}+\sin{\left(2\sqrt{\frac{2}{3}}\frac{k}{k_{\rm eq}}\sqrt{1+\left(\frac{\eta_{\rm rec}}{\eta_\star}\right)^2}\right)}\right.\nonumber\\&\left. \times \sin{\left(2\sqrt{\frac{2}{3}}\frac{k}{k_{\rm eq}}\sqrt{1+\left(\frac{\tilde \eta}{\eta_\star}\right)^2}\right)} \right]\Bigg\}.
\end{align}
Multiplying Eq. \eqref{deltarMDmatch1} with $\cos{(k\eta_{\rm rec}/\sqrt{3})}$, Eq. \eqref{deltarMDmatch2} with $\sin{(k\eta_{\rm rec}/\sqrt{3})}$, and subtracting the latter from the former, we get
\begin{align}
    A_{\rm r MD}(\bm{k}) &= A_{\rm r}(\bm{k})\left[\cos{\left(2\sqrt{\frac{2}{3}}\frac{k}{k_{\rm eq}}\sqrt{1+\left(\frac{\eta_{\rm rec}}{\eta_\star}\right)^2}\right)}\cos{\frac{k\eta_{\rm rec}}{\sqrt{3}}}\right.\nonumber\\&\left.+\frac{2\sqrt{2}}{k_{\rm eq}}\frac{\eta_{\rm rec}}{\eta_*^2\sqrt{1+\left(\frac{\eta_{\rm rec}}{\eta_\star}\right)^2}}\sin{\left(2\sqrt{\frac{2}{3}}\frac{k}{k_{\rm eq}}\sqrt{1+\left(\frac{\eta_{\rm rec}}{\eta_\star}\right)^2}\right)}\sin{\frac{k\eta_{\rm rec}}{\sqrt{3}}}\right]\nonumber\\&
    +B_{\rm r}(\bm{k})\left[\sin{\left(2\sqrt{\frac{2}{3}}\frac{k}{k_{\rm eq}}\sqrt{1+\left(\frac{\eta_{\rm rec}}{\eta_\star}\right)^2}\right)}\cos{\frac{k\eta_{\rm rec}}{\sqrt{3}}}\right.\nonumber\\&\left.-\frac{2\sqrt{2}}{k_{\rm eq}}\frac{\eta_{\rm rec}}{\eta_*^2\sqrt{1+\left(\frac{\eta_{\rm rec}}{\eta_\star}\right)^2}}\cos{\left(2\sqrt{\frac{2}{3}}\frac{k}{k_{\rm eq}}\sqrt{1+\left(\frac{\eta_{\rm rec}}{\eta_\star}\right)^2}\right)}\sin{\frac{k\eta_{\rm rec}}{\sqrt{3}}}\right]\nonumber\\&
    +\int_{\eta_\alpha}^{\eta_{\rm rec}} d \Tilde{\eta} \, \Tilde{\eta}  \mathcal{Q}_{\bm{k}}(\tilde{\eta})\Bigg\{\frac{k_{\rm eq}}{k\eta_*^2}\sqrt{\frac{3}{2}}\cos{\frac{k\eta_{\rm rec}}{\sqrt{3}}}\nonumber\\&\times \left[\sin{\left(2\sqrt{\frac{2}{3}}\frac{k}{k_{\rm eq}}\sqrt{1+\left(\frac{\eta_{\rm rec}}{\eta_\star}\right)^2}\right)}\cos{\left(2\sqrt{\frac{2}{3}}\frac{k}{k_{\rm eq}}\sqrt{1+\left(\frac{\tilde \eta}{\eta_\star}\right)^2}\right)}\right.\nonumber\\&\left.-\cos{\left(2\sqrt{\frac{2}{3}}\frac{k}{k_{\rm eq}}\sqrt{1+\left(\frac{\eta_{\rm rec}}{\eta_\star}\right)^2}\right)}\sin{\left(2\sqrt{\frac{2}{3}}\frac{k}{k_{\rm eq}}\sqrt{1+\left(\frac{\tilde \eta}{\eta_\star}\right)^2}\right)} \right]\nonumber\\&-\frac{2\sqrt{3}\eta_{\rm rec}}{k\eta_*^4\sqrt{1+\left(\frac{\eta_{\rm rec}}{\eta_\star}\right)^2}} \sin{\frac{k\eta_{\rm rec}}{\sqrt{3}}}\left[\cos{\left(2\sqrt{\frac{2}{3}}\frac{k}{k_{\rm eq}}\sqrt{1+\left(\frac{\eta_{\rm rec}}{\eta_\star}\right)^2}\right)}\right.\nonumber\\&\left. \times\cos{\left(2\sqrt{\frac{2}{3}}\frac{k}{k_{\rm eq}}\sqrt{1+\left(\frac{\tilde \eta}{\eta_\star}\right)^2}\right)}+\sin{\left(2\sqrt{\frac{2}{3}}\frac{k}{k_{\rm eq}}\sqrt{1+\left(\frac{\eta_{\rm rec}}{\eta_\star}\right)^2}\right)}\right.\nonumber\\&\left. \times\sin{\left(2\sqrt{\frac{2}{3}}\frac{k}{k_{\rm eq}}\sqrt{1+\left(\frac{\tilde \eta}{\eta_\star}\right)^2}\right)} \right]\Bigg\}.
\end{align}
Before concluding this Section, we make the following comment. As we immediately note, in the analytical analysis discussed in this work, the solutions for
$1/\eta_{\rm rec}<k<k_{\rm eq}$ are missing (e.g., see the gray area Fig. \ref{Figure}). One possible way we can overcome this problem could be the following prescription. Let us define the general solutions, both for matter and radiation, in this way
\begin{align}
   \delta^{(2)}_{A }(\eta \ge \eta_{\rm rec}, \bm{k}) =  {\left[\delta^{(2)}_{A \bm{k}}(k<1/\eta_{\rm rec}, \eta \ge \eta_{\rm rec}) +\left(k/k_{\rm eq}\right)^{n_A}\delta^{(2)}_{A \bm{k}}(k > k_{\rm eq}, \eta \ge \eta_{\rm rec}) \right]\over 1 + \left(k/k_{\rm eq}\right)^{n_A}}\;,
\end{align}
with $A =\{ {\rm m},{\rm r}\}$ and $n_A>0$ (e.g. $n_A \simeq 2$). Here, clearly,  $k<1/\eta_{\rm rec}$ indicates the solutions obtained above which do not contain the homogeneous solutions %in which modes enters in the horizon only when $\eta > \eta_{\rm rec}$
and for $k > k_{\rm eq}$ we are considering  solutions in which the modes entered the horizon before equality. Obviously this guess has to be tested numerically.

Here we present Eqs. \eqref{finaldelta-iota} (with \eqref{Green-iota}), and \eqref{deltarmd} as the fourth and final result of the paper.

\section{Summary}\label{disc}

Nowadays the detection of the primordial GW background is one of the main interests of cosmology and, recently, it has been shown that the tensor-induced scalar modes produced by GWs from inflation can give an important GW signature on cosmic structures \cite{PhysRevLett.129.091301}.
In particular,
\cite{PhysRevLett.129.091301} studied 
a novel mechanism for generating matter density perturbations based upon the non-linear evolution of primordial tensor modes which generates matter-density perturbations and its power spectrum for modes entering in the matter dominated period of evolution.
In this second paper we have explored analytically this  mechanism  both for matter and radiation density perturbations %based upon the non-linear evolution of primordial tensor modes 
and, in particular, during matter and radiation epochs.

Here, first of all, we extend the treatment to smaller scales, which enter the horizon during a radiation dominated period, during which there are two matter components contributing to the energy density of the Universe and we compute also the radiation-density perturbations produced by these tensor-induced scalar modes.
% our matter density contrast com- pletely mimics the linear one on the sub-horizon scales.

Starting from a comoving (with CDM frame) and time-orthogonal gauge, %, our calculation is effectively,
we have shown that we can safely implement the metric perturbations in the synchronous gauge from the very beginning.
 In fact, using a perturbative expansion up to the second order in which the source term consists only of linear tensor perturbations, we have shown that the perturbations of scalar density can be directly re-expressed and re-formulated in the synchronous gauge.

%\begin{table}[h]
 %   \centering
  %  \begin{tabular}{ |p{3cm}||p{3cm}|l| }
   %      \hline
 %\multicolumn{3}{|c|}{Radiation domination} \\
 %%Dominant  component& Deep RD & Intermediate phase\\
 %%In background  & radiation& radiation +CDM\\
%Perturbation&   $\delta_{\rm r}^{(2)}$ & $\delta_{\rm m}^{(2)}$\\
 %\hline
   % \end{tabular}
    %\caption{The dominant component between CDM and radiation in the background and perturbation sector \DB{add here also the dark matter (+ dark energy) epoch. Then add in which section or appendix it is discussed that solutions}}
   % \label{tab}
%\end{table}

\begin{table}[h]
    \centering
    \begin{tabular}{ |p{2.6cm}|| p{2.6cm} || p{2.6cm}||p{2.6cm}||p{2.6cm}| }
         \hline
 \multicolumn{5}{|c|}{Phases of evolution} \\
 \hline
 Dominant \quad component & \centering{Deep radiation epoch} \quad \quad (Section \ref{purerd})& \centering{Radiation epoch in sub-horizon scales} \quad \quad\quad\quad(Section \ref{subrd}) & \centering{ Intermediate regime (matter+radiation) for $1/\eta_{\rm rec} \lesssim k \lesssim 1/\eta_{\rm eq}$ (no analytical solutions)} & {\centering{After}}\quad \quad\quad \quad recombination \quad \quad
 (Section \ref{matt})\\
 \hline
 In background  & \centering{radiation} & \centering{radiation} & \centering{radiation + CDM} & \quad CDM + DE \\
% \hline & & \\
 \hline \centering{Dominant perturbation component} &   \centering{$\delta_{\rm r}^{(2)}$} &  \centering{$ \delta_{\rm m}^{(2)}$} &  \centering{$\delta_{\rm m}^{(2)}~~\&~~\delta_{\rm r}^{(2)}$}  &  $~~~~~~~~\delta_{\rm m}^{(2)}$ \\
 \hline
    \end{tabular}
    \caption{The dominant component in different phases of evolution in the background and perturbation sector (see also Fig. \ref{Figure}).}
    \label{tab}
\end{table}

Then we started focusing on the radiation period in which we split the epoch into two phases according to the relative importance of background and perturbed quantities of the components. In Table \ref{tab}, we briefly summarised schematically the dominant component between CDM and radiation in the background and perturbation sector.

The first phase, also called the epoch of deep radiation, begins after the end of inflation, when the radiation dominates both the background and the perturbation sector.
During this period, in this paper, we have calculated the radiation and matter perturbation solutions which are sourced by primordial GWs.

Looking at these contributions, we observed that the matter perturbation grows faster than the radiation one. 
Due to this mechanism, still during the radiation era, a subsequent second phase  has begun, i.e. when the matter perturbation grows sufficiently to overcome the radiation perturbation, becoming the main contributor to Einstein’s field equations.

Note that only the first phase was presented in \cite{Wang:2019zhj, Doring:2021gue}, whereas our study shows a full solution of tensor-induced density contrast in the radiation domination. In this second phase,  we  have obtained a Meszaros' like equation with a source term quadratic in GWs. In obtaining the expressions, we have focused on the subhorizon regime, as our effect is non-existent on the superhorizon. 
Now, in order to obtain the initial conditions for the second phase, we matched the solutions from the two phases at the junction. The initial conditions for the first phase can be ignored because the effect is suppressed by the fact that all these modes enter within the horizon scale only during the radiation epoch. In this way we have derived full solution of the tensor-sourced scalar modes entering the horizon starting from the end of inflation to matter epoch for modes $k>k_{\rm eq}$. 

Finally, in Section \ref{matt}, we have extended the analysis of  \cite{PhysRevLett.129.091301}, computing the tensor contribution to matter and radiation perturbations at late times, i.e. from $\eta_{\rm eq} \ll \eta_{\rm rec} \le \eta \le \eta_0$,
%$\eta_0 \ge \eta \ge \eta_{\rm rec} \gg \eta_{\rm eq}$, 
both for  $k<1/\eta_{\rm rec}$  and $k > k_{\rm eq}$, where $\eta_0$ is the conformal time today and $\eta_{\rm rec}$ at recombination epoch.
% the radiation contribution  matter dominated period of evolution.

In this paper, for the sake of simplicity, the effects of baryons have been ignored. This contribution will be explored and analysed in a future study.  
Another important future research direction could be to calculate the power spectrum, following the approach used in \cite{PhysRevLett.129.091301}, 
considering also the matter contribution obtained in this paper, i.e. for $k>k_{\rm eq}$.
In addition it is also very interesting to analyse and evaluate the corrections in the CMB anisotropies due to radiation perturbations,  at all scales, induced by the energy density fluctuation of gravitational radiation.

Finally, a comparison between this analytical work with a proper numerical analysis will be very useful.  In particular, let us stress that, for $1/\eta_{\rm rec} \le k \lsim 1/\eta_{\rm eq}$, we are not able to make use of our analytical prescription and a numerical approach is needed.
All of these projects are left for a future work.

%\acknowledgments
\section*{Acknowledgements}
%We thank A. Ricciardone for useful comments and discussions.
DB, NB, SG, SM acknowledge support from the COSMOS network (www.cosmosnet.it) through the ASI (Italian Space Agency) Grants 2016-24-H.0 and 2016-24-H.1-2018.
%A.R. and 
PB and AR acknowledge funding from Italian Ministry of Education, University and Research (MIUR) through the ``Dipartimenti di eccellenza'' project Science of the Universe. SG acknowledges the financial support from the INFN InDark project. 

\begin{appendices}

\numberwithin{equation}{section}

\section{Tensor induced vector and tensor modes}
 \label{Appendix-pert}

As we pointed out in the main text of the paper, in this work we analyze and study the scalar modes sourced by the linear tensors. Here, for completeness, we briefly show the equations containing the second order vector and tensor modes,  sourced by the same. These equations can also be found at \cite{Wang:2019zhj}. Appending second order vector and tensors to the metric Eq. \eqref{line}, we have
\begin{align}
\begin{split}
    \gamma_{ij}&= \delta_{ij}+ \gamma_{ij}^{(1)}+\frac{ \gamma_{ij}^{(2)}}{2},\\
    %&=  \delta_{ij}+ \chi_{ij}^{(1)}-\phi^{(2)}\delta_{ij}+\frac{1}{2}\Big( D_{ij}\chi^{(2)||}+\chi^{\perp (2)}_{ij} +\chi^{T (2)}_{ij}\Big) ,
    &=  \delta_{ij}+ \chi_{ij}^{(1)}-\phi^{(2)}\delta_{ij}+\frac{1}{2}\Big( D_{ij}\chi^{(2)||}+\partial_i \chi^{\perp (2)}_{j}+\partial_j \chi^{\perp (2)}_{i} +\chi^{T (2)}_{ij}\Big) ,
    \end{split}\\
    \gamma^{ij}&= \delta^{ij}- \chi^{ij(1)}+\phi^{(2)}\delta^{ij}-\frac{1}{2}\Big( D^{ij}\chi^{(2)||}+\partial^i \chi^{\perp (2)j}+\partial^j \chi^{\perp (2)i}  +\chi^{T (2)}_{ij}\Big)+\chi^{ik(1)}{\chi_{k}}^{j(1)}\,.
\end{align}
%\DB{Here and below is WRONG the definition of vector perturbation of the metric... Please use the definition of Sabino paper, e.g. astro-ph/9707278 }
Second order vector and tensor perturbations satisfy $\partial^i  \chi^{\perp (2)}_{i} =0$, $\partial^i \chi^{T (2)}_{ij}=\chi^{T (2)}_{ii}=0$. 
Due to the gauge chosen , the observers are comoving with the CDM, ${u_{\rm m}}_\mu=-a \delta^0_\mu$, and the components of the energy-momentum tensor for the matter does not contain any vector mode.  Here only the radiation tensor has an additional term. Indeed, the four-vector of the radiation is defined as
$$u_{{\rm r}i}=av_{{\rm r} i} =a\big(v_{{\rm r},i}+v^{\perp}_{{\rm r}i}\big),$$
$${u_{\rm r}}^i=\frac{1}{a}{v_{\rm r}}^i  =\frac{1}{a}\big({v_{\rm r}}^{,i}+ v^{\perp i}_{\rm r}\big).$$
If we consider also vector and tensor perturbations sourced by tensors, the conservation equations of matter remains the same, but radiation gains a new component of the energy-momentum tensor, $ T^{0i}_{\rm r}=4\overline{\rho}_{\rm r}/3a^2\big({v_{\rm r}}^{,i}+ v^{\perp i}_{\rm r}\big)$, which ends up modifying the momentum conservation equation for it
\begin{equation}
     4 \big({v^{(2)}_{{\rm r},i}}'+{v^{\perp (2)}_{{\rm r}i}}' \big)+ \delta_{{\rm r},i}^{(2)}=0\;.
\end{equation}
We can immediately conclude that 
$v^{\perp (2)}_{{\rm r} i}$ is constant in time.
The energy constraint remains the same as (\ref{00th}), but the momentum constraint has new components
       % \begin{equation}
        %   {\phi^{(2)}}'_{,i}-\frac{1}{2}\chi^{jk}\chi'_{ki,j}+\frac{1}{4}\big(D_{ij}{\chi^{||(2),j}}'+{\chi^{\perp (2),j}}'_{ij}\big)+\frac{1}{2}\chi^{jk}\chi'_{jk,i}+\frac{1}{4}{\chi^{jk}}'\chi_{jk,i}=-2\mathcal{H}^2 \big(v^{(2)}_{\rm r,i}+v^{\perp(2)}_{\rm ri}\big),
    %    \end{equation}
         \begin{equation}
           {\phi^{(2)}}'_{,i}+\frac{1}{4}\big(D_{ij}{\chi^{||(2),j}}'+\nabla^2{\chi^{\perp (2)}_i}'\big)=-2\mathcal{H}^2 \big(v^{(2)}_{{\rm r},i}+v^{\perp(2)}_{{\rm r}i}\big)+\frac{1}{2}\chi^{jk}{\chi_{jk,i}}'+\frac{1}{4}{\chi^{jk}}'\chi_{jk,i}-\frac{1}{2}\chi^{jk}{\chi_{ki,j}}',
        \end{equation}
        and the  $ij-$th equation reads
        \begin{align} \label{ijvt}
        & \frac{1}{4}\big(D_{ij}{\chi^{||(2)}}''+\partial_i {\chi^{\perp (2)}_j}''+\partial_j {\chi^{\perp (2)}_i}''+{\chi^{T (2)}}''_{ij}\big)+\frac{\mathcal{H}}{2}\big(D_{ij}{\chi^{||(2)}}'+\partial_i {\chi^{\perp (2)}_j}'+\partial_j {\chi^{\perp (2)}_i}'+{\chi^{T (2)}}'_{ij}\big)\nonumber\\&-\frac{1}{4}\nabla^2\chi^{T (2)}_{ij}+\frac{1}{12}\nabla^2 D_{ij}\chi^{||(2)}-\frac{1}{18}\nabla^2 \nabla^2 \chi^{||(2)}\delta_{ij}+2\mathcal{H}{\phi^{(2)}}'\delta_{ij}+{\phi^{(2)}}''\delta_{ij}+\frac{1}{2}D_{ij}\phi^{(2)}\nonumber\\&-\frac{1}{3}\nabla^2 \phi^{(2)}\delta_{ij}
    -\frac{1}{2}\chi^{kl}(\chi_{lj,ik} +\chi_{il,jk}-\chi_{ij,lk}-\chi_{kl,ij}) +\frac{1}{4}{\chi^{kl}}_{,j}\chi_{kl,i}-\frac{1}{2}\chi^{k,l}_j\chi_{li,k}+\frac{1}{2}\chi^{k,l}_j\chi_{ki,l}\nonumber\\&-\frac{3}{8}\chi^{kl,p}\chi_{kl,p}\delta_{ij}+\frac{1}{4}\chi^{kp,l}\chi_{lp,k}\delta_{ij}-\frac{1}{2}{\chi^{k}}'_j {\chi_{ki}}'+\frac{3}{8}{\chi^{kl}}'{\chi_{kl}}'\delta_{ij}=\frac{\mathcal{H}^2}{2} \delta_{\rm r}^{(2)}\delta_{ij}.
        \end{align}
 Trace-less part of (\ref{ijvt}) now gives
 \begin{align}\label{tr-lesssvt}
    & D_{ij}\phi^{(2)}+\frac{1}{2} \big(D_{ij}{\chi^{||(2)}}''+\partial_i {\chi^{\perp (2)}_j}''+\partial_j {\chi^{\perp (2)}_i}''+{\chi^{T (2)}}''_{ij}\big)+\mathcal{H}\big(D_{ij}{\chi^{||(2)}}'+\partial_i {\chi^{\perp (2)}_j}'+\partial_j {\chi^{\perp (2)}_i}'\nonumber\\&+{\chi^{T (2)}}'_{ij}\big)-\frac{1}{2}\nabla^2\chi^{T (2)}_{ij}+ \frac{1}{6}\nabla^2 D_{ij}\chi^{||(2)}
    -\chi^{kl}(\chi_{lj,ik} +\chi_{il,jk}-\chi_{ij,lk}-\chi_{kl,ij}) +\frac{1}{2}{\chi^{kl}}_{,j}\chi_{kl,i}\nonumber\\&-\chi^{k,l}_j\chi_{li,k}+\chi^{k,l}_j\chi_{ki,l}-{\chi^{k}}'_j {\chi_{ki}}-\frac{1}{3}\chi^{kl}\nabla^2\chi_{kl}\delta_{ij}+\frac{1}{3}{\chi^{kl}}'{\chi_{kl}}\delta_{ij}-\frac{1}{2}\chi^{kl,p}\chi_{kl,p}\delta_{ij}\nonumber\\&+\frac{1}{3}\chi^{kp,l}\chi_{lp,k}\delta_{ij}=0.
 \end{align}
 In \eqref{tr-lesssvt}, we have a coupled system of tensor-sourced scalar, vector, and tensor perturbations. This equation is the generalization of Eq. \eqref{trl}. To obtain independent equations for vector and tensors, we apply  $3\nabla^{-2}\nabla^{-2}\partial_i\partial_j$ to  \eqref{tr-lesssvt} \cite{Wang:2019zhj}. The result is the evolution equation for the scalar $\chi^{||(2)}$
 \begin{align}\label{chiscalar}
     &{\chi^{||(2)}}''+2\mathcal{H} {\chi^{||(2)}}'+\frac{1}{3}\nabla^2 {\chi^{||(2)}}+2\phi^{(2)}=-\frac{3}{8}\chi^{kl}\chi_{kl}-\nabla^{-2}\left({\chi^{kl}}'{\chi_{kl}}-\frac{1}{2}\chi^{kl,p}\chi_{kp,l}\right)\nonumber\\&+\frac{1}{4}\chi^{kl}\nabla^2\chi_{kl}
     +3\nabla^{-2}\nabla^{-2}\left({\chi^{kl,p}}'{\chi_{kp,l}}'-\frac{1}{2}\chi^{kl}\nabla^2\nabla^2\chi_{kl}-\frac{1}{2}\chi^{kl,p}\nabla^2\chi_{kl,p}+\chi^{kl,p}\nabla^2\chi_{kp,l}\right).
 \end{align}

 Applying $\nabla^{-2}\Big\{\partial^i\left[\eqref{tr-lesssvt}-\frac{1}{2}D_{ij}\eqref{chiscalar}\right]+(i\leftrightarrow j)\Big\}$, we obtain
 %\begin{align}\label{chivector}
  %   &\left(\partial_i {\chi^{\perp (2)}_j}''+\partial_j {\chi^{\perp (2)}_i}''\right)+2\mathcal{H} \left(\partial_i {\chi^{\perp (2)}_j}'+\partial_j {\chi^{\perp (2)}_i}'\right)= 2\partial_i\nabla^{-2} \left({\chi^{kl}}'{\chi_{jk,l}}'-\frac{1}{2}\chi^{kl}\nabla^2\chi_{kl,j}+\chi^{kl}\nabla^2\chi_{jk,l}\right)\nonumber\\&
   %  -2\partial_i\partial_j\nabla^{-2}\nabla^{-2}\left({\chi^{kl,p}}'{\chi_{kp,l}}'-\frac{1}{2}\chi^{kl}\nabla^2\nabla^2\chi_{kl}-\frac{1}{2}\chi^{kl,p}\nabla^2\chi_{kl,p}+\chi^{kl,p}\nabla^2\chi_{kp,l}\right)+(i\leftrightarrow j)
 %\end{align}
 \begin{align}\label{chivector}
   & {\chi^{\perp (2)}_i}''+2\mathcal{H} {\chi^{\perp (2)}_i}'= 2\nabla^{-2} \left[\left({\chi^{kl}}'{\chi_{ik,l}}'-\frac{1}{2}\chi^{kl}\nabla^2\chi_{kl,i}+\chi^{kl}\nabla^2\chi_{ik,l}\right)\right.\nonumber\\&\left.+\partial_i\nabla^{-2}\left({\chi^{kl,j}}'{\chi_{jk,l}}'-\frac{1}{2}\chi^{kl,j}\nabla^2\chi_{kl,j}-\frac{1}{2}\chi^{kl}\nabla^2\nabla^2\chi_{kl}+\chi^{kl,j}\nabla^2\chi_{jk,l}\right)\right]\nonumber\\&
    -4\partial_i\nabla^{-2}\nabla^{-2}\left({\chi^{kl,p}}'{\chi_{kp,l}}'-\frac{1}{2}\chi^{kl}\nabla^2\nabla^2\chi_{kl}-\frac{1}{2}\chi^{kl,p}\nabla^2\chi_{kl,p}+\chi^{kl,p}\nabla^2\chi_{kp,l}\right)
 \end{align}
 as the evolution equation of tensor-sourced vector perturbation. Furthermore, application of $\left[\eqref{tr-lesssvt}-\frac{1}{2}D_{ij}\eqref{chiscalar}-\eqref{chivector}\right]$ gets us the evolution equation for tensor-sourced tensor modes
 \begin{align}
      &{\chi^{T (2)}}''_{ij}+2\mathcal{H} {\chi^{T (2)}}'_{ij}-\nabla^2 {\chi^{T (2)}}_{ij}= -\frac{5}{8}\chi^{kl}\chi_{kl,ij}-\frac{1}{4}\chi^{kl,p}\chi_{kp,l}\delta_{ij}-\frac{1}{8}\chi^{kl}_{,i}\chi_{kl,j}+\frac{3}{8}\chi^{kl,p}\chi_{kl,p}\delta_{ij}\nonumber\\&
      -\chi^{k,l}_j\chi_{ki,l}+\chi^{l,k}_j\chi_{ki,l}+\frac{1}{4}\chi^{kl}\nabla^2\chi_{kl}\delta_{ij}+{\chi^l_i}'{\chi_{lj}}'-\frac{1}{2}{\chi^{kl}}'{\chi_{kl}}'\delta_{ij}+\chi^{kl}\chi_{kj,il}+\chi^{kl}\chi_{ki,jl}\nonumber\\&-\chi^{kl}\chi_{ij,kl}
      +\frac{1}{2}\nabla^{-2}\left({\chi^{kl,p}}'{\chi_{kp,l}}'+\chi^{kl,p}\nabla^2\chi_{kp,l}-\frac{1}{2}\chi^{kl}\nabla^2\nabla^2\chi_{kl}-\frac{1}{2}\chi^{kl,p}\nabla^2\chi_{kl,p}\right)\delta_{ij}\nonumber\\&
      +\left[\partial_i\nabla^{-2}\left(\frac{1}{2}\chi^{kl}\nabla^2\chi_{kl,j}-{\chi^{kl}}'{\chi_{jk,l}}'-\chi^{kl}\nabla^2\chi_{jk,l}\right)+(i\leftrightarrow j)\right]\nonumber\\&
      +\frac{1}{2}\partial_i \partial_j \nabla^{-2}\left({\chi^{kl}}'{\chi_{kl}}'-\frac{1}{2}\chi^{kl,p}\chi_{kp,l}-\frac{1}{4}\chi^{kl}\nabla^2\chi_{kl}\right)\nonumber\\&
      +\frac{1}{2}\partial_i \partial_j \nabla^{-2}\nabla^{-2} \left({\chi^{kl,p}}'{\chi_{kp,l}}'+\chi^{kl,p}\nabla^2\chi_{kp,l}-\frac{1}{2}\chi^{kl}\nabla^2\nabla^2\chi_{kl}-\frac{1}{2}\chi^{kl,p}\nabla^2\chi_{kl,p}\right).
 \end{align}
 From these equations we are able to describe the dynamics of vector and tensor contribution from end of inflation until today. We refer a detailed analysis of these equations and their solutions to a future paper.

  \section{General solution during deep radiation dominance, without subhorizon approximation}\label{vrd}

In the second part of Section \ref{purerd} we only focused on the regime where $k\eta \gg 1$, i.e when the modes were sub-Hubble during the first phase of the radiation era.
However, it is useful to obtain the general solution for any $k$. Let start again with Eq. \eqref{ueq}
\begin{equation}
    u^{''(2)}_{{\rm r}\bm{k}}+\Big(\frac{k^2}{3}-2\mathcal{H}^2\Big)u^{(2)}_{{\rm r}\bm{k}}= S_{\bm{k}},
\end{equation}
and using again the definition $\tau=k\eta$, this equation takes the form
\begin{equation}
   \tau^2 \frac{d^2u^{(2)}_{{\rm r}\bm{k}}}{d\tau^2}+\Big(\frac{\tau^2}{3}-2\Big)u^{(2)}_{{\rm r}\bm{k}}=  \tau^2 \frac{S_{\bm{k}}}{k^2},
\end{equation}
where  $S_{\bm{k}}$ is given by Eq. \eqref{S_k-eta}.
Now, we note that the solutions of the homogeneous part of this equation can easily be obtained  if we consider the following second order differential equation
$$x^2 {d^2 y \over d x^2 }(x)+[a^2x^2-n(n+1)]y(x)=0, \, \quad \quad {\rm with}\quad \quad n=0,1,2... \;,$$
or, equivalently,
$$y(x) x^{n+1}= \left(x^3 \frac{d}{dx}\right)^n\left(\frac{C_1 \cos{ax}+C_2 \sin{ax}}{x^{2n-1}}\right)\,,$$
where  $C_1$ and $C_2$ are two constants.
Then if $y=u^{(2)}_{\rm r\bm{k}}$, $x=\tau$, $a=1/\sqrt{3}$ and $n=1$, we get the homogeneous solutions
$$\frac{1}{\tau} \cos{\frac{\tau}{\sqrt{3}}}+\frac{1}{\sqrt{3}}\sin{\frac{\tau}{\sqrt{3}}} \quad \quad {\rm and} \quad \quad \frac{1}{\tau} \sin{\frac{\tau}{\sqrt{3}}}-\frac{1}{\sqrt{3}}\cos{\frac{\tau}{\sqrt{3}}}\,.$$
%\begin{multline}
 %   u^{(2)}_{\rm r\bm{k}}=C_1 (\bm{k})\left(\frac{1}{\tau} \cos{\frac{\tau}{\sqrt{3}}}+\frac{1}{\sqrt{3}}\sin{\frac{\tau}{\sqrt{3}}}\right)+C_2 (\bm{k})\left(\frac{1}{\tau} \sin{\frac{\tau}{\sqrt{3}}}-\frac{1}{\sqrt{3}}\cos{\frac{\tau}{\sqrt{3}}}\right).
%\end{multline}
Discarding the initial conditions according to the reasoning mentioned in the main text, we have 
\begin{align}
    u^{(2)}_{{\rm r}\bm{k}}(\tau)&=-\frac{\sqrt{3}}{k^2\tau} \int_{\tau_{\rm in}}^\tau  d\tilde{\tau}S_{\bm{k}}(\tilde{\tau})\left[\left( \sin{\frac{\tilde{\tau}}{\sqrt{3}}}  \cos{\frac{\tau}{\sqrt{3}}}  - \cos{\frac{\tilde{\tau}}{\sqrt{3}}} \sin{\frac{\tau}{\sqrt{3}}} \right) \left(\tau+\frac{3}{\tilde{\tau}}\right)\right.\nonumber\\&\left.
    +\sqrt{3}\left( \cos{\frac{\tilde{\tau}}{\sqrt{3}}}  \cos{\frac{\tau}{\sqrt{3}}}  +\sin{\frac{\tilde{\tau}}{\sqrt{3}}} \sin{\frac{\tau}{\sqrt{3}}} \right)\left(\frac{\tau}{\tilde{\tau}}-1\right)
    \right].
\end{align}
Then the perturbation $ v^{(2)}_{{\rm r} \bm{k}}$ becomes 
\begin{align}\label{detailv}
   v^{(2)}_{{\rm r} \bm{k}}(\tau)&= \frac{3}{k^3\tau}\int_{\tau_{\rm in}}^\tau d\tilde{\tau}\, \left(\tilde{\tau}+\frac{6}{\tilde{\tau}}\right)S_{\bm{k}}(\tilde{\tau}) \nonumber\\&-\left(\frac{2}{\tau}\cos{\frac{\tau}{\sqrt{3}}}+\frac{1}{\sqrt{3}}\sin{\frac{\tau}{\sqrt{3}}}\right)\int_{\tau_{\rm in}}^\tau d\tilde{\tau}\, \frac{9\cos{\frac{\tilde{\tau}}{\sqrt{3}}}+3\sqrt{3}\tilde{\tau} \sin{\frac{\tilde{\tau}}{\sqrt{3}}}}{k^3\tilde{\tau}} S_{\bm{k}}(\tilde{\tau}) \nonumber\\&-\left(\frac{2}{\tau}\sin{\frac{\tau}{\sqrt{3}}}-\frac{1}{\sqrt{3}} \cos{\frac{\tau}{\sqrt{3}}}\right)\int_{\tau_{\rm in}}^\tau  d\tilde{\tau}\, \frac{9\sin{\frac{\tilde{\tau}}{\sqrt{3}}}-3\sqrt{3}\tilde{\tau} \cos{\frac{\tilde{\tau}}{\sqrt{3}}}}{k^3\tilde{\tau}} S_{\bm{k}}(\tilde{\tau}).
\end{align}
Therefore, using the relation \eqref{momr}, we obtain
\begin{align}\label{detailr}
  %  \delta_{\rm r}^{(2)}(\bm{k},\eta)&=-4 v^{(2)'}_{\rm r}(\bm{k},\eta)\nonumber\\
   % \Rightarrow  
  \delta_{\rm r}^{(2)}(\bm{k},\tau)&=\frac{12}{k^2\tau^2}\int_{\tau_{\rm in}}^\tau d\tilde{\tau}\, \left(\tilde{\tau}+\frac{6}{\tilde{\tau}}\right)S_{\bm{k}}(\tilde{\tau}) \nonumber\\&-4\left(\frac{2}{\tau^2}\cos{\frac{\tau}{\sqrt{3}}}+\frac{2}{\sqrt{3}\tau}\sin{\frac{\tau}{\sqrt{3}}}-\frac{1}{3}\cos{\frac{\tau}{\sqrt{3}}}\right)\int_{\tau_{\rm in}}^\tau d\tilde{\tau}\, \frac{9\cos{\frac{\tilde{\tau}}{\sqrt{3}}}+3\sqrt{3}\tilde{\tau} \sin{\frac{\tilde{\tau}}{\sqrt{3}}}}{k^2\tilde{\tau}} S_{\bm{k}}(\tilde{\tau}) \nonumber\\&-4\left(\frac{2}{\tau^2}\sin{\frac{\tau}{\sqrt{3}}}-\frac{2}{\sqrt{3}\tau}\cos{\frac{\tau}{\sqrt{3}}}-\frac{1}{3}\sin{\frac{\tau}{\sqrt{3}}}\right)\int_{\tau_{\rm in}}^\tau d\tilde{\tau}\, \frac{9\sin{\frac{\tilde{\tau}}{\sqrt{3}}}-3\sqrt{3}\tilde{\tau} \cos{\frac{\tilde{\tau}}{\sqrt{3}}}}{k^2\tilde{\tau}} S_{\bm{k}}(\tilde{\tau}).
\end{align}

For $\delta_{\rm m}^{(2)}(\bm{k},\tau)$, we use the relation \eqref{vmatt}. In \eqref{vmatt}, the first term (${v^{(2)}_{{\rm r}\bm{k}}}'$) can be readily obtained from $\delta_{\rm r}^{(2)}(\bm{k},\tau)$ expression above (using relation \eqref{momr}), and  the integral in the  second term is (in terms of $\tau$)
\begin{align}
    \int^{\tau}_{\tau_{\rm in}} d\tilde \tau \; v^{(2)}_{{\rm r}\bm{k}}(\tilde \tau) &= \frac{3 }{k^3} 
  \int_{\tau_{\rm in}}^\tau d\tilde{\tau}\, \ln{\frac{\tau}{\tilde{\tau}}} \left(\tilde{\tau}+\frac{6}{\tilde{\tau}}\right)S_{\bm{k}}(\tilde{\tau}) \nonumber\\&
  -   \int^{\tau}_{\tau_{\rm in}} d\tilde{\tau}\frac{S_{\bm{k}}(\tilde{\tau})}{k^3\tilde{\tau}} \left[9\left(\cos{\frac{\tau}{\sqrt{3}}} \cos{\frac{\tilde{\tau}}{\sqrt{3}}}+ \sin{\frac{\tau}{\sqrt{3}}} \sin{\frac{\tilde{\tau}}{\sqrt{3}}}\right)\right.\nonumber\\&\left.+3\sqrt{3}\tilde{\tau}\left(\cos{\frac{\tau}{\sqrt{3}}} \sin{\frac{\tilde{\tau}}{\sqrt{3}}}- \sin{\frac{\tau}{\sqrt{3}}} \cos{\frac{\tilde{\tau}}{\sqrt{3}}}\right)\right]\nonumber\\&
  -\int_{\tau_{\rm in}}^\tau d\tilde{\tau}\, \frac{9\cos{\frac{\tilde{\tau}}{\sqrt{3}}}+3\sqrt{3}\tilde{\tau} \sin{\frac{\tilde{\tau}}{\sqrt{3}}}}{k^3\tilde{\tau}} S_{\bm{k}}(\tilde{\tau})\times 2\int_{\tau_{\rm in}}^\tau d\tilde{\tau}\,\frac{\cos{\frac{\tilde{\tau}}{\sqrt{3}}}}{\tilde{\tau}}\nonumber\\&-\int_{\tau_{\rm in}}^\tau d\tilde{\tau}\, \frac{9\sin{\frac{\tilde{\tau}}{\sqrt{3}}}-3\sqrt{3}\tilde{\tau} \cos{\frac{\tilde{\tau}}{\sqrt{3}}}}{k^3\tilde{\tau}} S_{\bm{k}}(\tilde{\tau})\times 2\int_{\tau_{\rm in}}^\tau d\tilde{\tau}\,\frac{\sin{\frac{\tilde{\tau}}{\sqrt{3}}}}{\tilde{\tau}}\nonumber\\&
  + \int^{\tau}_{\tau_{\rm in}} d\tilde{\tau}\frac{S_{\bm{k}}(\tilde{\tau})}{k^3\tilde{\tau}} \left[2\left(9\cos{\frac{\tilde{\tau}}{\sqrt{3}}}+3\sqrt{3}\tilde{\tau} \sin{\frac{\tilde{\tau}}{\sqrt{3}}}\right)\int_{\tau_{\rm in}}^{\Tilde{\tau}} \frac{d\vardbtilde{\tau}}{\vardbtilde{\tau}}\cos{\frac{\vardbtilde{\tau}}{\sqrt{3}}}\right.\nonumber\\&\left.+2\left(9\sin{\frac{\tilde{\tau}}{\sqrt{3}}}-3\sqrt{3}\tilde{\tau} \cos{\frac{\tilde{\tau}}{\sqrt{3}}}\right)\int_{\tau_{\rm in}}^{\Tilde{\tau}} \frac{d\vardbtilde{\tau}}{\vardbtilde{\tau}}\sin{\frac{\vardbtilde{\tau}}{\sqrt{3}}}-9\right],
\end{align}
where the expression of $v^{(2)}_{{\rm r}\bm{k}}$ from Eq. \eqref{detailv} has been used.
%\DB{Please say how! From which equation?}\PB{Done}
Combining both the terms, we arrive at 
 \begin{align}\label{detailm}
     \delta_{\rm m}^{(2)}(\bm{k},\tau)&= \frac{3}{k^2}\int_{\tau_{\rm in}}^\tau d\tilde{\tau}\, \left[\left(\frac{3}{\tau^2}-\ln{\frac{\tau}{\tilde{\tau}}}\right) \left(\tilde{\tau}+\frac{6}{\tilde{\tau}}\right)+\frac{3}{\tilde{\tau}}\right]S_{\bm{k}}(\tilde{\tau}) \nonumber\\&-3\left(\frac{2}{\tau^2}\cos{\frac{\tau}{\sqrt{3}}}+\frac{2}{\sqrt{3}\tau}\sin{\frac{\tau}{\sqrt{3}}}\right)\int_{\tau_{\rm in}}^\tau d\tilde{\tau}\, \frac{9\cos{\frac{\tilde{\tau}}{\sqrt{3}}}+3\sqrt{3}\tilde{\tau} \sin{\frac{\tilde{\tau}}{\sqrt{3}}}}{k^2\tilde{\tau}} S_{\bm{k}}(\tilde{\tau}) \nonumber\\&-3\left(\frac{2}{\tau^2}\sin{\frac{\tau}{\sqrt{3}}}-\frac{2}{\sqrt{3}\tau}\cos{\frac{\tau}{\sqrt{3}}}\right)\int_{\tau_{\rm in}}^\tau d\tilde{\tau}\, \frac{9\sin{\frac{\tilde{\tau}}{\sqrt{3}}}-3\sqrt{3}\tilde{\tau} \cos{\frac{\tilde{\tau}}{\sqrt{3}}}}{k^2\tilde{\tau}} S_{\bm{k}}(\tilde{\tau})\nonumber\\&
      +2\int_{\tau_{\rm in}}^\tau d\tilde{\tau}\, \frac{9\cos{\frac{\tilde{\tau}}{\sqrt{3}}}+3\sqrt{3}\tilde{\tau} \sin{\frac{\tilde{\tau}}{\sqrt{3}}}}{k^2\tilde{\tau}} S_{\bm{k}}(\tilde{\tau})\left[\int_{\tau_{\rm in}}^\tau d\tilde{\tau}\,\frac{\cos{\frac{\tilde{\tau}}{\sqrt{3}}}}{\tilde{\tau}}-\int_{\tau_{\rm in}}^{\tilde{\tau}} d\vardbtilde{\tau}\,\frac{\cos{\frac{\vardbtilde{\tau}}{\sqrt{3}}}}{\vardbtilde{\tau}}\right]\nonumber\\&+2\int_{\tau_{\rm in}}^\tau d\tilde{\tau}\, \frac{9\sin{\frac{\tilde{\tau}}{\sqrt{3}}}-3\sqrt{3}\tilde{\tau} \cos{\frac{\tilde{\tau}}{\sqrt{3}}}}{k^2\tilde{\tau}} S_{\bm{k}}(\tilde{\tau})\left[\int_{\tau_{\rm in}}^\tau d\tilde{\tau}\,\frac{\sin{\frac{\tilde{\tau}}{\sqrt{3}}}}{\tilde{\tau}}-\int_{\tau_{\rm in}}^{\tilde{\tau}} d\vardbtilde{\tau}\,\frac{\sin{\frac{\vardbtilde{\tau}}{\sqrt{3}}}}{\vardbtilde{\tau}}\right],
\end{align}
which is the expression for CDM density contrast in deep radiation domination, without subhorizon approximation. In \cite{Wang:2019zhj}, only radiation perturbation was studied. As a result, only \eqref{detailv} and \eqref{detailr} was derived there. Here we re-obtain them, along with the matter perturbation \eqref{detailm}, which was missing in \cite{Wang:2019zhj}.
\section{Calculation of $\delta_r$ in the second phase}\label{appdeltar}

This appendix section is devoted to the computation of the radiation-density perturbations produced by  tensor-induced scalar modes during the second phase of the radiation epoch. As we pointed out in the main text, in this second phase we have modes well inside the Hubble radius, i.e. $k\eta \gg 1$,  and with  $y\delta_{\rm m}^{(2)} > 2\delta_{\rm r}^{(2)}$. In this case, the matter perturbation is the main contributor to Einstein's field equations. In Section \ref{subrd}, we obtained a new Meszaros equation due to GWs contribution which describes the dynamics of $\delta_{\rm m}^{(2)}$. Then, using
\eqref{radc} and \eqref{momr}, we are able to write second order differential equation for $\delta_{\rm r}^{(2)}$. Indeed, in Fourier space, we have 
\begin{align}
   {\delta_{{\rm r}\bm{k}}^{(2)}}''+\frac{k^2}{3} \delta_{{\rm r}\bm{k}}^{(2)} &= \frac{4}{3} {\delta_{{\rm m}\bm{k}}^{(2)}}''\,,
\end{align}
which, in terms of the variable $y$, turns out
\begin{align}
 %  {\delta_{\rm r}^{(2)}}''-\frac{4}{3}(\chi^{(1)ij}{\chi_{ij}^{(1)}}')'+\frac{4}{3}\nabla^2 {v_{\rm r}^{(2)}}'-4{\phi^{(2)}}''&=0,\\
   \label{labelapp}
    \mathcal{H}y \left[\mathcal{H}y \frac{d^2}{dy^2}+  \left(\mathcal{H}+y \frac{d\mathcal{H}}{dy}\right)\frac{d}{dy}\right]\delta_{{\rm r}\bm{k}}^{(2)}+\frac{k^2}{3} \delta_{{\rm r}\bm{k}}^{(2)}&=\frac{4}{3}   \mathcal{H}y \left[\mathcal{H}y \frac{d^2}{dy^2}+  \left(\mathcal{H}+y \frac{d\mathcal{H}}{dy}\right)\frac{d}{dy}\right]\delta_{{\rm m}\bm{k}}^{(2)}\;.
\end{align}
(As we also pointed out in the main part of the paper, here $\delta_{{\rm r}\bm{k}}^{(2)}(\eta)=\delta_{\rm r}^{(2)}(\bm{k}, \eta)$ and $\delta_{{\rm m}\bm{k}}^{(2)}(\eta)=\delta_{\rm m}^{(2)}(\bm{k}, \eta)$.)
We note immediately that  the source term on the right-hand side depends on $\delta_{\rm m}^{(2)}$.
Now, using
\begin{align}
   \frac{d\mathcal{H}}{dy}=\frac{1}{\mathcal{H}y} \mathcal{H}'=-\frac{\mathcal{H}}{2y}\frac{2+y}{1+y}
\end{align}
%with \eqref{labelapp} gives us the coupled equation of $\delta_{\rm r}^{(2)}$ and $\delta_{\rm m}^{(2)}$
and the definition of $k_{\rm eq}\equiv \mathcal{H}_{\rm eq}$, Eq. \eqref{labelapp} reads as follows
\begin{align}
   %   \frac{d^2\delta_{\rm r}^{(2)}}{dy^2} +\frac{1}{2(y+1)}\frac{d\delta_{\rm r}^{(2)}}{dy}+\frac{k^2}{3 \mathcal{H}^2y^2}\delta_{\rm r}^{(2)}&=\frac{4}{3} \frac{d^2\delta_{\rm m}^{(2)}}{dy^2}+\frac{2}{3(y+1)}\frac{d\delta_{\rm m}^{(2)}}{dy},\\
      \label{subdeltareq} %\Rightarrow
       \frac{d^2\delta_{{\rm r}\bm{k}}^{(2)}}{dy^2} +\frac{1}{2(y+1)}\frac{d\delta_{{\rm r}\bm{k}}^{(2)}}{dy}+\frac{k^2}{k^2_{\rm eq}}\frac{2}{3(y+1)}\delta_{{\rm r}\bm{k}}^{(2)}&=\frac{4}{3} \frac{d^2\delta_{{\rm m}\bm{k}}^{(2)}}{dy^2}+\frac{2}{3(y+1)}\frac{d\delta_{{\rm m}\bm{k}}^{(2)}}{dy}.
\end{align}

At this stage it is useful changing the variable $y\to w=\sqrt{1+y}$. In this case Eq. \eqref{subdeltareq} becomes
\begin{equation}\label{deltarw}
    \frac{d^2\delta_{{\rm r}\bm{k}}^{(2)}}{dw^2} +\frac{8}{3}\frac{k^2}{k^2_{\rm eq}}\delta_{{\rm r}\bm{k}}^{(2)}= \mathcal{Q}_{\bm{k}}(w)\,,
\end{equation}
where $$\mathcal{Q}_{\bm{k}}(w)=\frac{4}{3} \frac{d^2\delta_{{\rm m}\bm{k}}^{(2)}}{dw^2}.$$

Writing the solution of  $\delta_{\rm m}^{(2)}$ from Eq. \eqref{subdel} w.r.t. the variable $w$, we have
\begin{align}
    \delta_{\rm m}^{(2)}(\bm{x},w)&= \left(w^2-\frac{1}{3}\right) P_1(\bm{x}) +\left[\left(w^2-\frac{1}{3}\right)\ln{\frac{w+1}{w-1}}-2w\right] P_2(\bm{x}) \nonumber\\&+\frac{1}{4}\int_{w_\alpha}^w \frac{d \tilde{w}}{\tilde w} \, G\left(w^2-1, \tilde{w}^2-1\right)  \frac{d \chi^{ij}}{d\Tilde{w}}\frac{d \chi_{ij}}{d\Tilde{w}},
    \end{align}
    in configuration space, and
    \begin{align}
    \delta_{\rm m}^{(2)}(\bm{k},w)&= \left(w^2-\frac{1}{3}\right) P_1(\bm{k}) +\left[\left(w^2-\frac{1}{3}\right)\ln{\frac{w+1}{w-1}}-2w\right] P_2(\bm{k}) \nonumber\\&+\frac{1}{4}\int_{w_\alpha}^w \frac{d \Tilde{w}}{\tilde w} \, G\left(w^2-1, \tilde{w}^2-1\right)   F_3(\bm{k},\tilde w)\,, 
\end{align}
in Fourier space. Here $ F_3(\bm{k},w) $ is related to $ F_1(\bm{k},y)$ via the following relation 
   \begin{align}\label{F3}
F_3(\bm{k},w)=4(1+y) F_1(\bm{k},y)\,
\end{align}
and 
\begin{align}
   G\left(w^2-1, \tilde{w}^2-1\right) &=-\frac{1}{4}\Tilde{w}(\tilde{w}^2-1) \left[6\Big( \tilde w (3 w^2-1)-w(3\tilde{w}^2-1)\Big)
   \right.\nonumber\\ &\left.
   -(3\tilde{w}^2-1) (3 w^2-1)\ln{\frac{(\tilde{w}+1)(w-1)}{(\tilde{w}-1)(w+1)}}\right]\,.%\nonumber\\
\end{align}

Then, the source term of Eq. \eqref{deltarw} can be written in the following way
\begin{align}
    \mathcal{Q}_{\bm{k}}(w)%=\frac{4}{3} \frac{d^2\delta_{{\rm m}\bm{k}}^{(2)}}{dw^2}%\nonumber\\
    &= \frac{4}{3} \left[2P_1(\bm{k})+P_2(\bm{k})\left(\frac{4w(5-3w^2)}{3 (w^2-1)^2}+2\ln{\frac{w+1}{w-1}}\right)\right]\nonumber\\&
    -\frac{1}{2}\left[\left(\ln{\frac{w+1}{w-1}} +\frac{2w(5-3w^2)}{3 (w^2-1)^2}\right) \int_{w_\alpha}^w d \tilde{w} \,(\tilde{w}^2-1)(3\tilde{w}^2-1) %\frac{d \chi^{ij}}{d\Tilde{w}}\frac{d \chi_{ij}}{d\Tilde{w}}
    F_3(\bm{k}, \tilde{w})\right.\nonumber\\&\left. + \int_{w_\alpha}^w d\tilde{w}\,(\tilde{w}^2-1)  \left(6\tilde{w}+(3\tilde{w}^2-1)\ln{\frac{\tilde{w}-1}{\tilde{w}+1}}\right)%\frac{d \chi^{ij}}{d\Tilde{w}}\frac{d \chi_{ij}}{d\Tilde{w}}
    F_3(\bm{k},\tilde{w})-\frac{4}{3}\,F_3(\bm{k},w)\right],%\frac{d \chi^{ij}}{dw}\frac{d \chi_{ij}}{dw}\right].
\end{align}
%where $F_1$ is given by Eq. \eqref{Fone}.
%The homogeneous solution is
%\begin{equation}
 %   \delta_{\rm r, homo}^{(2)} =A_r(\bm{k}) \cos{\left(2\sqrt{\frac{2}{3}}\frac{k}{k_{\rm eq}}w\right)}+B_r(\bm{k}) \sin{\left(2\sqrt{\frac{2}{3}}\frac{k}{k_{\rm eq}}w\right)},
%\end{equation}
%The source term of \eqref{deltarw} becomes 
%\begin{multline}
%    \mathcal{Q}_{\bm{k}}(w)= \frac{4}{3} \left[2P_1(\bm{k})+P_2(\bm{k})\left(\frac{4w(5-3w^2)}{3 (w^2-1)^2}+2\ln{\frac{w+1}{w-1}}\right)\right]\\
%    -\frac{1}{2}\left[\left(\ln{\frac{w+1}{w-1}} +\frac{2w(5-3w^2)}{3 (w^2-1)^2}\right) \int_{w_\alpha}^w (\tilde{w}^2-1)(3\tilde{w}^2-1) \frac{d \chi^{ij}}{d\Tilde{w}}\frac{d \chi_{ij}}{d\Tilde{w}}d \tilde{w}\right.\\\left. + \int_{w_\alpha}^w (\tilde{w}^2-1) \frac{d \chi^{ij}}{d\Tilde{w}}\frac{d \chi_{ij}}{d\Tilde{w}} \left(6\tilde{w}+(3\tilde{w}^2-1)\ln{\frac{\tilde{w}-1}{\tilde{w}+1}}\right)d\tilde{w}\right.\\\left.+(w^2-1)(3w^2-1)\frac{d \chi^{ij}}{dw}\frac{d \chi_{ij}}{dw}\frac{(3w^2-2)^2-5}{(3w^2-2)^2-1}\right].
%\end{multline}
and the full solution of \eqref{deltarw} reads
\begin{equation}
    \delta_{\rm r}^{(2)}(\bm{k},w) =A_{\rm r}(\bm{k}) \cos{\left(2\sqrt{\frac{2}{3}}\frac{k}{k_{\rm eq}}w\right)}+B_{\rm r} (\bm{k}) \sin{\left(2\sqrt{\frac{2}{3}}\frac{k}{k_{\rm eq}}w\right)} + \int_{w_\alpha}^w d \Tilde{w}\, G_{\rm r} (w, \tilde{w}) \mathcal{Q}_{\bm{k}}(\tilde{w})\,, 
\end{equation}
where the Green's function of the above relation is defined as
\begin{align}
    G_{\rm r} (w, \tilde{w}) =&\frac{k_{\rm eq}}{2k}\sqrt{\frac{3}{2}}\left[\sin{\left(2\sqrt{\frac{2}{3}}\frac{k}{k_{\rm eq}}w\right)}\cos{\left(2\sqrt{\frac{2}{3}}\frac{k}{k_{\rm eq}}\tilde{w}\right)}\right.\nonumber\\&\left.-\cos{\left(2\sqrt{\frac{2}{3}}\frac{k}{k_{\rm eq}}w\right)}\sin{\left(2\sqrt{\frac{2}{3}}\frac{k}{k_{\rm eq}}\tilde{w}\right)} \right].
\end{align}
Now, going back to the variable $y$, and naming the perturbation as $\delta_{\rm r (Tmesz)}^{(2)}(\bm{k},y)$, it becomes
\begin{align}\label{pert_r-secphase}
   \delta_{\rm r (Tmesz)}^{(2)}(\bm{k},y) &= A_{\rm r}(\bm{k}) \cos{\left(2\sqrt{\frac{2}{3}}\frac{k}{k_{\rm eq}}\sqrt{1+y}\right)}+B_{\rm r}(\bm{k}) \sin{\left(2\sqrt{\frac{2}{3}}\frac{k}{k_{\rm eq}}\sqrt{1+y}\right)}\nonumber\\& + \int_{y_\alpha}^y \frac{d \Tilde{y}}{2\sqrt{1+y}} \, G_r\left(\sqrt{1+y}, \sqrt{1+\tilde{y}}\right)\; \mathcal{Q}_{\bm{k}}\left(\sqrt{1+\tilde{y}}\right),
\end{align}
where 
\begin{align}
    G_{\rm r}\left(\sqrt{1+y}, \sqrt{1+\tilde{y}}\right)=&\frac{k_{\rm eq}}{2k}\sqrt{\frac{3}{2}}\left[\sin{\left(2\sqrt{\frac{2}{3}}\frac{k}{k_{\rm eq}}\sqrt{1+y}\right)}\cos{\left(2\sqrt{\frac{2}{3}}\frac{k}{k_{\rm eq}}\sqrt{1+\tilde{y}}\right)}\right.\nonumber\\&\left.-\cos{\left(2\sqrt{\frac{2}{3}}\frac{k}{k_{\rm eq}}\sqrt{1+y}\right)}\sin{\left(2\sqrt{\frac{2}{3}}\frac{k}{k_{\rm eq}}\sqrt{1+\tilde{y}}\right)} \right]
\end{align}
and 
\begin{align}
    \mathcal{Q}_{\bm{k}}\left(\sqrt{1+y}\right)=& \frac{4}{3} \left[2P_1(\bm{k})+P_2(\bm{k})\frac{4(2-3y)\sqrt{1+y}+6y^2\ln{\frac{2+y+2\sqrt{1+y}}{y}}}{3 y^2}\right]\nonumber\\&
    -\frac{1}{2}\left[\frac{4(2-3y)\sqrt{1+y}+6y^2\ln{\frac{2+y+2\sqrt{1+y}}{y}}}{3 y^2} \int_{y_\alpha}^y\tilde{y}\sqrt{1+\tilde{y}} (2+3\tilde{y})%\frac{d \chi^{ij}}{d\Tilde{y}}\frac{d \chi_{ij}}{d\Tilde{y}}d \tilde{y}
    ~ F_1(\bm{k},\tilde{y})\right.\nonumber\\&\left. + 2\int_{y_\alpha}^y \tilde{y} \sqrt{1+\tilde{y}}%\frac{d \chi^{ij}}{d\Tilde{y}}\frac{d \chi_{ij}}{d\Tilde{y}}
    ~ F_1(\bm{k},\tilde{y})\left(6\sqrt{1+\tilde{y}}+(2+3\tilde{y})\ln{\frac{2+\Tilde{y}-2\sqrt{1+\Tilde{y}}}{\Tilde{y}}}\right)d\tilde{y}\right.\nonumber\\&\left.-\frac{16(1+y)}{3}%\frac{d \chi^{ij}}{dy}\frac{d \chi_{ij}}{dy}
    F_1(\bm{k},y)\right]\,,
\end{align}
where we used Eq. \eqref{F3}.
 The coefficients $A_{\rm r}$ and $B_{\rm r}$ 
 can be determined  exactly the same way as the coefficients $P_1$ and $P_2$ of $\delta_{\rm m}^{(2)}$. Following the discussion related to $\delta_{\rm m}^{(2)}$, presented in Section \ref{match}, 
 the perturbation $\delta_{\rm r}^{(2)}$ and its derivatives have to be continuous throughout evolution and, in particular, at  $y=y_\alpha$ .
 In other words, the following matching condition must be satisfied 
 \begin{align}\label{deltarmatch}
  \delta_{\rm r (DRe)}^{(2)}(\bm{k},\tau_\alpha)&=  \delta_{\rm r (Tmesz)}^{(2)}(\bm{k},y_\alpha),\\
\label{deltardotmatch}
     {\left({d \over d y}{ \delta}_{\rm r (DRe)}^{(2)}(\bm{k},\tau)\right)}\Bigg|_{\tau_\alpha}&=    {\left({d \over d y}{\delta}_{\rm r (Tmesz)}^{(2)}(\bm{k},y)\right)}\Bigg|_{y_\alpha},
\end{align}
Also for radiation contribution, we have defined $ \delta_{\rm r (DRe)}^{(2)}(\bm{k},\tau)$ as the radiation perturbation solution during the deep radiation era [i.e. Eq. \eqref{radpertshort} or, equivalently, \eqref{deltary}], while 
$ \delta_{\rm r (Tmesz)}^{(2)}(\bm{k},y)$ is the solution obtained in Eq. \eqref{pert_r-secphase}. 

%\begin{align}\label{deltarmatch}
%  \delta_{\rm r (DRe)}^{(2)}(\bm{k},y_\alpha)&=  \delta_{\rm r (Tmesz)}^{(2)}(\bm{k},y_\alpha),\\\label{deltardotmatch}
%     \dot{ \delta}_{\rm r (DRe)}^{(2)}(\bm{k},y)|_{y_\alpha}&=  \dot{\delta}_{\rm r (Tmesz)}^{(2)}(\bm{k},y)|_{y_\alpha}.
%\end{align}
Using Eq. \eqref{deltary}, the first condition, Eq. \eqref{deltarmatch}, gives us
\begin{align}\label{deltarmatch1}
   &    \frac{12}{k^2y_\alpha^2 } \int^{y_\alpha}_{y_{\rm in}} d \tilde{y}\, \tilde{y}\,S_{\bm{k}}\left({\eta_*\tilde{y} \over 2}\right)
      -\frac{2\sqrt{3}\eta_*}{k}\left[\sin{\frac{k\eta_*y_\alpha}{2\sqrt{3}}}+\frac{2\sqrt{3}}{k\eta_*y_\alpha}\cos{\frac{k\eta_*y_\alpha}{2\sqrt{3}}}\right] \nonumber\\& \times \int^{y_\alpha}_{y_{\rm in}} d \tilde{y} \, \cos{\frac{k\eta_*\tilde{y}}{2\sqrt{3}}}\, S_{\bm{k}}\left({\eta_*\tilde{y} \over 2}\right)
      +\frac{2\sqrt{3}\eta_*}{k}\left[\cos{\frac{k\eta_*y_\alpha}{2\sqrt{3}}}-\frac{2\sqrt{3}}{k\eta_*y_\alpha}\sin{\frac{k\eta_*y_\alpha}{2\sqrt{3}}}\right] \nonumber\\
      & \times \int^{y_\alpha}_{y_{\rm in}}d \tilde{y} \,\sin{\frac{k\eta_*\tilde{y}}{2\sqrt{3}}}\, S_{\bm{k}}\left({\eta_*\tilde{y} \over 2}\right)=A_{\rm r}(\bm{k}) \cos{\left(2\sqrt{\frac{2}{3}}\frac{k}{k_{\rm eq}}\sqrt{1+y_\alpha}\right)} \nonumber\\
      &+B_{\rm r}(\bm{k}) \sin{\left(2\sqrt{\frac{2}{3}}\frac{k}{k_{\rm eq}}\sqrt{1+y_\alpha}\right)},
\end{align}
and the second condition, Eq. \eqref{deltardotmatch}, becomes 
\begin{align}\label{deltarmatch2}
    & -\frac{24}{k^2y_\alpha^3 } \int^{y_\alpha}_{y_{\rm in}}d \tilde{y} \, \tilde{y}\,S_{\bm{k}}\left({\eta_*\tilde{y} \over 2}\right)\nonumber\\&
      -\left[\left(\eta_*^2-\frac{12}{k^2y_\alpha^2 }\right)\cos{\frac{k\eta_*y_\alpha}{2\sqrt{3}}}-\frac{2\sqrt{3}\eta_*}{ky_\alpha}\sin{\frac{k\eta_*y_\alpha}{2\sqrt{3}}}\right]  \int^{y_\alpha}_{y_{\rm in}}d \tilde{y} \, \cos{\frac{k\eta_*\tilde{y}}{2\sqrt{3}}}\, S_{\bm{k}}\left({\eta_*\tilde{y} \over 2}\right) \nonumber\\&
      -\left[\left(\eta_*^2-\frac{12}{k^2y_\alpha^2 }\right)\sin{\frac{k\eta_*y_\alpha}{2\sqrt{3}}}+\frac{2\sqrt{3}\eta_*}{ky_\alpha}\cos{\frac{k\eta_*y_\alpha}{2\sqrt{3}}}\right]   \int^{y_\alpha}_{y_{\rm in}}d \tilde{y} \,\sin{\frac{k\eta_*\tilde{y}}{2\sqrt{3}}}\, S_{\bm{k}}\left({\eta_*\tilde{y} \over 2}\right)\nonumber\\&=2\sqrt{\frac{2}{3}}\frac{k}{k_{\rm eq}} \left[-A_{\rm r}(\bm{k}) \sin{\left(2\sqrt{\frac{2}{3}}\frac{k}{k_{\rm eq}}\sqrt{1+y_\alpha}\right)}+B_{\rm r}(\bm{k}) \cos{\left(2\sqrt{\frac{2}{3}}\frac{k}{k_{\rm eq}}\sqrt{1+y_\alpha}\right)}\right].
\end{align}
Multiplying Eq. \eqref{deltarmatch1} with $$\sin{\left(2\sqrt{\frac{2}{3}}\frac{k}{k_{\rm eq}}\sqrt{1+y_\alpha}\right)}\,,$$   Eq. \eqref{deltarmatch2} with $$\left(2\sqrt{\frac{2}{3}}\frac{k}{k_{\rm eq}}\right)^{-1}\cos{\left(2\sqrt{\frac{2}{3}}\frac{k}{k_{\rm eq}}\sqrt{1+y_\alpha}\right)}\,$$ and adding them, we obtain 
\begin{align}
    B_r(\bm{k})=& -\frac{6\sqrt{6}k_{\rm eq}}{k^3y_\alpha^3 } \cos{\left(2\sqrt{\frac{2}{3}}\frac{k}{k_{\rm eq}}\sqrt{1+y_\alpha}\right)} \int^{y_\alpha}_{y_{\rm in}}d \tilde{y} \, \tilde{y}\,S_{\bm{k}}\left({\eta_*\tilde{y} \over 2}\right) \nonumber\\
    & -\frac{\sqrt{3}}{2\sqrt{2}}\frac{k_{\rm eq}}{k}\cos{\left(2\sqrt{\frac{2}{3}}\frac{k}{k_{\rm eq}}\sqrt{1+y_\alpha}\right)}\Bigg\{\left[\left(\eta_*^2-\frac{12}{k^2y_\alpha^2 }\right)\cos{\frac{k\eta_*y_\alpha}{2\sqrt{3}}}-\frac{2\sqrt{3}\eta_*}{ky_\alpha}\sin{\frac{k\eta_*y_\alpha}{2\sqrt{3}}}\right] \nonumber\\
      & \times  \int^{y_\alpha}_{y_{\rm in}}d \tilde{y} \, \cos{\frac{k\eta_*\tilde{y}}{2\sqrt{3}}}\, S_{\bm{k}}\left({\eta_*\tilde{y} \over 2}\right)
       + \left[\left(\eta_*^2-\frac{12}{k^2y_\alpha^2 }\right)\sin{\frac{k\eta_*y_\alpha}{2\sqrt{3}}}+\frac{2\sqrt{3}\eta_*}{ky_\alpha}\cos{\frac{k\eta_*y_\alpha}{2\sqrt{3}}}\right] \nonumber \\
      &\times
      \int^{y_\alpha}_{y_{\rm in}}d \tilde{y} \,\sin{\frac{k\eta_*\tilde{y}}{2\sqrt{3}}}\, S_{\bm{k}}\left({\eta_*\tilde{y} \over 2}\right)\Bigg\}+\frac{12}{k^2y_\alpha^2 } \sin{\left(2\sqrt{\frac{2}{3}}\frac{k}{k_{\rm eq}}\sqrt{1+y_\alpha}\right)}\int^{y_\alpha}_{y_{\rm in}}d \tilde{y} \, \tilde{y}\,S_{\bm{k}}\left({\eta_*\tilde{y} \over 2}\right)\nonumber\\
      & -\frac{2\sqrt{3}\eta_*}{k}\sin{\left(2\sqrt{\frac{2}{3}}\frac{k}{k_{\rm eq}}\sqrt{1+y_\alpha}\right)}\left[\sin{\frac{k\eta_*y_\alpha}{2\sqrt{3}}}+\frac{2\sqrt{3}}{k\eta_*y_\alpha}\cos{\frac{k\eta_*y_\alpha}{2\sqrt{3}}}\right] \nonumber\\
      & \times \int^{y_\alpha}_{y_{\rm in}}d \tilde{y} \, \cos{\frac{k\eta_*\tilde{y}}{2\sqrt{3}}}\, S_{\bm{k}}\left({\eta_*\tilde{y} \over 2}\right)
      +\frac{2\sqrt{3}\eta_*}{k}\sin{\left(2\sqrt{\frac{2}{3}}\frac{k}{k_{\rm eq}}\sqrt{1+y_\alpha}\right)} \nonumber\\
      & \times \left[\cos{\frac{k\eta_*y_\alpha}{2\sqrt{3}}}-\frac{2\sqrt{3}}{k\eta_*y_\alpha}\sin{\frac{k\eta_*y_\alpha}{2\sqrt{3}}}\right] \int^{y_\alpha}_{y_{\rm in}}d \tilde{y} \,\sin{\frac{k\eta_*\tilde{y}}{2\sqrt{3}}}\, S_{\bm{k}}\left({\eta_*\tilde{y} \over 2}\right)\nonumber\\&
      = \frac{6\sqrt{2}}{k^2y_\alpha^2 }\left[\sqrt{2}\sin{\left(2\sqrt{\frac{2}{3}}\frac{k}{k_{\rm eq}}\sqrt{1+y_\alpha}\right)}-\frac{\sqrt{3}k_{\rm eq}}{ky_\alpha} \cos{\left(2\sqrt{\frac{2}{3}}\frac{k}{k_{\rm eq}}\sqrt{1+y_\alpha}\right)}\right]\nonumber \\
      & \times \int^{y_\alpha}_{y_{\rm in}}d \tilde{y} \, \tilde{y}\,S_{\bm{k}}\left({\eta_*\tilde{y} \over 2}\right)
      -\frac{\sqrt{3}}{k}\Bigg\{\frac{k_{\rm eq}}{2\sqrt{2}}\cos{\left(2\sqrt{\frac{2}{3}}\frac{k}{k_{\rm eq}}\sqrt{1+y_\alpha}\right)} 
      \nonumber\\
      & \times \left[\left(\eta_*^2-\frac{12}{k^2y_\alpha^2 }\right)\cos{\frac{k\eta_*y_\alpha}{2\sqrt{3}}}-\frac{2\sqrt{3}\eta_*}{ky_\alpha}\sin{\frac{k\eta_*y_\alpha}{2\sqrt{3}}}\right]
      + 2\eta_*\sin{\left(2\sqrt{\frac{2}{3}}\frac{k}{k_{\rm eq}}\sqrt{1+y_\alpha}\right)}
      \nonumber\\
      & \times \left[\sin{\frac{k\eta_*y_\alpha}{2\sqrt{3}}}+\frac{2\sqrt{3}}{k\eta_*y_\alpha}\cos{\frac{k\eta_*y_\alpha}{2\sqrt{3}}}\right] \Bigg\} \int^{y_\alpha}_{y_{\rm in}}d \tilde{y} \, \cos{\frac{k\eta_*\tilde{y}}{2\sqrt{3}}}\, S_{\bm{k}}\left({\eta_*\tilde{y} \over 2}\right)\nonumber\\&
      - \frac{\sqrt{3}}{k}\Bigg\{\frac{k_{\rm eq}}{2\sqrt{2}}\cos{\left(2\sqrt{\frac{2}{3}}\frac{k}{k_{\rm eq}}\sqrt{1+y_\alpha}\right)}  \left[\left(\eta_*^2-\frac{12}{k^2y_\alpha^2 }\right)\sin{\frac{k\eta_*y_\alpha}{2\sqrt{3}}}+\frac{2\sqrt{3}\eta_*}{ky_\alpha}\cos{\frac{k\eta_*y_\alpha}{2\sqrt{3}}}\right]\nonumber\\&- 2\eta_*\sin{\left(2\sqrt{\frac{2}{3}}\frac{k}{k_{\rm eq}}\sqrt{1+y_\alpha}\right)}\left[\cos{\frac{k\eta_*y_\alpha}{2\sqrt{3}}}-\frac{2\sqrt{3}}{k\eta_*y_\alpha}\sin{\frac{k\eta_*y_\alpha}{2\sqrt{3}}}\right] \Bigg\}\nonumber \\
      &\times \int^{y_\alpha}_{y_{\rm in}}d \tilde{y} \, \sin{\frac{k\eta_*\tilde{y}}{2\sqrt{3}}}\, S_{\bm{k}}\left({\eta_*\tilde{y} \over 2}\right).
\end{align}
Similarly, multiplying Eq. \eqref{deltarmatch1} with $$\cos{\left(2\sqrt{\frac{2}{3}}\frac{k}{k_{\rm eq}}\sqrt{1+y_\alpha}\right)}\,,$$   Eq. \eqref{deltarmatch2} with $$\left(2\sqrt{\frac{2}{3}}\frac{k}{k_{\rm eq}}\right)^{-1}\sin{\left(2\sqrt{\frac{2}{3}}\frac{k}{k_{\rm eq}}\sqrt{1+y_\alpha}\right)}\,,$$ and subtracting the latter from the former, we obtain 
\begin{align}
    A_{\rm r} (\bm{k})= & \frac{6\sqrt{6}k_{\rm eq}}{k^3y_\alpha^3 } \sin{\left(2\sqrt{\frac{2}{3}}\frac{k}{k_{\rm eq}}\sqrt{1+y_\alpha}\right)} \int^{y_\alpha}_{y_{\rm in}}d \tilde{y} \, \tilde{y}\,S_{\bm{k}}\left({\eta_*\tilde{y} \over 2}\right)\nonumber\\
    & +\frac{\sqrt{3}}{2\sqrt{2}}\frac{k_{\rm eq}}{k}\sin{\left(2\sqrt{\frac{2}{3}}\frac{k}{k_{\rm eq}}\sqrt{1+y_\alpha}\right)} \Bigg\{\left[\left(\eta_*^2-\frac{12}{k^2y_\alpha^2 }\right)\cos{\frac{k\eta_*y_\alpha}{2\sqrt{3}}}-\frac{2\sqrt{3}\eta_*}{ky_\alpha}\sin{\frac{k\eta_*y_\alpha}{2\sqrt{3}}}\right] \nonumber\\
    & \times \int^{y_\alpha}_{y_{\rm in}}d \tilde{y} \, \cos{\frac{k\eta_*\tilde{y}}{2\sqrt{3}}}\, S_{\bm{k}}\left({\eta_*\tilde{y} \over 2}\right)  + \left[\left(\eta_*^2-\frac{12}{k^2y_\alpha^2 }\right)\sin{\frac{k\eta_*y_\alpha}{2\sqrt{3}}}+\frac{2\sqrt{3}\eta_*}{ky_\alpha}\cos{\frac{k\eta_*y_\alpha}{2\sqrt{3}}}\right] \nonumber \\
    & \times \int^{y_\alpha}_{y_{\rm in}}d \tilde{y} \,\sin{\frac{k\eta_*\tilde{y}}{2\sqrt{3}}}\, S_{\bm{k}}\left({\eta_*\tilde{y} \over 2}\right)\Bigg\}+\frac{12}{k^2y_\alpha^2 } \cos{\left(2\sqrt{\frac{2}{3}}\frac{k}{k_{\rm eq}}\sqrt{1+y_\alpha}\right)}\int^{y_\alpha}_{y_{\rm in}}d \tilde{y} \, \tilde{y}\,S_{\bm{k}}\left({\eta_*\tilde{y} \over 2}\right)\nonumber\\
      & -\frac{2\sqrt{3}\eta_*}{k}\cos{\left(2\sqrt{\frac{2}{3}}\frac{k}{k_{\rm eq}}\sqrt{1+y_\alpha}\right)}\left[\sin{\frac{k\eta_*y_\alpha}{2\sqrt{3}}}
      +\frac{2\sqrt{3}}{k\eta_*y_\alpha}\cos{\frac{k\eta_*y_\alpha}{2\sqrt{3}}}\right] \nonumber\\
      & \times \int^{y_\alpha}_{y_{\rm in}}d \tilde{y} \, \cos{\frac{k\eta_*\tilde{y}}{2\sqrt{3}}}\, S_{\bm{k}}\left({\eta_*\tilde{y} \over 2}\right)
      +\frac{2\sqrt{3}\eta_*}{k}\cos{\left(2\sqrt{\frac{2}{3}}\frac{k}{k_{\rm eq}}\sqrt{1+y_\alpha}\right)}\nonumber\\
      &\times\left[\cos{\frac{k\eta_*y_\alpha}{2\sqrt{3}}}-\frac{2\sqrt{3}}{k\eta_*y_\alpha}\sin{\frac{k\eta_*y_\alpha}{2\sqrt{3}}}\right] \int^{y_\alpha}_{y_{\rm in}}d \tilde{y} \,\sin{\frac{k\eta_*\tilde{y}}{2\sqrt{3}}}\, S_{\bm{k}}\left({\eta_*\tilde{y} \over 2}\right)\nonumber\\
       = & \frac{6\sqrt{2}}{k^2y_\alpha^2 }\left[\sqrt{2}\sin{\left(2\sqrt{\frac{2}{3}}\frac{k}{k_{\rm eq}}\sqrt{1+y_\alpha}\right)}+\frac{\sqrt{3}k_{\rm eq}}{ky_\alpha} \sin{\left(2\sqrt{\frac{2}{3}}\frac{k}{k_{\rm eq}}\sqrt{1+y_\alpha}\right)}\right]\nonumber\\
       &\times \int^{y_\alpha}_{y_{\rm in}}d \tilde{y} \, \tilde{y}\,S_{\bm{k}}\left({\eta_*\tilde{y} \over 2}\right) +\frac{\sqrt{3}}{k}\Bigg\{\frac{k_{\rm eq}}{2\sqrt{2}}\sin{\left(2\sqrt{\frac{2}{3}}\frac{k}{k_{\rm eq}}\sqrt{1+y_\alpha}\right)}  \nonumber\\
       &\times \left[\left(\eta_*^2-\frac{12}{k^2y_\alpha^2 }\right)\cos{\frac{k\eta_*y_\alpha}{2\sqrt{3}}}-\frac{2\sqrt{3}\eta_*}{ky_\alpha}\sin{\frac{k\eta_*y_\alpha}{2\sqrt{3}}}\right]\nonumber\\
      &+ 2\eta_*\cos{\left(2\sqrt{\frac{2}{3}}\frac{k}{k_{\rm eq}}\sqrt{1+y_\alpha}\right)}\left[\sin{\frac{k\eta_*y_\alpha}{2\sqrt{3}}}+\frac{2\sqrt{3}}{k\eta_*y_\alpha}\cos{\frac{k\eta_*y_\alpha}{2\sqrt{3}}}\right] \Bigg\} \nonumber \\
      &\times \int^{y_\alpha}_{y_{\rm in}}d \tilde{y} \, \cos{\frac{k\eta_*\tilde{y}}{2\sqrt{3}}}\, S_{\bm{k}}\left({\eta_*\tilde{y} \over 2}\right) + \frac{\sqrt{3}}{k}\Bigg\{\frac{k_{\rm eq}}{2\sqrt{2}}\cos{\left(2\sqrt{\frac{2}{3}}\frac{k}{k_{\rm eq}}\sqrt{1+y_\alpha}\right)} \nonumber \\
       & \times \left[\left(\eta_*^2-\frac{12}{k^2y_\alpha^2 }\right)\sin{\frac{k\eta_*y_\alpha}{2\sqrt{3}}}+\frac{2\sqrt{3}\eta_*}{ky_\alpha}\cos{\frac{k\eta_*y_\alpha}{2\sqrt{3}}}\right]\nonumber\\&- 2\eta_*\sin{\left(2\sqrt{\frac{2}{3}}\frac{k}{k_{\rm eq}}\sqrt{1+y_\alpha}\right)}\left[\cos{\frac{k\eta_*y_\alpha}{2\sqrt{3}}}-\frac{2\sqrt{3}}{k\eta_*y_\alpha}\sin{\frac{k\eta_*y_\alpha}{2\sqrt{3}}}\right] \Bigg\} \nonumber\\
       &\times \int^{y_\alpha}_{y_{\rm in}}d \tilde{y} \, \sin{\frac{k\eta_*\tilde{y}}{2\sqrt{3}}}\, S_{\bm{k}}\left({\eta_*\tilde{y} \over 2}\right).
\end{align}
In this case we are able to  obtain the solution of $ \delta_{\rm r }^{(2)}$ from end of inflation to CDM epoch. 

\section{ Setup of the initial conditions}\label{phi0}

This Appendix is devoted to addressing some issues relating to the initial conditions, i.e. at the end of Inflation, used in this paper, adopted in Section \ref{Section2}. First of all, let us focus on Eqs. \eqref{all00}, \eqref{allijtr} and \eqref{chi_ij-1}.
Substituting $\nabla^2 (\phi^{(2)}+\nabla^2{\chi^{||(2)}}/6)$ from \eqref{all00} in \eqref{allijtr}, we have 
\begin{align}
& {\phi^{(2)}}''+ \mathcal{H}{\phi^{(2)}}'-\frac{1}{3}\mathcal{H}\chi^{kl}{\chi_{kl}}'+\frac{1}{3}\chi^{kl}\nabla^2\chi_{kl}+\frac{1}{6}{\chi^{kl}}'{\chi_{kl}}'=\frac{4 \pi G a^2}{3} \left(2\delta_{\rm r}^{(2)}\overline{\rho}_{\rm r}+\delta_{\rm m}^{(2)}\overline{\rho}_{\rm m}\right)\,.
        \end{align} 
The fourth additive term may be rewritten by replacing $\nabla^2\chi_{kl}$ with ${\chi_{ij}}''+2\mathcal{H}{\chi_{ij}}'$, see Eq. \eqref{chi_ij-1}, and  the above equation can be written in the following way
\begin{align}\label{deq}
& {\phi^{(2)}}''+ \mathcal{H}{\phi^{(2)}}'+\frac{1}{3}\mathcal{H}\chi^{kl}{\chi_{kl}}'+\frac{1}{3}\chi^{kl}{\chi_{kl}}''+\frac{1}{6}{\chi^{kl}}'{\chi_{kl}}'=\frac{4 \pi G a^2}{3} \left(2\delta_{\rm r}^{(2)}\overline{\rho}_{\rm r}+\delta_{\rm m}^{(2)}\overline{\rho}_{\rm m}\right)\,.
        \end{align} 
In this paper, we are setting the initial conditions at the end of inflation. Before discussing them, two important observations are in order.
\begin{itemize}
    \item In several expressions considered in this work [see, e.g., the above Eq. \eqref{deq}], in each  source  term defined in  Fourier space, we have a loop integral which runs at all scales (or, equivalently, we are integrating over the whole frequency range of GW modes). However, at $\eta=\eta_{\rm in}$, all GW modes that we are interested in here are already outside the the horizon scale $1/{\cal H}(\eta_{\rm in})$. This means that, if $\bm k$ corresponds to the induced scalar modes and ${\bm q}$ is the loop momentum, the loop integral is truncated on horizon scales and, consequently, $q$ and  $|{\bm k}-{\bm q}|$ of  tensor perturbations cannot be larger than ${\cal H}(\eta_{\rm in})$ (see also Fig. \ref{Figure}). In conclusion, at initial time  GW modes will be frozen outside the horizon. This point is crucial for the below discuss.
    \item As we already pointed out in the main text, assuming initial adiabatic conditions and the synchronous comoving gauge fixed here, the induced scalar modes will be zero because the contribution will come only after horizon entry, i.e. when GW tensor perturbations start oscillating \cite{Watanabe:2006qe}. This imply that, at $\eta_{\rm in}$ and $k < {\cal H}(\eta_{\rm in})$, $\delta_{\rm m}^{(2)}$, $\delta_{\rm r}^{(2)}$, $v^{(2)}_{{\rm r}}$ and ${v^{(2)}_{\rm r}}'$ can be set to zero.
\end{itemize}
      As a result of these comments,  at $\eta=\eta_{\rm in}$, 
      %following the argument above, then, 
      ${\chi^{kl}}_0'$ and ${\chi^{kl}}_0''$  can set to zero
      in \eqref{deq}. 
      (As we already pointed out in the main text, also here the subscript '$0$' denotes the initial conditions, i.e. the end of inflation when $\eta=\eta_{\rm in}$.) Then, \eqref{deq} reads
        \begin{equation}\label{phi0eq}
            {\phi_0^{(2)}}''+ \mathcal{H}{\phi_0^{(2)}}'=0\;.
        \end{equation}
%        Note that here we have neglected $\nabla^2 \phi_0$ (precisely, we have assumed that, in Fourier space, $|k^2\phi_0| \ll |\mathcal{H}{\phi_0^{(2)}}'| $)  because the scalar modes are well outside the horizon.
        In this case, Eq. \eqref{phi0eq} suggests choosing
        %which is consistent with
        $\phi_0=\rm const.$ (in time). 
        This conclusion can be further justified and confirmed if we also look at Eqs. (\ref{deltam-evol}) and (\ref{radc}) at $\eta=\eta_{\rm in}$. 
        %Following all the arguments till now, we can now proceed to \eqref{all0i} to get the relation
        Now let us examine $\nabla^2 (\phi^{(2)}+\nabla^2{\chi^{||(2)}}/6)$ in Fourier space, at the initial time. Following all the arguments made so far, from Eq. \eqref{all0i}, we can get directly the relation
        \begin{equation}\label{d1}
           \phi^{(2)}_{\bm{k} 0}-\frac{k^2}{6}{\chi_{\bm{k} 0}^{||(2)}}={\cal F}(\bm{k}),
        \end{equation}
      where  ${\cal F}(\bm{k})$, being a constant of time, can be derived  explicitly  in configuration space from Eq. \eqref{all00}. Defining  ${\cal F}(\bm{x})$  as its Fourier inverse, from \eqref{all00}, we find
        \begin{equation}
          {\cal F}(\bm{x})=\frac{\nabla^{-2}}{4}\left(\chi_0^{ik,l}\chi_{0li,k}-\frac{3}{2}\chi_0^{kl,i}\chi_{0kl,i}\right).
        \end{equation}
        Note that the above results have been obtained in whole generality. However there is a residual ambiguity which could be related to the gauge chosen here in this work. 
        For instance, one could fix this ambiguity imposing that $\phi^{(2)}_0=0$. Then 
        \begin{equation}
        \chi_0^{||(2)}=\frac{3}{2}\nabla^{-4}\left(\chi_0^{ik,l}\chi_{0li,k}-\frac{3}{2}\chi_0^{kl,i}\chi_{0kl,i}\right)\;.
        \end{equation}
       This concludes the discussion related to the issue of how to set the initial conditions of the paper. 
        %%\begin{comment}  
      %  \section{About $h_{ij}$ during DM+DE}
        
       % I might suggest using the following approximate solution
       % \begin{equation}
       % \chi_{ij}(\eta) \sim \left[C_1^{\pm} + C_2^{\pm} \int_{a_{\rm rec}}^{a(\eta)} {d\tilde a \over \tilde { \cal H} \tilde a^3} \right] \exp{(\pm i k \eta)}\;.
       % \end{equation}
        
      %  In this case, using a correct initial condition which connects this approximate solution with the Watanabe \& Komatsu solutions (a particular $\eta$ such that $\eta_{\rm erc}<\eta \ll \eta_\Lambda$), maybe we can obtain a new $\cal T$....
        
      %  If you do not like $\exp{(\pm i k \eta)}$ we can use $\sin$ and $\cos$ relations...
        
      %  \end{comment}
\end{appendices}

\printbibliography
\end{document}